\documentclass[aps,prd,twocolumn,nofootinbib,preprintnumbers,floatfix]{revtex4-1}

\usepackage[T1]{fontenc}
\usepackage[normalem]{ulem}
\usepackage[caption=false]{subfig}
\usepackage{enumerate}  
\usepackage{enumitem}  
\usepackage[colorlinks,citecolor=blue,urlcolor=blue,linkcolor=blue]{hyperref}
\usepackage{amsmath,amssymb,graphicx,bigstrut,bm}
\usepackage{slashed}
\usepackage{mathtools}
\usepackage{refcount}
\usepackage[usenames,dvipsnames,svgnames,table]{xcolor}
\usepackage{footmisc} 

\begin{document}
\newcommand{\del}{\partial}
\def\hhref#1{[\href{http://arxiv.org/abs/#1}{arXiv:#1}]} 
\def\hhhref#1{\href{http://arxiv.org/abs/#1}{arXiv:#1}} 
\newcommand{\nn}{\nonumber}
\newcommand{\alphaIR}{\alpha_\textsc{ir}}
\newcommand{\alphaWW}{\alpha_\textsc{ww}}
\newcommand{\ltextfrac}[2]{\raisebox{-0.2mm}{\large $\frac{#1}{#2}$}}
\newcommand{\Ltextfrac}[2]{\raisebox{-0.4mm}{\Large $\frac{#1}{#2}$}}
\newcommand{\smallfrac}[2]{\mbox{\small $\displaystyle \frac{#1}{#2}$}}
\newcommand{\sbiggl}{\mbox{\small $\displaystyle \biggl\{ $}}
\newcommand{\sbiggr}{\mbox{\small $\displaystyle \biggr\} $}}
\newcommand{\ccdot}{\hspace*{-0.3mm}\cdot\hspace*{-0.3mm}}
\newcommand{\hsp}[1]{\hspace*{#1 mm}}

\newcommand*\blue[1]{\textcolor{blue}{#1}}
\newcommand*\red[1]{\textcolor{red}{#1}}
\newcommand*\purple[1]{\textcolor{BlueViolet}{#1}}
\newcommand*\green[1]{\textcolor{OliveGreen}{#1}}
\setlength{\unitlength}{1mm}
\preprint{SI-HEP-2018-09, QFET-2018-05}
\preprint{ADP-18-3/T1051}

\title{Crawling technicolor}
\author{O.~Cat\`{a}}
\email[]{oscar.cata@uni-siegen.de}
\affiliation{Theoretische Physik 1, Universit\"at Siegen,\\
Walter-Flex-Stra\ss e 3, D-57068 Siegen, Germany}

\author{R.~J.~Crewther}
\email[]{rodney.crewther@adelaide.edu.au}
\affiliation{CSSM and
ARC Centre of Excellence for Particle Physics at the Tera-scale,  \\
Department of Physics, University of Adelaide,
Adelaide SA 5005 Australia}

\author{Lewis~C.~Tunstall}
\email[]{tunstall@itp.unibe.ch}
\affiliation{Albert Einstein Center for Fundamental Physics, Institute for Theoretical
Physics,  \\
University of Bern, Sidlerstrasse 5, CH--3012 Bern, Switzerland}

\begin{abstract}
    We analyze the Callan-Symanzik equations when scale
    invariance at a nontrivial infrared (IR) fixed point $\alphaIR$
    is realized in the Nambu-Goldstone (NG) mode. As a result, Green's
    functions at $\alphaIR$ do not scale in the same way as for the conventional
    Wigner-Weyl (WW) mode. This allows us to propose a new mechanism
    for dynamical electroweak symmetry breaking where the running
    coupling $\alpha$ ``crawls'' towards (but does not pass) $\alphaIR$
    in the exact IR limit. The NG mechanism at $\alphaIR$ implies the
    existence of a massless dilaton $\sigma$, which becomes massive for IR
    expansions in $\epsilon \equiv \alphaIR - \alpha$ and is identified
    with the Higgs boson. Unlike ``dilatons'' that are close to a WW-mode
    fixed point or associated with a Coleman-Weinberg potential, our
    NG-mode dilaton is genuine and hence naturally light. Its (mass)$^2$ is
    proportional to $\epsilon \beta'(4+\beta')F_\sigma^{-2}
    \langle\hat{G}^2\rangle_{\text{vac}}$, where $\beta'$ is the (positive)
    slope of the beta function at $\alphaIR$, $F_\sigma$ is the dilaton decay
    constant and $\langle\hat{G}^2\rangle_{\text{vac}}$ is the technigluon
    condensate. Our effective field theory for this works because it respects
    Zumino's consistency condition for dilaton Lagrangians. We
    find a closed form of the Higgs potential with $\beta'$-dependent
    deviations from that of the Standard Model. Flavor-changing neutral
    currents are suppressed if the crawling region $\alpha \lesssim
    \alphaIR$ includes a sufficiently large range of energies above the TeV
    scale. In Appendix A, we observe that, contrary to folklore,
    condensates protect fields from decoupling in the IR limit.
\end{abstract}

\maketitle
\tableofcontents

\vfill

\section{WW or NG mechanism at fixed points?}
\label{WWorNG}
The discovery of the Higgs boson has focussed attention on strongly
coupled electroweak theories that can produce a light scalar. Crawling
technicolor (TC) is a new proposal for this.

The main idea of crawling TC is that there is a conformal limit of dynamical
electroweak theory at which the Higgs boson corresponds to a zero-mass
dilaton. This differs fundamentally from recent work on ``dilatonic''
walking gauge theories \cite{Appel10,Yam11,Appel13,Yam14,Golt16} in
that we have a true dilaton: it does not decouple in the relevant
conformal limit.

Modern approaches to the conformal properties of field theories depend
on a key assertion from long ago: renormalization destroys the
conformal invariance of a theory at all couplings $\alpha$ except at
fixed points where the $\psi$ function of Gell-Mann and Low or the
related $\beta$ function of Callan and Symanzik (CS) vanishes.

At a fixed point, exact conformal invariance corresponds to the limit
$\theta^\mu_\mu \to 0$, where $\theta_{\mu\nu}$ is the energy-momentum
tensor (improved \cite{CCJ70} when scalar fields are present). Like
other global symmetries, this symmetry can be realized in two
ways~\cite{MGM69}:
\begin{enumerate}[label=(\arabic*)]
\item  The Wigner-Weyl (WW) mode, where conformal symmetry is
  manifest, Green's functions exhibit power-law behavior,  and all
  particle masses go to zero;
\item  The Nambu-Goldstone (NG) mode, where there is a massless
  scalar boson of the NG type (a genuine dilaton) that allows other
  masses to be nonzero.
\end{enumerate}
There are no theoretical grounds for preferring one mode over the
other: consistent model field theories that exhibit scale invariance
in either the WW or NG mode exist. The choice ultimately
depends on phenomenological requirements.

Dilaton Lagrangians were invented long ago \cite{Nambu68,Isham70b,
Ell70,Zum70,Ell71}. They were used recently to construct chiral-scale
perturbation theory \cite{CT1,CT2,CT3} for three-flavor quantum
chromodynamics (QCD) with a nonperturbative infrared (IR) fixed point.

Nevertheless, most theoretical discussions of IR fixed points, such as
all work on dynamical electroweak symmetry breaking since 1997 \cite{Appel97},
implicitly assume that the WW mode of exact scale invariance is realized
at the fixed point. This is natural if perturbation theory is the guide,
since the NG mode is necessarily nonperturbative. This choice was also
influenced by Wilson's pioneering work on ultraviolet (UV) fixed
points~\cite{KGW71}. As he noted in footnote 21 of Ref.~\cite{KGW71},
the NG scaling mode is a phenomenological possibility but he had no
way of applying his methods to that case. Accordingly, he designed his
theoretical framework for the WW mode and required \cite{WK74} that
the nonlocality of rescaled interaction Hamiltonians be short range.
Subsequent observations in lattice QCD of long-range effects such as
pions, which are not an obvious consequence of Wilson's method, indicate that a
self-consistent procedure to replace the Wilsonian framework when
dynamics chooses the NG scaling mode may not be necessary after all.%
\footnote{Wilson's framework has recently been used to analyze the NG
mode at a UV fixed point in the $O(N)$ model in three dimensions
\cite{Mar17}. In practice, the NG mode is more practical for IR fixed
points because soft-dilaton theorems are derived from low-energy
expansions.}

There is extensive theoretical and phenomenological interest in the
possibility that $\alpha$ runs to an IR fixed point in non-Abelian
gauge theories. Investigations of this type should be distinguished
according to the manner in which conformal symmetry is realized.

The WW mode is associated with the \emph{conformal window}, where the
signal for a fixed point is the scaling of Green's functions. For $N_f$
fermion gauge triplets, WW-mode fixed points are seen in lattice
studies~\cite{Appel08,Appel09,Del10,Del14,DeGrand} in the range
$9 \lessapprox N_f \leqslant 16.$ The lower edge of the conformal
window is thought to lie between $N_f = 8$ and $N_f = 12$, with the
value $N_f = 12$ being debated currently \cite{Lin15,Fod16,Has16}. At a
WW-mode fixed point $\alphaWW$, massive particles and all types of NG
bosons are forbidden.

\begin{figure*}[t]
\centering\includegraphics[scale=0.69]{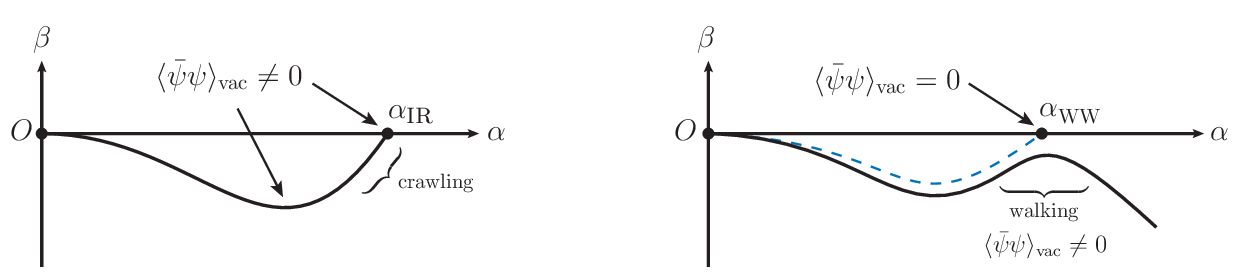}
\caption{Crawling and walking scenarios for the TC $\beta$ function
in $SU(3)$ gauge theories with $N_f$ Dirac flavors. Our proposal is
shown in the left diagram: for small $N_f$ values outside the conformal
window, a fermion condensate $\langle\bar\psi\psi\rangle_\text{vac}
\neq 0$ forms at nonzero coupling and remains nonvanishing at the
IR fixed point $\alphaIR$ (NG mode). The solid curve in the right
diagram is for a walking gauge theory on the lower edge of the
conformal window (large $N_f$ consistent with $\langle\bar{\psi}\psi
\rangle_\text{vac} \not= 0$). The dashed curve is for a theory with
a still larger $N_f$ inside the conformal window; it has an IR fixed
point $\alphaWW$ where scale invariance is manifest (WW mode):
$\langle\bar{\psi}\psi\rangle_\text{vac} = 0$. The gauge coupling
$\alpha$ of the solid curve walks past $\alphaWW$ and continues into
its IR region $\alpha \gg \alphaWW$, where its $|\beta|$ is
assumed to be large and (say) linear.}
\label{fig:beta}
\end{figure*}

The NG mode corresponds to small values of $N_f$ \emph{outside}
the conformal window. Much of this article is devoted to explaining
why this possibility is so often overlooked. In particular, i) the
lattice results above are not applicable because Green's functions do
not scale at a fixed point in the NG mode (Secs.\ \ref{NGsolutions} and
\ref{lattice}), and ii) neither confinement nor dimensional
transmutation can be used to prove anything about the IR running of
$\alpha$ (Sec.\ \ref{NGsolutions} and Appendix \ref{decouple}).
Indeed, there have been many attempts (reviewed in Ref.~\cite{Deur16}) to
find IR fixed points for small $N_f$, but the outcome is unclear:
there is no reliable theory of nonperturbative gauge theory beyond
the lattice, and lattice investigations of IR behavior for small $N_f$
are in their infancy. The signal for a fixed point in the NG mode would be
either $\alpha$ tending to a constant value $\alphaIR$, or better
(given the scheme dependence of $\alpha$), the presence of a light
scalar particle (a pseudodilaton $\sigma$) with $M_{\sigma}^2$
linearly dependent on the techniquark mass $m_\psi$ as the TC limit
$m_\psi \to 0$ is approached. Then conformal symmetry is hidden, so
particle masses and scale condensates such as%
\footnote{A misconception that fermion condensates decouple in
the IR limit has crept into the literature; reasons why that idea fails
are given in Appendix \ref{decouple}. Having the chiral condensate
act as a scale condensate was proposed for strong interactions in
Refs.~\cite{Ell70,Cre70}.  This was later extended to QCD in chiral-scale
perturbation theory \cite{CT1,CT2,CT3}, of which crawling TC is a
technicolored analogue. Reference~\cite{Golt16} cited \cite{CT1,CT2,CT3}
as forerunners for their TC theory, but the IR fixed point considered
in Ref.~\cite{Golt16} is actually in the WW mode, as in walking
TC. \label{IRcond}}
$\langle\bar{\psi}\psi\rangle_\text{vac}$ can be generated dynamically
in the conformal limit $\alpha \rightharpoondown \alphaIR$,
as in the left-hand diagram of Fig.\ \ref{fig:beta}.

The result is a new theoretical possibility which we call
``crawling technicolor''. The Higgs boson corresponds to the
dilaton of the scaling NG mode at $\alphaIR$. Its small mass is
due to the proximity of $\alpha$ to $\alphaIR$ at the Standard
Model (SM) energy scale, which is IR relative to the TC scale. Unlike all other
Higgs-boson theories, crawling TC is an expansion about a limit
$\dot{D} \to 0$ with $D|\text{vac}\rangle \not= 0$, where $D$
generates dilatations. This theory is unique in being an expansion
about a genuine scaling limit in the NG mode with a genuine dilaton.
It has its own phenomenology (Secs.\ \ref{New}--\ref{conclusions}),
distinct from all others.

Unlike WW-mode fixed points, it is not possible to understand this
scenario from a perturbative point of view. Already, small-$N_f$
lattice studies have shown that the dynamics of gauge bosons and
fermions with zero Lagrangian (or ``current'') masses can drive
$\alpha$ while producing hadronization, a nonperturbative effect.
These dynamical variables do not drop out of the analysis because
TC hadrons are created. They will remain the basic variables
of any nonperturbative method to show that $\alphaIR$ exists,
irrespective of whether one has written down an effective low-energy
theory or not. The effective theory will be the result, not the cause,
of the existence of $\alphaIR$.

The dynamical setting of our theory requires an analysis of the
CS equations near IR fixed points in the NG mode, something that,
to the best of our knowledge, has not been attempted before.
This topic is introduced in Sec.\ \ref{NGsolutions}.

The main result is that conventional scaling equations are replaced
by soft-dilaton theorems. That is why NG-mode scale invariance
produces scale-dependent amplitudes. It reinforces a key point
made above: lattice investigations of the conformal window
\cite{Appel08,Appel09,Del10,DeGrand,Has16,Appel97}
assume a power-law behavior for Green's functions at fixed points,
so they \emph{do not} exclude NG-mode fixed points occurring
at small $N_f$ values \cite{CT2,Del14}.

Section \ref{New} introduces crawling TC, a new dynamical mechanism
for electroweak theory. As indicated above, we assume the existence
of an IR fixed point $\alphaIR$ in the NG mode at which \emph{both}
electroweak and conformal symmetry are hidden.\footref{IRcond}
This happens if there is a fermion condensate
$\langle\bar\psi\psi\rangle_\text{vac}\neq 0$ at $\alphaIR$.
Crawling TC differs from the standard theory---walking TC---in several
key ways that are summarized in Fig.\ \ref{fig:beta} (left diagram
for crawling TC and right diagram for walking TC). In crawling TC, the
Higgs boson is identified as the pseudodilaton for $\alpha \lesssim
\alphaIR$. Unlike other ``dilaton'' proposals, our pseudodilaton does
\emph{not} decouple at $\alphaIR$ and so we can legitimately
argue that it is naturally light for $\alpha \lesssim \alphaIR$. An
explicit formula is derived for the pseudodilaton mass in terms of
the parameters of the underlying gauge theory at the fixed point;
this follows from a direct application of the CS analysis of Sec.\
\ref{NGsolutions}.

This leads to the following general observations (Sec.\ \ref{notscalons}):
\begin{enumerate}[label=(\arabic*)]
\item  A careful distinction must be made between a theory like crawling TC
where \emph{exact} scale invariance is realized in the NG mode, and a large
class of theories based on scalons \cite{scalon}. Scalons are not genuine
dilatons because the scale-invariant limit in which they become exactly
massless is in the WW mode characteristic of an unconstrained
polynomial Lagrangian.
\item  In a Lagrangian formalism, a scaling NG mode is possible if a real
scalar field $\chi$ that scales homogeneously obeys the scale-invariant
constraint $\chi > 0$, e.g.\ when written $\sim \exp(\sigma/F_\sigma)$
in terms of an unconstrained field $\sigma$. However, a scaling NG mode is
guaranteed to exist only if \emph{amplitudes} are shown to depend on
dimensionful constants in the scale-invariant limit.
\item  In 1970, Zumino (on page 472 of \cite{Zum70}) observed that
dilaton Lagrangians are consistent only if $\phi^4$ interactions
disappear in the limit of scale invariance. That avoids problems with
the conformal NG mode of $\lambda\phi^4$ theory found by Fubini 6 years
later \cite{Fub76}---a point largely overlooked since then.
\item  Zumino's condition is stable under NG-mode renormalization
of the nonlinear theory, where NG bosons couple via derivative interactions.
\end{enumerate}

Following brief remarks about phenomenology in Sec.~\ref{comments},
the construction of the low-energy effective field theory (EFT) for
crawling TC is considered in Sec.~\ref{EFT}.  The resulting EFT
looks like an electroweak chiral Lagrangian \cite{Appel80,Long80,Buch12}
with a generic Higgs-like scalar field $h$
\cite{Feru92,Bagg94,Koul93,Burg99,Wang06,Grin07,Contino10,
Alon12,Buch13,Appel17,Appel17a}, but in our theory, the NG mode for
exact scale invariance requires us to constrain $h$ and verify that
the equivalence theorem permits our change of field variables
$\sigma \to h$. As a result, we obtain a closed form for the Higgs
potential as a function of $h$. It differs from the SM Higgs potential
by terms depending on $\beta'$.

Section \ref{lattice} contains a discussion of signals for NG-mode fixed points
which may be seen in lattice investigations. In particular, we note that
observations \cite{Aoki14,Aoki16,Appel16,Appel19} of a light scalar
particle  for $N_f= 8$ flavors may indicate the presence of an NG-mode
IR fixed point in the $N_f = 8$ theory. Prompted by recent work 
\cite{Golt16,Appel17,Appel17a} on ``dilaton-based'' potentials, we    
consider testing our Higgs potential on the lattice in order to
determine $\beta'$.

The main text concludes in Sec.\ \ref{conclusions} with a brief review
of the key points and an analysis showing that the effects of
flavor-changing neutral currents (FCNCs) can be naturally
suppressed  in crawling TC.

There are five appendices. Appendix \ref{decouple} shows that
the assertion\footref{IRcond} that condensates decouple in the IR
limit is underivable and contradicts QCD. Appendix \ref{chiral} 
reviews the original current-algebraic approach to soft-pion theorems 
and their extension to scale \cite{Mack68,Cre70,Carr71} and conformal 
\cite{RJC71,PdiV16,PdiV17} symmetry. Appendix \ref{technigluon}
examines how gluon and technigluon condensates may be defined
without relying on perturbative subtractions. Appendix \ref{app:ward}
describes the NG-mode scale-invariant world at $\alphaIR$: most
amplitudes depend on dimensionally transmuted masses, but coefficient
functions in short-distance expansions are shown to obey the same
scaling and conformal rules as leading singularities in WW-mode theories.
Finally, Appendix \ref{app:scaling} reviews formulas for the anomalous
dimension of the trace-anomaly operator.

In standard terminology, a symmetry realized in the NG mode is said to be
``spontaneously broken.'' As noted by Dashen long ago~\cite{Dash69},
this can be misleading: a global symmetry in the NG mode is \emph{hidden},
not broken. Similarly, the term ``electroweak symmetry breaking''
misleads, since gauge-invariant physical quantities are necessarily
invariant under the global chiral subgroup of the local gauge group.
Of course, all of this is well known. However, in crawling TC, we have
to deal with the NG mode not just for chiral invariance but also
for the less familiar case of scale invariance as well. In this paper, we
take care to avoid the terms ``spontaneous'' and ``electroweak
symmetry breaking'' because, in a scaling context, they are so easily
confused with explicit symmetry breaking.

Throughout, the gauge constant $g$ and coupling $\alpha = g^2/(4\pi)$
refer to TC. Our notation for the gluon and electroweak gauge fields
will be $G^A_\mu, W^a_\mu, B_\mu$, with $g_s$, $g_w$, $g_w^\prime$ and
$G^A_{\mu\nu}, W^a_{\mu\nu}, B_{\mu\nu}$ for the corresponding coupling
constants and field-strength tensors. To indicate TC fields, we will
add a hat, i.e.\ $\hat{G}^A_\mu$ and $\hat{G}^A_{\mu\nu}$. For symbols
like $\theta_{\mu\nu}$ and the dilaton $\sigma$, we will let the context
distinguish between TC and QCD. Dilaton decay constants are $F_\sigma$
for TC and $f_\sigma$ for QCD \cite{CT1,CT2,CT3}:
\begin{align}
\langle\text{vac}|\theta_{\mu\nu}|\sigma(q)\rangle^{}_{\textsc{tc}}
&= \bigl(F_\sigma\bigl/3\bigr)\bigl(q_\mu q_\nu - g_{\mu\nu}q^2\bigr)\,,
\nonumber \\
\langle\text{vac}|\theta_{\mu\nu}|\sigma(q)\rangle^{}_{\textsc{qcd}}
&= \bigl(f_\sigma\bigl/3\bigr)
 \bigl(q_\mu q_\nu - g_{\mu\nu}q^2\bigr) \,.
\label{Fsigma}                                            
\end{align}
The phases of $|\sigma\rangle_{\textsc{tc,\,qcd}}$ are chosen such that
$F_\sigma$ and $f_\sigma$ are positive.

\section{NG-mode solutions of the CS equations}
\label{NGsolutions}
The basic idea of this section is to understand the CS equation as a
Ward identity for scale transformations near an IR fixed point in the NG
mode. The method is similar to the original non-Lagrangian procedure
for analyzing chiral condensates; see Appendix \ref{chiral} for a review.

Let us begin with TC where the Lagrangian is chiral $SU(N_f)_L \times
SU(N_f)_R$ symmetric. For scale transformations, the relevant operator
is the divergence of the dilatation current
${\cal D}_\mu =  x^\alpha \theta_{\alpha\mu}$. It is governed by the
trace anomaly \cite{Mink76,Adler77,Niels77,Coll77}, which for massless
fermion fields takes the form
\begin{equation}
\del^\mu{\cal D}_\mu
=\, \theta^\mu_\mu\,
=\, \frac{\beta(\alpha)}{4\alpha}\Bigl\{\hat{G}_{\mu\nu}^{A}\hat{G}^{A\mu\nu}
-\bigl\langle \hat{G}_{\mu\nu}^{A}\hat{G}^{A\mu\nu}\bigr\rangle_\text{vac}
\Bigr\}\,,
\label{anom2}
\end{equation}
where $\langle \hat{G}^2 \rangle_\text{vac}$ is the technigluon
condensate $\langle\text{vac}|\hat{G}^2|\text{vac}\rangle$ and
$|\text{vac}\rangle$ is the nonperturbative vacuum state. We apply
Eq.~(\ref{anom2}) at zero momentum transfer, where there is a
standard prescription%
\footnote{This is an example of the renormalized action principle.
The simplest version of it \cite{BM77} is for minimal schemes such
as dimensional renormalization, where gauge-invariant composite
operators have block-diagonal renormalization matrices.
See the discussion in Ref.~\cite{Spiri84}.}
for a connected insertion of the renormalized action into Green's functions:
\begin{equation}\label{prescr}
i\left.\alpha\frac{\del}{\del\alpha}\right|_{\mu,\,\vec{\cal J}}
\longleftrightarrow\, -\, \frac{1}{4}
 \left.\int\!d^4x\,\hat{G}^2\right|_\text{conn} \,.
\end{equation}
Here $\alpha$ is the renormalized TC coupling, $\mu$ is the
renormalization scale, and $\vec{\cal J}$ are source functions for
renormalized spectator operators $\{{\cal O}_n\}$. This prescription
is valid provided that each ${\cal O}_n$ is constructed from covariant
derivatives but is otherwise $\alpha$ independent in the following sense.

Briefly, ignoring details of gauge fixing and ghosts, the rule
(\ref{prescr}) is a consequence \cite{Novikov81} of absorbing
the bare coupling constant $g_B$ into the functional measure
\begin{equation}
{\cal D}\hat{A}_\mu{}^{}_B \to {\cal D}\hat{\cal A}_\mu{}^{}_B
\ , \quad \hat{\cal A}_\mu{}^{}_B = g_B \hat{A}_\mu{}^{}_B \,.
\end{equation}
Then all operators $\bigl({\cal O}_n\bigr)_B$ constructed from covariant
derivatives alone, including 
\begin{equation}
\hat{\cal G}_{\mu\nu}{}^{}_B = g_B \hat{G}_{\mu\nu}{}^{}_B \,,
\end{equation}
are $g_B$ independent. In the action, all dependence on $g_B$ appears
as a constant source $1/g^2_B$ for $-\frac{1}{4}\hat{\cal G}^2_B$. A
textbook argument \cite{Collins84} relates terms linear in the sources
$1/g^2_B$ and $\vec{\cal J}_{\! B}$ to their renormalized
counterparts:
\begin{align}\label{action}
\int&\!d^4x\,\Bigl\{-\smallfrac{1}{4g_B^2}\hat{\cal G}^2_B
+ \vec{\cal J}_{\! B}(x)\cdot\vec{\cal O}_B(x)\Bigr\} \notag \\
&= \int\!d^4x\,\Bigl\{-\smallfrac{1}{4g^2}\hat{\cal G}^2
+ \vec{\cal J}(x)\cdot\vec{\cal O}(x)\Bigr\} + O\bigl({\cal J}^2\bigr) \,.
\end{align}
Then the rule follows from $\alpha = g^2/(4\pi)$. The term $O({\cal J}^2)$
represents subtractions of quadratic or higher order in ${\cal J}$ for
multiple insertions of the composite operators ${\cal O}_n$.

In our analysis, the operator in the technigluon condensate appears as
a spectator, so an $\alpha$-independent choice such as
\begin{equation}\label{indep}
{\cal O} = \smallfrac{1}{4\pi^2}\hat{\cal G}^2
= \smallfrac{\alpha}{\pi}\hat{G}^2
\end{equation}
is appropriate when using Eq.~(\ref{prescr}); (the normalization is that
originally chosen \cite{SVZ_JETPlett78} for the gluon condensate). In
Appendix \ref{technigluon}, we show that there is a multiplicatively
renormalizable version of $\cal O$, i.e.\ one which does not mix with
the identity operator $I$. These twin requirements are essential if ambiguities
in the definitions of gluon and technigluon condensates are to be avoided.

Consider the CS equation for (say) the vacuum expectation value (VEV)
of ${\cal O}(x)$:
\begin{equation}
\left\{\mu\frac{\del\ }{\del\mu} + \beta(\alpha)
\frac{\del\;\ }{\del\alpha} + \gamma_{\cal O}(\alpha)\right\}
\bigl\langle\text{vac}\bigl|{\cal O}(0)\bigr|\text{vac}\bigr\rangle
= 0 \,.
\label{CS}
\end{equation}
Let us move the $\beta\del/\del\alpha$ term to the right-hand side of
this formula. Then Eqs.~(\ref{anom2}) and (\ref{prescr}) imply
that the right-hand side is given by a suitably renormalized zero-momentum
insertion of $-i\theta^\mu_\mu$:
\begin{align}
&\left\{\mu\frac{\del\ }{\del\mu} + \gamma^{}_{\cal O}(\alpha)\right\}
\bigl\langle\text{vac}\bigl|{\cal O}(0)\bigr|\text{vac}
\bigr\rangle \notag \\
&\hspace{3mm} = - i\,
 \lim_{q \to 0}\int\!\! d^{4} x\, e^{iq\cdot x}\,
\text{T}\bigl\langle\text{vac}\bigl|\theta^\mu_\mu(x)
{\cal O}(0)\bigr|\text{vac}\bigr\rangle_\text{subtr}  \,.
\label{RHS}
\end{align}
The notation $\langle\rangle_\text{subtr}$ indicates that
small-$x$ singularities have been subtracted to renormalize
the answer minimally with a counterterm of order $\theta^\mu_\mu$;
it will not affect our conclusions. The result (\ref{RHS}) remains
valid [modulo $O({\cal J}^2)$ terms in (\ref{action})] for a
product $\prod\!\cal O$ if $\gamma_{\prod\!\cal O}(\alpha)$ is the
sum of $\gamma$ functions of individual spectator operators.
Note that the limit $q \to 0$ in Eq.~(\ref{RHS}) is taken for
$\theta^\mu_\mu \not=  0$ when there are no massless states to
which $\theta^\mu_\mu$ can couple.

Having taken the limit $q \to 0$, what happens to the
right-hand side of Eq.~(\ref{RHS}) if there is an IR fixed point which
allows a second limit%
\footnote{Care must be taken with the order of limits, as noted in
Appendix \ref{chiral} for the chiral case. The analysis, but not the
final answer, depends on which limit is taken first.}
$\theta^\mu_\mu \to 0$ to be taken?

The standard procedure is to set all amplitudes involving
$\theta^\mu_\mu$ to zero. In effect, this \emph{assumes} that  there
is no NG \mbox{mechanism},
i.e.\ that scale invariance is realized in the WW mode:
\begin{equation}
\left\{ \mu\frac{\partial}{\partial\mu} + \gamma_{\cal O}(\alphaWW)
\right\} \bigl\langle\text{vac} \bigl| {\cal O}(0)\bigr|
\text{vac}\bigr\rangle_\textsc{ww} = 0\hsp{0.3}.
\label{CS WW}
\end{equation}
Then the theory at a WW fixed point $\alphaWW$ is \emph{manifestly} scale
and conformal invariant. Green's functions scale according to power laws, with
$\mu$ dependence reduced to trivial factors $\mu^{-\gamma_{\cal O}(\alphaWW)}$.
There is no mass gap, so particles (if they exist) are massless. Dimensional
transmutation does not occur. In particular, fermions cannot condense
at $\alphaWW$ if scale invariance is in the WW mode. Instead, it must be
\emph{assumed} that fermion condensation is possible \emph{only} when
scale symmetry is explicitly broken. For example, in walking gauge theories
\cite{Appel97}, $\alpha$ is thought to vary rapidly after it walks past
$\alphaWW$ because, by assumption, a large $\theta^\mu_\mu$ is necessary for
the region where $\langle\bar{\psi}\psi\rangle_{\text{vac}} \not= 0$
(Fig.\ \ref{fig:beta}, right diagram).

If scale invariance is realized in the NG mode at $\alphaIR$, as we
propose, there are amplitudes for which the right-hand side of
Eq.~(\ref{RHS}) \emph{does not vanish} at $\alphaIR$ as
$\theta^\mu_\mu \to 0$. That can occur if the sum over physical states
$|n\rangle$ in the dispersion integral for
T$\langle\theta^\mu_\mu(x){\cal O}(0)\rangle_\text{subtr}$
includes the exchange of a pseudodilaton $\sigma$:
\begin{equation}
I = \sum_n |n\rangle\langle n|
= |\sigma\rangle\langle \sigma|
+ \sum_{n \not= \sigma}|n\rangle\langle n|\,.
\end{equation}
Here $\sum_{n \not= \sigma}$ includes multi-NG boson states and
states containing non-NG particles; the latter have invariant mass
$M_\text{non-NG} \not=0$ in the scale-symmetry limit $M_\sigma \to 0$.
The exchange of $\sigma$ produces a pole term
\begin{align}
\int&\! d^{4} x \,e^{iq\cdot x}\,\text{T}\bigl\langle\text{vac}
\bigl|\theta^\mu_\mu(x) {\cal O}(0)\bigr|\text{vac}
\bigr\rangle_\text{subtr}^{\sigma\;\text{pole}} \notag \\
&= \langle\text{vac}|\theta_\mu^\mu(0)|\sigma(q)\rangle
\frac{i}{q^2 - M_\sigma^2} \langle\sigma(q)|{\cal O}(0)|\text{vac}\rangle\,,
\end{align}
which does not depend on the subtraction procedure. Taking the limit
$q \to 0$ with $M_\sigma \not= 0$, we see that the zero-momentum
propagator
\begin{equation}
i\bigl/\bigl(q^2 - M_\sigma^2\bigr)\bigr|_{q = 0} = -i \bigl/M_\sigma^2 \,,
\end{equation}
cancels the $M^2_\sigma$ dependence of the matrix element
\begin{equation}
\langle \text{vac} |\theta^{\mu}_{\mu} | \sigma\rangle
= -F_\sigma M_{\sigma}^2\,,
\label{PCDC}
\end{equation}
where $F_\sigma$ is defined in Eq.~(\ref{Fsigma}). In the scale-invariant 
limit, $F_\sigma$ remains nonzero because $\sigma$ is a
dilaton, and so Eq.~(\ref{RHS}) implies our key result:
\begin{align}
\sbiggl \mu\smallfrac{\del\ }{\del\mu} &+
\gamma^{}_{\cal O}(\alpha) \sbiggr \bigl\langle\text{vac}
\bigl|{\cal O}(0)\bigr|\text{vac}\bigr\rangle_\textsc{ng} \notag \\
&\to F_\sigma \bigl\langle \sigma(q=0)\bigl|{\cal O}(0)
\bigr|\text{vac}\bigr\rangle_{\textsc{ng}} \,,
\quad \theta^\mu_\mu \to 0\hsp{0.3}.
\label{result}
\end{align}
States $|n \not= \sigma\rangle$ do not affect this result: at most, relative to
the $\sigma$-pole term, their contributions are $O(M_\sigma^2 \ln M_\sigma)$
for two-dilaton states and $O(M_\sigma^2)$ for other states, including the
subtraction. Equation (\ref{result}) remains valid if $\cal O$ is replaced
by unordered products $\prod_n{\cal O}_n(y_n)$ of operators
${\cal O}_n$ with scaling functions $\gamma_n(\alpha)$,
\begin{align}
\biggl\{ \mu & \frac{\del\ }{\del\mu} + \mbox{\large $\sum\limits_n$}
\gamma_n(\alphaIR) \biggr\}\Bigl\langle\text{vac}\Bigl|
\mbox{$\prod\limits_n$}\,{\cal O}_n(y_n)\Bigr|\text{vac}
\Bigr\rangle_\textsc{ng} \notag \\
&\quad= F_\sigma \Bigl\langle \sigma(q=0)\Bigl|\mbox{$\prod\limits_n$}
\,{\cal O}_n(y_n)\Bigr|\text{vac}\Bigr\rangle_\textsc{ng}
\,, \quad \alpha \rightharpoondown \alphaIR
\label{result2}
\end{align}
provided that light-like momenta in $0^{++}$ channels with $\sigma$ poles are
avoided.

Two features of Eqs. (\ref{result}) and (\ref{result2}) are unfamiliar:
\begin{enumerate}[label=(\arabic*)]
\item  They are soft-meson theorems which have not been derived directly
from an effective Lagrangian. That reflects the fact that effective
Lagrangians for scale invariance were not constructed with the CS
equation in mind.
\item  The CS equation cannot be formulated at $\alpha =\alphaIR$ in
the presence of a dilaton. Results such as Eq. (\ref{result})
refer to the limit $\alpha \rightharpoondown \alphaIR$.
\end{enumerate}
The rest of this section examines the peculiarities of IR fixed points
in the NG scaling mode.

The most important point is that the world at $\alphaIR$ is \emph{not}
the same as the physical world on {\mbox{$0 < \alpha < \alphaIR$}}.
In particular, short-distance behavior at $\alphaIR$ is not governed by
asymptotic freedom because $\alpha$ is fixed: it cannot run towards the
origin $\alpha = 0$.%
\footnote{In chiral-scale perturbation theory for three-flavor QCD
\cite{CT1,CT2}, the asymptotic value $R_{\textsc{ir}}$ of the
Drell-Yan ratio for $e^+ e^-$ annihilation at $\alpha_s =
\alpha_s{}^{}_\textsc{ir}$ is not the same as the QCD value
$R_\textsc{uv} = 2$. The most recent estimate is $2.4 \lesssim
R_\textsc{ir} \lesssim 3.1$ \cite{CT3}.\label{Drell-Yan}}
The theory at $\alphaIR$ is exactly scale invariant but in the NG
mode, amplitudes may be complicated functions of dynamically transmuted
scales. Exceptions are coefficient functions of operator product
expansions at short distances, which are manifestly scale and
conformal covariant; the proof (Appendix \ref{app:ward}) is similar
to that for chiral symmetry \cite{Bernard75}.

Consider what happens at $\alphaIR$ when the conserved dilatation
current ${\cal D}_\mu$ carries momentum $q \not= 0$ in a scaling Ward
identity (Appendix \ref{A.3}), and then the limit $q \to 0$ is taken,
i.e.\ \emph{after} the limit of scale invariance $\theta^\mu_\mu \to
0$. That yields a soft-dilaton formula
\begin{equation}
F_\sigma\langle\sigma(q = 0)|{\cal O}(0)|\text{vac}\rangle_{\textsc{ng}}
= d_{\cal O}\langle\text{vac}|{\cal O}(0)|\text{vac}\rangle_{\textsc{ng}}
\label{soft_dil}
\end{equation}
where
\begin{equation}
d_{\cal O} = \text{dynamical dimension of $\cal O$ at $\alphaIR$}.
\end{equation}
In an effective Lagrangian formalism for dilatons with $\cal O$
represented by an external effective operator
\begin{align}
{\cal O}_{\text{eff}} &= \langle{\cal O}\rangle_{\text{vac}}
\exp\bigl(d_{\cal O}\sigma\bigl/F_\sigma\bigr) \notag \\
&= \langle{\cal O}\rangle_{\text{vac}}
\Bigl\{1 + d_{\cal O}\sigma/F_\sigma + O(\sigma^2)\Bigr\} \,,
\end{align}
Eq.~(\ref{soft_dil}) arises from the term linear in $\sigma$. The
K\"{a}ll\'{e}n-Lehmann representation requires $d_{\cal O}\geqslant 1$
for all local operators ${\cal O} \not=I$ \cite{KGW69}, so every
soft-$\sigma$ amplitude $\langle \sigma|{\cal O}|\text{vac}\rangle$
which does not vanish in the limit $\theta^\mu_\mu \to 0$ corresponds
to a \emph{scale condensate} $\langle{\cal O}\rangle_{\text{vac}}
\not= 0$ (and similarly for ${\cal O} \to \prod_n\!{\cal O}_n$).
Not all scale condensates are chiral condensates, but if
$\langle\bar\psi\psi\rangle_\text{vac} \not= 0$, the vacuum at
$\alphaIR$ breaks both chiral \emph{and} scale invariance.

Connecting this with the physical region involves a subtlety: in
contrast with UV fixed points, the dynamical dimension
of an operator may change at an IR fixed point. That is
because operator dimension is determined by short-distance behavior:
in the physical region $0 < \alpha < \alphaIR$, asymptotic freedom
requires it to take its canonical value
\begin{equation}\label{can_dim}
\text{dynamical dimension of $\cal O$} = d_{\cal O}^{\text{can}}\ ,
\quad 0 < \alpha < \alphaIR
\end{equation}
up to renormalized Schwinger terms (Appendix \ref{B.1}), whereas
$d_{\cal O}$ is determined by the short-distance properties of the
world at $\alphaIR$ (Appendix \ref{app:ward}). 

In the limit of scale invariance at $\alphaIR$, there is a
continuum of vacua related by scale transformations. In the first half
of Appendix \ref{app:ward}, we explain why physics does \emph{not}
depend on which vacuum is chosen. For $0 < \alpha < \alphaIR$,
scale invariance is broken explicitly, and there is a unique vacuum
to which quantities like $\gamma_{\cal O}(\alpha)$ refer.

The relation between $d^{}_{\cal O}$ and $d^{\text{can}}_{\cal O}$
can be easily seen by considering the connected two-point function
\begin{equation}
\Delta^+(x) = \bigl\langle\text{vac}\bigl|
{\cal O}(x){\cal O}(0)\bigr|\text{vac}
\bigr\rangle_{\textsc{ng},\,\text{conn}}
\label{2pt}
\end{equation}
at short distances $x \sim 0$, where the effects of dimensional
transmutation are nonleading. In the physical region
$0 < \alpha < \alphaIR$, we have
\begin{align}
\Delta^+(x) \sim \{\mbox{constant}\}
\bigl(x^2\bigr)^{-d^{\text{can}}_{\cal O}}\bigl(\ln(\mu^2
x^2)\bigr)^{2\gamma_1/\beta_1} \,,
\end{align}
where
\begin{equation} \label{one-loop}
\beta(\alpha) \sim -\beta_1\alpha^2 \quad\mbox{and}\quad
\gamma_{\cal O}(\alpha) \sim \gamma_1\alpha\ , \quad\alpha \to 0
\end{equation}
define the one-loop coefficients $\beta_1 > 0$ and $\gamma_1 > 0$
for an asymptotically free theory. At $\alphaIR$, according to
Eq.~(\ref{result2}), $\Delta^+(x)$ satisfies the relation
\begin{align}
\Bigl\{\mu\smallfrac{\del\ }{\del\mu} + 2\gamma_{\cal O}&(\alphaIR)\Bigr\}
\Delta^+(x) \nn \\
&= F_\sigma \bigl\langle \sigma\bigl|{\cal O}(x){\cal O}(0)
\bigr|\text{vac}\bigr\rangle^{}_{\textsc{ng},\,\text{conn}}\,.
\label{result3}
\end{align}
Consider contributions to each side from the operator product
expansion for ${\cal O}(x){\cal O}(0)$. Clearly, the term proportional
to the dimension-0 identity operator $I$ that contributes to the
left-hand side of Eq.~(\ref{result3}) dominates the leading
contribution to the right-hand side from a dimension $\geqslant 1$
operator $\not= I$:
\begin{equation}
\sbiggl \mu\smallfrac{\del\ }{\del\mu} + 2\gamma^{}_{\cal O}(\alphaIR)
\sbiggr \Delta^+(x) \sim 0 \,.
\end{equation}
So the leading singularity of $\Delta^+$ has $\mu$ dependence
$\sim \mu^{-2\gamma_{\cal O}(\alphaIR)}$. Since $x$ is the only other
dimensionful quantity%
\footnote{By convention, the normalization of composite field operators
excludes dimensionful factors, so $d^{\text{can}}_{\cal O}$
is the engineering dimension of $\cal O$.}
which can appear in the result, we have for $\alpha = \alphaIR$
\begin{align}
\Delta^+(x) \sim \{\mbox{constant}\}
\bigl(x^2\bigr)^{-d^{\text{can}}_{\cal O}}
\bigl(\mu^2 x^2\bigr)^{-\gamma_{\cal O}(\alphaIR)} \,,
\end{align}
which corresponds to
\begin{equation}
d_{\cal O} = d^{\text{can}}_{\cal O} + \gamma^{}_{\cal O}(\alphaIR) \,.
\label{dim}
\end{equation}

The role of dimensional transmutation requires some discussion. Often
it is regarded as a one-loop phenomenon which, \emph{in that context},
breaks scale invariance explicitly, as Coleman and Weinberg
\cite{Cole73} discovered for scalar quantum electrodynamics (QED)
before the discovery of asymptotic freedom. But in non-Abelian
gauge theories, the one-loop approximation makes sense \emph{only} in
the UV limit, where there is a dimensionally transmuted scale
$\Lambda_{\textsc{qcd/tc}}$ which normalizes arguments of UV logarithms
$\ln\bigl(q^2/\Lambda^2_{\textsc{qcd/tc}}\bigr)$ as $q \to \infty$.
Of course, dimensional transmutation persists outside the UV region
because it is necessary to incorporate nonperturbative effects like
fermion condensation into QCD and (by analogy) TC. Chiral perturbation
theory and EFTs for TC are low-energy expansions with their own dimensionally
transmuted scales
\begin{equation} \label{chiscales}
\Lambda_{\chi\textsc{pt}} =  4\pi f_\pi \mbox{ or } 4\pi F_\pi,
\mbox{ and non-NG masses}
\end{equation}
which have nothing to do with $\Lambda_{\textsc{qcd/tc}}$. 
They normalize arguments of IR logarithms
\begin{equation}
\ln\bigl(q\cdot q' \mbox{ or }\{m \mbox{ or }M\}_\pi^2/\Lambda^2_{\chi\textsc{pt}}\bigr)
\label{IRlogs}
\end{equation}
for $q \cdot q' \sim \{m\ \mbox{or}\ M\}^2_\pi \sim 0$.
In chiral-scale perturbation theory or crawling TC, dimensional
transmutation persists at $\alpha_s{}^{}_\textsc{ir}$ or $\alphaIR$
through dependence on the dilaton decay constants $4\pi f_\sigma$ or
$4\pi F_\sigma$. There are \emph{no} theoretical reasons, beyond a
disregard for old but well-established work on the scaling NG mode
for strong interactions \cite{Carr71}, to suppose that
dimensional transmutation a) necessarily ``turns itself off'' as a fixed
point is approached, or b) prevents an NG-mode fixed point from forming
anywhere outside the conformal window, with scale invariance hidden and
not explicitly broken.%
\footnote{As observed in a ``note added'' in Ref.~\cite{CT2}, footnote 20
of Ref.~\cite{Yam14} missed these points. Contrary to footnote 8 of
Ref.~\cite{Bando}, it is \emph{not possible} to deduce anything about
IR fixed points from the one-loop formula for the beta function.}
At present, lattice calculations are the only guide:
\begin{enumerate}[label=(\arabic*)]
\item If $N_f$ is large enough, WW-mode IR fixed points with
  manifestly scale-invariant Green's functions are observed, in
  agreement with Eq.~(\ref{CS WW}). The results define what is meant
  by the conformal window \cite{Del10}.
\item For smaller values of $N_f$ outside the conformal window, where
  dimensional transmutation occurs, it remains to be seen if there are
  NG-mode IR fixed points \cite{Del14}. If so, scale invariance is
  not manifest because of Eqs.~(\ref{result}) and (\ref{result2}).
  Signals for this on the lattice will be considered in Sec.\ \ref{lattice}.
\end{enumerate}

Let us recall how, despite the absence of fermion mass terms in
the Lagrangian, dimensional transmutation can arise in massless QCD
and TC. Observable constants $\cal M$ with dimensions of mass, such as
decay constants and non-NG masses, are permitted because renormalization
group (RG) invariance
\begin{equation}
\left\{\mu\frac{\del\ }{\del\mu}
+ \beta(\alpha)\frac{\del\;\ }{\del\alpha}\right\}{\cal M} = 0
\label{dimtrans}
\end{equation}
is consistent with $\cal M$ being proportional to the sole scale
in the theory, the renormalization scale $\mu$:
\begin{equation}
{\cal M}
= \mu \exp \bigg\{-\int^\alpha_{\kappa_{\cal M}} dx\bigl/\beta(x)\bigg\} \,,
\quad 0 < \kappa_{\cal M} < \alphaIR \,.
\label{M_inv}
\end{equation}
Here $\kappa_{\cal M}$ is a dimensionless constant which depends on
$\cal M$ but not on $\alpha$ or $\mu$. As is well known~\cite{SVZ79},
the nonperturbative nature of $\cal M$ can be verified by considering
the limit $\alpha \sim 0$ at fixed $\mu$: from Eq.~(\ref{one-loop}),
there is an essential singularity due to the factor
$\exp\{-1/(\beta_1\alpha)\}$ which ensures the absence of a Taylor
series in $\alpha$.

In the IR limit, two points of view are possible. One, to be discussed
below, is to treat amplitudes $\cal A$ as functions of $\alpha, \mu$
and various momenta $\{p\}$ and consider  what happens as $\alpha$
tends to $\alphaIR$. The other is to note that, since observable
constants $\cal M$ are annihilated by the CS differential operator
in Eq.~(\ref{dimtrans}), they act as constants of integration in
CS equations for amplitudes, i.e.\ the CS equations allow any
dependence on $\cal M$ consistent with engineering dimensions.
Therefore, if an amplitude is observable and hence RG invariant,
its dependence on $\alpha$ and $\mu$ can be entirely replaced by a
dependence on the transmuted masses $\cal M$ alone. This matters:
we want to apply approximate scale invariance to physical amplitudes,
so the limit $\theta^\lambda_\lambda \to 0$ is taken at fixed $\cal
M$, not fixed $\mu$.

Alternatively, as shown by the analysis leading to Eq.~(\ref{dim}),
it can be useful to consider amplitudes depending on operators $\cal O$
such as $(\alpha/\pi)\hat{G}^2$ which are not RG invariant. Such
amplitudes can be treated as functions of $\cal M$ and $\mu$, with
residual dependence on $\mu$ being retained in the scale-invariant
theory at the fixed point. For example, the amplitude $\langle{\cal O}
\rangle_{\text{vac}}$ appearing in Eqs.~(\ref{result}) and
(\ref{soft_dil}) can be written as
\begin{align}
\bigl\langle\text{vac}\bigl|{\cal O}(0)
 &\bigr|\text{vac}\bigr\rangle_{\textsc{ng}} \sim c^{}_{\cal O}  {\cal M}^{d^{\text{can}}_{\cal O}}
\bigl({\cal M}\bigl/{\mu}\bigr)^{\gamma^{}_{\cal O}}
= c_{\cal O} {\cal M}^{d_{\cal O}}\bigr/\mu^{\gamma^{}_{\cal O}}
\label{M_dep}
\end{align}
for $\theta^\lambda_\lambda \to 0$, where the dimensionless constant
$c_{\cal O}$ does not depend on $\mu$ or $\cal M$.

The result (\ref{M_dep}) must not be confused with hyperscaling
relations in mass-deformed theories \cite{DD+Z10, DD+Z11, DD+Z14},
where the conformal invariance of a gauge theory 
in the WW mode is explicitly broken by a mass term 
$-m\bar{\psi}\psi$ in the Lagrangian. Hyperscaling 
is a property of the scaling WW mode. All VEVs
$\langle{\cal O}\rangle_\text{vac}$ and ``hadron'' masses
$M_\text{inside}$ inside the conformal window%
\footnote{In walking TC, there must be a phase transition
at the sill of the conformal window \cite{Appel_LSD10} that causes
fermions to condense and hence create NG and non-NG technihadrons
outside the conformal window, but the distinction between spectra
inside and outside the window is usually left unclear. We
reserve the term ``condensate'' for VEVs which are nonzero
in a symmetry limit such as $m \to 0$.\label{phase}}
scale with \emph{fractional} powers of the Lagrangian parameter $m$:
\begin{align}\label{hyper}
\langle{\cal O}\rangle_\text{vac}
&\sim m^{\mbox{\footnotesize
 ${\{d^\text{can}_{\cal O} + \gamma_{\cal O}(\alphaWW)\}/\{1 +
            \gamma_m(\alphaWW)\}}$}}\,, \notag \\
M_\text{inside} &\sim m^{\mbox{\footnotesize
 ${1/\{1 + \gamma_m(\alphaWW)\}}$}} \,.
\end{align}

Not surprisingly, both $\langle{\cal O}\rangle_\text{vac}$ and
$M_\text{inside}$ in Eq.~(\ref{hyper}) vanish in the limit $m \to 0$ of
scale invariance. That is not the case for Eq.~(\ref{M_dep}) because,
unlike $m$, $\cal M$ is \emph{not} a variable current fermion mass
in a Lagrangian. Rather, $\cal M$ is a fixed non-Lagrangian constant
associated with a condensate, such as a decay constant or a non-NG
technihadron mass arising from nonzero \emph{constituent} masses
of $m=0$ fermions (Appendix \ref{decouple}). The key property of a
scaling NG-mode fixed point is that amplitudes depend on the
\emph{nonzero} scales ${\cal M}$ in the scale-invariant limit
$\theta^\mu_\mu \to 0$ (Appendix \ref{app:ward}), so unlike 
Eq.~(\ref{hyper}), the right-hand side of Eq.~(\ref{M_dep}) does not vanish
in that limit. Furthermore, if we add $-m\bar{\psi}\psi$ to the
Lagrangian, nonzero results such as Eq.~(\ref{M_dep}) are corrected by
terms \emph{linear} in $m$, as expected in chiral perturbation theory
or its chiral-scale extension \cite{CT1,CT2,CT3}; fractional powers of
$m$ are \emph{never} seen. There is no such thing as hyperscaling in
crawling TC.

Now let us check what happens when amplitudes are treated as functions
of $\alpha, \mu, \{p\}$ as $\alpha$ tends to $\alphaIR$ at fixed $\mu$.
Comparison with the $\beta\del/\del\alpha$ term in the CS equation (\ref{CS})
shows that the right-hand side of Eq.~(\ref{result}) arises from
the singular $\alpha$ dependence of the condensate as $\alpha$ 
approaches $\alphaIR$ for fixed $\mu$:
\begin{align}
\frac{\del\ }{\del\alpha}\bigl\langle\text{vac}\bigl|
{\cal O}(0)\bigr|\text{vac}\bigr\rangle_\textsc{ng}
\sim \frac{1}{\alphaIR - \alpha}\frac{F_\sigma}{\beta'}
\bigl\langle\sigma\bigl|{\cal O}(0)\bigr|\text{vac}
\bigr\rangle_{\textsc{ng}}\,.
\label{sing}
\end{align}
This singularity is to be expected.%
\footnote{Perhaps this could be exploited in searches for NG-mode
fixed points on the lattice (Sec.\ \ref{lattice}).}
The operator $\del/\del\alpha$ inserts $\cal O$ at zero-momentum
transfer, so a pole due to a zero-mass particle (the dilaton) coupled
to $\cal O$ will produce a singular result. Note that it is
\emph{only} at the fixed point that this is allowed. A singularity or  
lack of smoothness in the $\alpha$ dependence of \emph{any} 
amplitude 
within the interval $0 < \alpha < \alphaIR$ would be a disaster: it 
would indicate a lack of analyticity, such as a Landau pole, at a
finite space-like momentum.  

Similarly, the fixed-$\mu$ limit $\alpha \to \alphaIR$ applied
to Eq.~(\ref{M_inv}) is singular: 
\begin{equation}
{\cal M} \sim \mu\bigl(\alphaIR - \alpha\bigr)^{-1/\beta'}
\bigl\{\mbox{constant}\bigr\} \ , \quad\mbox{fixed } \mu\,.
\label{M_asy}
\end{equation}
Note that this implies
\begin{equation}
\frac{\del\ }{\del\alpha}{\cal M}^{d_{\cal O}} \sim
\frac{d_{\cal O}{\cal M}^{d_{\cal O}}}{(\alphaIR - \alpha)\beta'}
\end{equation}
and hence, from Eq.~(\ref{M_dep}), 
\begin{equation}
\frac{\del\ }{\del\alpha}\bigl\langle\text{vac}
\bigl|{\cal O}(0)\bigr|\text{vac}\bigr\rangle_\textsc{ng}
\sim \frac{d_{\cal O}}{(\alphaIR - \alpha)\beta'}
\bigl\langle\text{vac}\bigl|{\cal O}(0)\bigr|
\text{vac}\bigr\rangle_{\textsc{ng}}  \,,
\end{equation}
which shows that Eq.~(\ref{sing}) is consistent with the
soft-dilaton theorem (\ref{soft_dil}). 

Equation (\ref{M_asy}) implies that, for $\cal M$ to remain
finite in the scaling limit $\theta^\lambda_\lambda \to 0$,
$\mu$ tends to $0$ according to the rule
\begin{equation}
\mu \propto (\alphaIR - \alpha)^{1/\beta'} \,.
\end{equation}
Singularities are removed when the $\mu,\alpha$
dependence of amplitudes is eliminated in terms of physical
quantities.%
\footnote{This is similar to what happens in the large-$N_c$
limit of QCD, where the singularity $f_\pi \sim \sqrt{N_c}$ is
eliminated by writing everything in terms of the pion decay
constant $f_\pi$.}

\begin{figure}[t]
    \centering\includegraphics[scale=0.7]{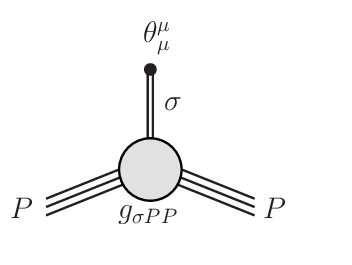}
    \caption{Generation of the mass $M_P$ of a non-NG particle $P$ via
    the dominant $\sigma$ pole in $\langle P|\theta_\mu^\mu|P\rangle$,
    where $-g_{\sigma PP}\bar{P}P$ defines the $\sigma PP$ coupling.
    In the scale-invariant limit $\theta_\mu^\mu\to 0$, $M_P$ remains
    nonzero.}
    \label{fig:g_sigPP}
    \end{figure}

A simple example of $\mu$ dependence being related to a soft-dilaton amplitude
is when ${\cal M}$ in Eq.~(\ref{M_inv}) is the mass $M_P$ of a non-NG particle
$P$. Then the scalar analogue \cite{MGM62,Carr71} of the Goldberger-Treiman
relation (Fig.~\ref{fig:g_sigPP}) applies:
\begin{equation}
\mu\frac{\del\ }{\del\mu} M_P = M_P
= F_\sigma g_{\sigma PP}  \,.
\label{sigmaPP}
\end{equation}

We close this section with a discussion of the analogue of the low-energy
theorem (\ref{result}) for QCD.  \pagebreak 
There the \mbox{relevant} equations have extra
terms because quarks $q$ have mass $m_q \neq 0$. When the trace
anomaly~\cite{Mink76,Adler77,Niels77,Coll77} (with the vacuum expectation
value subtracted)%
\footnote{For consistency, the $\gamma_m$ terms in Eqs.~(\ref{eqn:anomaly})
and (\ref{CSqcd}) must have opposite signs (unlike Ref.~\cite{CT2} where
conventions were changed during review). Here we choose the definition
$\gamma_m = - \mu\del\ln m_q/\del \mu$ \cite{Del13}. Then $\bar{q}q$
has dynamical dimension $3 -\gamma_m(\alpha_s{}^{}_\textsc{ir})$
at a QCD fixed point $\alpha_s{}^{}_\textsc{ir}$, and similarly for
$\bar{\psi}\psi$ in crawling TC, where the notation becomes
$\gamma_m(\alpha)$ and $\alphaIR$.
\label{gamma_m}}
\begin{align}\label{eqn:anomaly}
\theta^\mu_\mu
=\biggl\{&\frac{\beta(\alpha_s)}{4\alpha_s} G^A_{\mu\nu}G^{A\mu\nu}
+ \bigl(1 + \gamma_{m}(\alpha_s)\bigr)\sum_q
m_{q}\bar{q}q\biggr\}  \notag \\
 &- \Bigl\{\text{VEV}\Bigr\}
\end{align}
and the CS equation
\begin{widetext}
\begin{align}
\biggl\{\mu\frac{\del\ }{\del\mu}
+ \beta(\alpha_s)\frac{\del\;\ }{\del\alpha_s}
- \gamma_m(\alpha_s)\sum_q m_q \frac{\del\ \ }{\del m_q}
+ \gamma^{}_{\cal O}(\alpha_s)\biggr\} \bigl\langle\text{vac}
\bigl|{\cal O}(0)\bigr|\text{vac}\bigr\rangle = 0
\label{CSqcd}
\end{align}
are compared, we find
\begin{align}
\sbiggl \mu\smallfrac{\del\ }{\del\mu} + \gamma^{}_{\cal O}(\alpha_s)
\sbiggr \bigl\langle\text{vac}\bigl|{\cal  O}(0)\bigr|\text{vac}\bigr\rangle 
= i\, \lim_{q \to 0}\int\!\! d^{4} x\,e^{iq\cdot x}
\text{T}\Bigl\langle\text{vac}\Bigl|
\Bigl\{\textstyle{\sum\limits_q}m_q\bar{q}q(x)
- \theta^\mu_\mu(x)\Bigr\} {\cal O}(0)
\Bigr|\text{vac}\Bigr\rangle_{\text{subtr}}  \,.
\label{RHSqcd}
\end{align}
If heavy quarks have been decoupled, and the limit $m_q \to 0$ is
taken for the light quarks $q = u,d,s$ as the IR fixed point is
approached, dilaton pole terms from both $\theta^\mu_\mu$
and $\sum_q m_q\bar{q}q$ may survive the limit \cite{CT1,CT2,CT3}:
\begin{align}
\sbiggl\mu\smallfrac{\del\ }{\del\mu} + \gamma^{}_{\cal O}(\alpha_s)\sbiggr
\bigl\langle\text{vac}\bigl|{\cal O}(0)\bigr|\text{vac}\bigr\rangle
\to f_\sigma\bigl\langle\sigma\bigl|{\cal O}(0)\bigr|\text{vac}\bigr\rangle
\Bigl\{1 - \bigl(3 - \gamma_m(\alpha_s{}^{}_\textsc{ir})\bigr)
(f_\pi/f_\sigma)^2\Bigl(m^2_K + \ltextfrac{1}{2}
      m^2_\pi\Bigr)\Bigl/m_\sigma^2\Bigr\}\,.
\end{align}
\end{widetext}

\section{Crawling TC: Hidden electroweak-scale symmetry}
\label{New}

TC is based on the idea \cite{Wei76,Wei79,Suss79} that electroweak
symmetry ``breaking'' is the dynamical effect of a gauge theory which
resembles QCD but whose coupling becomes strong at scales of a
few TeV. The trigger for this effect is a techniquark
condensate $\langle\bar \psi\psi\rangle_\text{vac}\neq 0$. The
resulting technipions become the longitudinal components of the
$W^\pm$ and $Z^0$ bosons, while the masses and couplings of the other
technihadrons are estimated by scaling up QCD quantities, where the
electroweak scale $v \simeq 246$ GeV plays the role of the pion decay
constant $f_\pi \simeq 93$ MeV.

An attractive feature of TC is that the hierarchy problem is avoided:
the mechanism for mass generation does not rely on elementary
Higgs-like scalars. Instead, masses are generated dynamically through
dimensional transmutation \cite{Cole73}, as in QCD.

When TC was invented, the Particle Data Group (PDG) tables did
not include QCD scalar $J^{PC}=0^{++}$ resonances below $\approx 1$ GeV,%
\footnote{\label{one}The $\epsilon(700)$ was excluded from the PDG tables in
1974. Its successor $f_0(500)$ was first mentioned in 1996, but became a
well-defined resonance only in the 2008 tables.}
  so for many years, it was thought, by analogy with QCD, that TC scalar
  particles would not be seen below the TeV scale.

  There is now strong evidence for a light, broad $0^{++}$ resonance $f_0(500)$
  in the QCD meson spectrum with mass $m_{f_0} \simeq 441$ MeV
  \cite{Cap06,PDG,Pelaez} (evidence which seems to have been mostly overlooked in
  the TC literature), and also for a narrow Higgs boson $h$ at $m_h \simeq 125$
  GeV \cite{ATLAS,CMS}. Given these facts, can $h$ be the TC version of
  the $f_0(500)$? At first sight, the answer to this question is negative.
  An application of the scaling rules mentioned above requires the TC
  analogue $f_{0T}$ of $f_0$ to have a large mass \cite{Sann09}
  \begin{equation}
  m_{f_{0T}} \approx (v/f_\pi)\, m_{f_0} = O(\text{TeV}) \,;
  \label{f0 scaling}
  \end{equation}
  also, they seem to imply an $O$(TeV) width except for the fact that the
  $f_0(500)$ has plenty of phase space for its decay into two pions, whereas
  there are no technipions for $f_{0T}$ to decay into and (for a mass of
  125 GeV) no phase space for it to decay into $W^+ W^-$ or $Z^0 Z^0$. But it is
  evident that this estimate for the mass is much too large.

  A convincing explanation for why the observed mass $m_h \simeq 125$ MeV
  is so small relative to TeV scales is hard to find. That is a key problem
  shared by all theories of dynamical Higgs mass generation, including TC and
  its extensions. The most promising strategy is to suppose that the Higgs is
  a pseudo-NG (pNG) boson of a hidden symmetry. Then the mass acquired by the
  pNG boson due to explicit symmetry breaking is protected by the underlying
  symmetry~\cite{Wein72,GeoPais75}. A light Higgs mass can arise if
  explicit symmetry breaking is due to physics at the electroweak scale
  and hence small relative to the scale of dynamical symmetry breaking.

  In composite Higgs models~\cite{Kap84,KapGD84,GeoKG84,Geo84,Dug85}, where the
  hidden symmetry is internal, this mechanism is well understood: the Higgs
  boson and all would-be NG bosons are placed in the same multiplet of an
  extended group such as $SO(5)$~\cite{Aga04,Con06,Giudice07}. For a recent
  review of these models, see chapter III of Ref.~\cite{Csa15}.

  Our focus is on the main alternative: broken scale and conformal invariance
  with a ``dilatonic'' Higgs boson. A dilaton, or NG boson for conformal
  invariance, has the property that it couples to particle mass \cite{MGM62}.
  At first, this idea was applied to strong interactions, as reviewed in
  Ref.~\cite{Carr71}. A few years later, it was noted \cite{Ell76} that, in the SM,
  tree-level couplings of the Higgs field are dilaton-like, i.e.\ they couple to
  mass. The literature on dynamical Higgs bosons spawned by this observation is
  unfortunately not consistent about the meaning of ``dilaton'' and
  overlooks the need to \emph{hide} conformal invariance as it becomes
  exact.

  The clearest examples of this are walking TC theories with dilatonic
  modifications \cite{Appel10,Yam11,Appel13,Yam14,Golt16}.
  Consider the walking region shown in the right-hand diagram of
  Fig.~\ref{fig:beta}. The WW-mode fixed point lies within the conformal
  window where dilatons cannot exist, but it is supposed that in the
  walking region at the edge of the window, dynamics is affected by
  ``dilatons'' due to a field dependence $\sim \exp(\sigma/F_\sigma)$ in
  an effective Lagrangian. It is then argued (following a suggestion in
  Ref.~\cite{Diet05}) that these ``dilatons'' couple to an operator which is small
  near the scale-symmetry limit
  \begin{equation}
  \theta^\mu_\mu = O\bigl(\alpha - \alphaWW\bigr) \,,
  \end{equation}
  and so they have a small mass protected by scale symmetry at $\alphaWW$.

  The flaw in this argument becomes evident when the relation
  \begin{equation}\label{alphaWW}
  M_\sigma^2 F_\sigma = -\langle\text{vac}|\theta^\mu_\mu|\sigma\rangle
  = O(\alpha - \alphaWW)
  \end{equation}
  is considered. In walking TC, the so-called ``dilaton'' decouples from the
  theory as the WW-mode fixed point is approached,
  \begin{equation}
  F_\sigma \sim 0\ \mbox{ for }\ \alpha \sim \alphaWW
  \end{equation}
  because there can be no scales at $\alphaWW$. Therefore, \emph{no conclusion}
  can be drawn about $M_\sigma$ from Eq.~(\ref{alphaWW}). The only general
  theorem governing particle decoupling is that of Appelquist and Carazzone
  \cite{AC75} for \emph{heavy} particles.

  In crawling TC (left diagram in Fig.~\ref{fig:beta}), the IR fixed
  point is in the NG-mode, not the WW-mode. As noted above Eq.~(\ref{result}),
  the (pseudo)dilaton does \emph{not} decouple as the fixed point is approached,
  \begin{equation}
  F_\sigma \to \mbox{ constant } \not= 0\ \mbox{ as }\ \alpha \to \alphaIR
  \label{couple}
  \end{equation}
  so from
  \begin{equation}
  M_\sigma^2 F_\sigma = -\langle\text{vac}|\theta^\mu_\mu|\sigma\rangle
  = O(\alpha-\alphaIR) \,,
  \end{equation}
  we can safely conclude that $M_\sigma^2$ is $O(\alpha - \alphaIR)$ and
  hence small.

  A precise formula for the pseudodilaton mass can be obtained
  as an important application of Eq.~(\ref{result}). The result is an
  analogue of the Gell-Mann--Oakes--Renner relation
  \cite{GMOR} for $0^-$ mesons.

  To see this, consider the case ${\cal O} = \hat G^2$ with each side
  of Eq.~(\ref{result}) multiplied by the factor $\ltextfrac{1}{4}\beta/\alpha$.
  The result is
  \begin{equation}
  \frac{\beta(\alpha)}{4\alpha} \left\{ \mu\frac{\del\ }{\del\mu}
  + \gamma^{}_{\hat G^2}(\alpha) \right\}
  \bigl\langle\hat G^2\bigr\rangle_{\text{vac}}
  \to F_\sigma \bigl\langle \sigma\bigl|\theta_\mu^\mu\bigr|
  \text{vac}\bigr\rangle\,,
  \label{result4}
  \end{equation}
  where a simple derivation \cite{CT1,CT2,Spiri84,Grin89} (discussed
  in Appendix \ref{app:scaling}) implies
  \begin{equation}
  \gamma_{\hat G^2}(\alpha) = \beta'(\alpha) - \beta(\alpha) / \alpha\,,
  \quad \beta'(\alpha) = \del\beta(\alpha)/\del\alpha\,,
  \label{gamma}
  \end{equation}
  for the anomalous scaling function of $\hat G^2$. Equation (\ref{PCDC})
  implies that the right-hand side of Eq.~(\ref{result4}) is given by
  $-M_\sigma^2 F_\sigma^2$. For an IR expansion in $\epsilon = \alphaIR
  - \alpha \gtrsim 0$ about the fixed point, the left-hand side reads
  \begin{align}
  \frac{\beta(\alpha)}{4\alpha} &\left\{ \mu\frac{\del\ }{\del\mu}
  + \gamma^{}_{\hat G^2}(\alphaIR) \right\}
  \bigl\langle\hat G^2\bigr\rangle_{\text{vac}} \notag \\
  &\hspace*{10mm}= -\frac{\epsilon\beta'(4+\beta')}{4\alphaIR}
  \bigl\langle\hat G^2\bigr\rangle_{\text{vac}}
  + O(\epsilon^2)\,,
  \label{result5}
  \end{align}
  where the critical exponent $\beta' = \beta'(\alphaIR)$ is positive
  (Fig.~\ref{fig:beta}, left diagram) and we have used dimensional analysis
  to trade the $\mu\del/\del \mu$ term for the engineering dimension of
  $\langle\hat G^2\rangle_\text{vac}$.  Equations (\ref{result4}) and (\ref{result5})
  imply the desired mass relation
  \begin{equation}
  M_\sigma^2 = \frac{\epsilon\beta'(4+\beta')}{4\alphaIR F_\sigma^2}
  \bigl\langle\hat G^2\bigr\rangle_{\text{vac}} + O(\epsilon^2)\,,
  \label{MsigPCDC}
  \end{equation}
which exhibits the pseudo-NG nature of $\sigma$ explicitly.%
\footnote{A similar formula in Refs.~\cite{Appel10,Appel13} lacks the anomalous
dimension $4 + \beta'$. The main problem is that its derivation assumes
$\theta^\mu_\mu \sim 0$ near a WW fixed point $\alphaWW$, where condensates
tend to zero and $\sigma$ is not a pseudodilaton because it decouples
($F_\sigma \sim 0$).}
  The requirement $M_\sigma^2 \geq 0$ fixes the sign of the condensate:
  $\langle\hat G^2\rangle_\text{vac} \geq 0$.

  This mass is protected by scale invariance at $\alphaIR$
  because the condition (\ref{couple}) ensures that our dilaton is a
  genuine NG boson. That is what allows us to identify the pseudodilaton
  in the crawling region near $\alphaIR$ as the Higgs boson with mass
  much smaller than the TeV scale of TC.

  This conclusion also applies if the techniquarks are given a current mass
  $m_\psi$, as in the case of TC lattice simulations where an extrapolation to the
  chiral limit $m_\psi \to 0$ must be performed.  Since the fermion mass is an
  additional source of explicit scale symmetry breaking, the IR expansion in
  $\epsilon$ must be augmented by powers of $m_\psi$.

  Repeating the same steps that led to our mass formula (\ref{MsigPCDC}),
  but this time for the operator\footref{gamma_m}
  \begin{equation}\label{TCtrace}
  {\cal O} =\frac{\beta(\alpha)}{4\alpha} \hat{G}^A_{\mu\nu}\hat{G}^{A\mu\nu}
  + \bigl(1 + \gamma_{m}(\alpha)\bigr)\sum_\psi
  m_{\psi}\bar{\psi}\psi\,,
  \end{equation}
  we find
  \begin{align}
  &\tilde{M}_\sigma^2 \tilde{F}_\sigma^2
  = \frac{\epsilon\beta'(4+\beta')}{4\alphaIR}
  \langle\hat G^2\rangle_\text{vac} \nn \\
 &\quad -(3-\gamma_m)(1+\gamma_m) m_\psi \langle\bar{\psi}\psi
  \rangle_\text{vac}
  + O(\epsilon^2,\epsilon m_\psi, m_\psi^2)\,,
  \label{Msigfull}
  \end{align}
where $\tilde{M}_\sigma$ and $\tilde{F}_\sigma$ are the dilaton
mass and decay constant in the presence of $m_\psi$, $\gamma_m$ is
shorthand for $\gamma_m(\alphaIR)$, and we made use of the homogeneity
equation
\begin{equation}
\left\{\mu\frac{\del}{\del\mu} + m_\psi\frac{\del}{\del m_\psi}
- d_\mathcal{O}^\text{can} \right\}
\langle \text{vac}| {\cal O}(0) | \text{vac}\rangle = 0  \,.
\end{equation}
If $\langle\hat G^2\rangle_\text{vac}$ can be reliably estimated
(Appendix \ref{technigluon}), the leading-order result (\ref{Msigfull})
may be used to test candidate theories of crawling TC on the lattice;
see also Sec.\ \ref{lattice}.

Returning to the $m_\psi = 0$ case, we note that the explicit
scale symmetry breaking responsible for the dilaton mass arises
from renormalization and is entirely nonperturbative. That should be
contrasted with
\begin{enumerate}[label=(\arabic*)]
\item  the pion mass due to (chiral) symmetry breaking by
   current quark-mass terms in the bare QCD Lagrangian, and
\item  the ``scalon'' mass \cite{scalon} of a Coleman-Weinberg
  potential \cite{Cole73} generated by explicit scale breaking from
  one-loop renormalization of gauge theories whose tree-level
  amplitudes lack massive parameters.
\end{enumerate}

\section{Peculiarities of dilaton Lagrangians}
\label{notscalons}

Compared with chiral Lagrangians, the conformal case involves
some subtleties which caused problems when first encountered in 1969
\cite{Salam69,Isham70a}: the would-be NG bosons seemed to be massive
in the limit of conformal invariance.  By late 1970, these puzzles had
been resolved: just one NG field (the dilaton) is needed for the entire
conformal group \cite{Isham70b} (Appendix \ref{A.3}), and the class
of consistent dilaton Lagrangians is specified by Zumino's condition
\cite{Zum70}
\begin{equation}
\lambda = O(\epsilon)
\label{Bruno}\end{equation}
if there is a term $\lambda \phi^4$ in the potential. The unusual feature
of Eq.~(\ref{Bruno}) is the requirement that a symmetry-preserving
operator $\phi^4$ have a symmetry-breaking coefficient $\lambda$, i.e.
\begin{equation}
\lambda \to 0
\label{Bruno2}
\end{equation}
in the limit $\epsilon \to 0$ of conformal invariance.

We are revisiting this topic because the NG and WW scaling modes
are still being confused and Zumino's condition is not being respected.
This seems to stem from two 1976 papers, both of which a) referred to the
NG mode of conformal invariance but not to the 1969-1970 literature, and
b) have attracted a lot of interest since then:
\begin{enumerate}[label=(\arabic*)]
\item  Gildener and Weinberg \cite{scalon} used the term ``scalon'' to
 describe a scalar particle which couples to $\theta^\mu_\mu$ but
 where the limit $\theta^\mu_\mu \to 0$ is in the WW scaling mode. It is
 therefore not a dilaton, contrary to remarks in an early paragraph of
 Ref.~\cite{scalon} and to assertions in subsequent literature
 \cite{Meiss07,Chang07,Foot07,Gold08,Vecc10,Bell13,Bell14,Cor13}.
\item  Fubini's ``new approach'' to conformal invariance \cite{Fub76}
 is limited to $\lambda\phi^4$ and its generalizations. Therefore it
 cannot be used to disprove the existence of the NG mode for (exact)
 scale invariance, contrary to subsequent claims \cite{Bell13,Bell14,Cor13}.
\end{enumerate}

\subsection{Flat directions?}
\label{flat}
If a symmetry is realized in the NG mode, it follows that there are
directions in field space, one for each NG boson, for which the action
is flat. Often this is used as a shortcut to search for NG modes of
complex Lagrangians.

So, if a Lagrangian $\cal L$ is scale invariant, it is tempting to
suppose that, when the action is varied, a flat direction necessarily
corresponds to a dilaton. The classic counterexample is the Lagrangian
${\cal L}_{\text{free}} = \frac{1}{2}(\del\phi)^2$ for a massless spin-$0$
field $\phi$.

As is well known, $\phi$ describes a genuine NG boson, but that is for
invariance under field translations
\begin{equation}
\phi \to \phi + \{\mbox{constant}\},
\end{equation}
\emph{not} for scale transformations. The theory is exactly soluble
with amplitudes which \emph{do not depend on a scale}, so scale
invariance is realized in the \emph{WW mode} and $\phi$ is not a
dilaton. This is entirely different from exact scale invariance in the NG
mode, where amplitudes depend on a \emph{nonzero} dilaton decay
constant $F_\sigma$ and hence other dimensionful constants
(Appendix \ref{app:ward}).

If a scale-invariant $\cal L$ depends on many field components, there
can be many flat directions. One of them may be associated with the
NG mode of scale transformations, but not necessarily. If
amplitudes do not depend  on dimensionful constants in the
scale-invariant limit,  as in scalon theories (Sec.~\ref{GilWein}),
the theory is dilaton-free.

\subsection{Zumino's consistency condition}
\label{Zumino}
Zumino's condition (\ref{Bruno}) is necessary for scale invariance to
be realized in the NG mode. 

Its genesis was the work of Salam and Strathdee \cite{Salam69}, who
sought to extend the nonlinear theory of chiral Lagrangians to the
conformal case. They introduced the now-standard parametrization
\begin{equation}
\phi(x) = F_\sigma\exp\bigl\{\sigma(x)\bigl/F_\sigma\bigr\}
\label{param}
\end{equation}
for the scalar field $\phi$ in terms of a would-be dilaton field
$\sigma$ with the transformation property
\begin{equation}
\sigma \to\, \sigma - \frac{F_\sigma}{4}\ln\det\frac{\del x'}{\del x}
\ , \quad x \to  x' \mbox{ conformal.}
\end{equation}
(There was also a vector field $A_\mu$ for special conformal
transformations, but that was subsequently abandoned \cite{Isham70a}
in favor of $\del_\mu\sigma$.) Then, imitating the procedure for chiral
Lagrangians, they wrote down the most general Lagrangian consistent
with symmetry requirements,
\begin{align}
{\cal L} &= \ltextfrac{1}{2}(\del\phi)^2 - \lambda_0\phi^4
     + \widetilde{\cal L}(\phi\,;\rho)
\label{trial-L}\\
 &= \ltextfrac{1}{2}\bigl(\del\sigma\bigr)^2 e^{2\sigma/F_\sigma}
     - \kappa e^{4\sigma/F_\sigma} + \widetilde{\cal L}(\sigma,\rho)\,,
\label{trialL}\end{align}
with $\lambda_0 > 0$ in the scale-invariant limit [unlike
$\lambda$ in Eq.~(\ref{Bruno})]. Here $\rho(x)$ denotes chiral   
and non-NG matter fields and $\kappa = \lambda_0 F_\sigma^4$   
is a positive constant.

The result of applying these apparently general principles was
puzzling. When the $\phi^4$ term is expanded in
$\sigma$,\hspace*{-1mm}   
\begin{equation}
\kappa e^{4\sigma/F_\sigma}
 = \kappa + 4\kappa\sigma\bigl/F_\sigma + 8\kappa\sigma^2\bigl/F_\sigma^2
   + O(\sigma^3)
\end{equation}
the $O(\sigma^2)$ term seems to give the would-be dilaton a 
mass\hspace*{-1mm}    
\begin{equation}
m_\sigma \overset{?}{=} 4\sqrt{\kappa}\bigl/F_\sigma
\label{false}
\end{equation}
in the scale-invariant limit \cite{Salam69}. Terms in
\begin{equation}
\widetilde{\cal L} = \sum_d {\cal O}_d(\chi)e^{(4-d)\sigma/F_\sigma}
\end{equation}
cannot compensate for this: the dimension-$d$ operators ${\cal  O}_d$
do not have vacuum expectation values because of their dependence on
$\rho(x)$. A massive $\sigma$ \emph{cannot} be an NG boson, but could
its mass have arisen from a Higgs-style mechanism \cite{Salam69},
despite the fact that the conformal symmetry being investigated is
global, not local?

Zumino observed that these puzzles were symptoms of a more basic
problem: scale-invariant $\phi^4$ theories and the NG scaling
mode are not compatible. If one tries to use the parametrization
(\ref{param}) to force the theory into the scaling NG mode, a
low-energy expansion cannot be performed:
\begin{enumerate}[label=(\arabic*)]
\item  The requirement $\sigma \to 0$ as $x_\mu \to \infty$ for the
  fluctuation field $\sigma(x)$ produces infinite action if there is a
  term $\sim e^{4\sigma/F_\sigma}$ in $\cal L$.
\item  Modifying
\begin{equation}
  e^{4\sigma/F_\sigma} \to e^{4\sigma/F_\sigma} - 1
\label{modif}
\end{equation}
  is not allowed because the subtraction would violate scale invariance.
\end{enumerate}

The subtlety exposed by Zumino is that writing $\phi$ in terms of
$\sigma$ does not necessarily  force a theory into the NG scaling mode,
and, for $\lambda_0 \not= 0$ in the symmetry limit, it is 
\emph{not legitimate} to do so. That is connected with the fact that 
Eq.~(\ref{param}) constrains $\phi$:
\begin{equation}
\phi > 0 \,.
\label{phi_con}
\end{equation}
The conclusion (\ref{false}) is incorrect because it was derived without
first finding a minimum about which to expand in the unconstrained
field $\sigma$, and  $e^{4\sigma/F_\sigma}$ has no minimum for
finite variations of $\sigma$.%
\footnote{We have been asked if Zumino's condition, when extended
    to include gravity, is consistent with having  a cosmological constant
    $\Lambda \not= 0$. The discussion above concerns the limit
    $\epsilon \to 0$, but $\Lambda$ breaks scale invariance
    explicitly, so there is no contradiction. For example,
    Zumino's $O(\epsilon)$ example (\ref{FN}) would allow $\Lambda \not= 0$.}

Given that $\lambda$ must vanish for scale invariance in the NG mode,
why is there an apparent clash with the principle learned from chiral
Lagrangians that the most general Lagrangian consistent with symmetry
should be considered? The answer is that the principle needs to be more
carefully stated.  When constructing an effective Lagrangian, the most
general result consistent with symmetry \emph{and NG-mode}
requirements must be sought.

Consider any continuous symmetry, compact or noncompact. Then
the set of all possible Lagrangians consistent with the symmetry will
include a subset in the WW mode, another subset in an NG mode, and
others which cannot be expanded about a point in field space because of
a poor choice of field variables or Lagrangian coefficients. So, having
written down a ``general'' Lagrangian, it is necessary to check by hand
that it can be expanded in all NG fields about a stationary point. Only
then can it be treated as an effective Lagrangian for the desired NG mode(s).

Let us contrast noncompact scale symmetry with symmetry under global
compact $U(1)$ transformations $\varphi \to e^{i\theta}\varphi$ of a complex
spin-0 field. Consider the class of symmetric Lagrangians
\begin{equation}
{\cal L}_{A,B} = \ltextfrac{1}{2}|\del\varphi|^2 - A|\varphi|^2
     - B|\varphi|^4
\label{L_AB}\end{equation}
parametrized by constants $A,B$. If free-field theory is excluded, $B$
lies in the range $B > 0$. Then both modes of the theory are
determined by \emph{inequalities}, i.e.\ by continuous ranges of the
(mass)$^2$ $A$:
\begin{equation}
\text{NG mode: } A < 0\,,\quad \text{WW mode: } A \geqslant 0\,.
\label{modes1}\end{equation}
Thus, when the choice of coefficients in a chiral Lagrangian is said to
be ``arbitrary,'' there is an understanding that this is not entirely
so, especially for the model (\ref{L_AB}) where the familiar constraints
$A < 0$ and $B > 0$ apply. For the scale-invariant Lagrangian
(\ref{trialL}), the free-field case is avoided by requiring
$\widetilde{\cal L} \not= 0$. As we have seen, one of the two modes
of scale invariance is specified by an \emph{equality}:
\begin{equation}
\text{NG mode: } \lambda = 0\,,\quad \text{WW mode: } \lambda > 0\,.
\label{modes2}\end{equation}
This difference between (\ref{modes1}) and (\ref{modes2}) is hardly
surprising, given that degenerate minima for scale transformations
have to lie on a half-line to infinity in field space, unlike the
periodic orbits characteristic of compact group symmetries.

A feature shared by chiral and dilaton Lagrangians is that in the
symmetry limit, NG bosons do not interact at zero momentum:
\begin{equation}
\pi + \pi \not\to \pi + \pi \quad,\quad
\sigma + \sigma \not\to \sigma + \sigma \,.
\label{no-interact}
\end{equation}
In both cases, this follows from the flatness requirement for
degenerate minima. For dilatons, it is obviously  consistent
with Eq.~(\ref{Bruno2}).

Now let scale symmetry be broken explicitly by adding $O(\epsilon)$
terms to the Lagrangian (\ref{trialL}). Zumino observed that one of
these terms could be $e^{4\sigma/F_\sigma}$ with a coefficient
proportional to $\epsilon$, as in Eq.~(\ref{Bruno}), and that subtractions
such as Eq.~(\ref{modif}) are then allowed. By itself, $e^{4\sigma/F_\sigma}$
still does not allow a minimum at any finite value of $\sigma$, but
when combined with $d \not= 4$ terms which break scale symmetry
explicitly, the resulting dilaton potential
\begin{equation}
V(\sigma) = O(\epsilon)
\end{equation}
may have a minimum and produce a genuinely light dilaton:
$m_\sigma^2 = O(\epsilon)$. Zumino gave an example%
\footnote{For early work consistent with Eq.~(\ref{Bruno}), see
Ref.~\cite{Isham70b} [formula below Eq.~(3.11)] and 
Ref.~\cite{Ell70} [Eq.~(4.6)]. Compare Eqs.~(45)--(50) of 
Ref.~\cite{CT2}.}
\begin{equation}
V(\sigma) = \epsilon F_\sigma^4 \Bigl\{e^{2\sigma/F_\sigma} - 1\Bigr\}^2
\label{FN}\end{equation}
related to the model of Freund and Nambu \cite{Nambu68}; it implies
$m_\sigma^2 = 8\epsilon F_\sigma^2$ and $\lambda = \epsilon$.

There may be a concern that renormalization violates the constraint
$\lambda = 0$ in the limit of scale invariance. Here it is
important to distinguish loop expansions in WW and NG scaling
modes---they are not equivalent.

In a renormalizable theory with a $\lambda\phi^4$ interaction,
the loop expansion is a series in a finite set of coupling constants (including
$\lambda$) which mix under RG flow, to all orders in the expansion.
The perturbation series is obtained via small-field fluctuations such as
$\phi \sim 0$, as in  the WW scaling mode, where $\phi$ is unconstrained.
Then $\phi$ propagators can be formed and used to construct tree
and loop diagrams. Since $\phi^4$ counterterms occur, the point
$\lambda = 0$ is unstable under \emph{WW-mode} RG flow.

However, in the NG scaling mode, we are dealing with a nonrenormalizable loop
expansion in powers of NG-boson momenta $q \sim 0$ and explicit
symmetry breaking $\epsilon$, as in Appendix A of Ref.~\cite{CT2}. The
constraint (68) occurs at $\phi = 0$, so fluctuations $\phi \sim 0$
to form $\phi$ propagators are not allowed. Instead, we expand in the
unconstrained field $\sigma$ and form loops with $\sigma$ propagators
and vertices. The outcome resembles that for nonlinear chiral
theories \cite{Lam73b,Gass84,Gass85}: each new loop order produces
a new set of coupling constants because $\epsilon$-independent
counterterms  have more derivatives than before. All RG mixing of
coupling constants of a given order is $O(\epsilon)$, as in Eq.~(\ref{Bruno}).

For example, let $\sigma$ be coupled to the matrix field $U$
\cite{Georgi_book,Gass85} for chiral NG-bosons as follows \cite{CT1,CT2,CT3}
\begin{equation}
{\cal L}_0 = \Bigl\{\ltextfrac{1}{2}\bigl(\del\sigma\bigr)^2
 + \ltextfrac{1}{4}F^2_\pi\text{tr}\bigl(\del_\mu U\del^\mu U^\dagger\bigr)
 \Big\} e^{2\sigma/F_\sigma} + O(\epsilon) \,,
\label{L-0}\end{equation}
where $O(\epsilon)$ denotes terms which break scale and chiral
invariance explicitly, and for $\epsilon \to 0$, we have chosen
$\lambda = 0$, i.e.\ $\kappa = 0$ in Eq.~(\ref{trialL}). Then for
$\epsilon = 0$, all NG-boson interactions (dilatons and chiral bosons)
involve a field derivative, and so there can be no nonderivative
counterterms like mass counterterms $\delta m^2\{\sigma^2\mbox{ or }
\pi^2\}$ or four-point interactions $\delta\lambda\{\sigma^4\mbox{ or }
\pi^4\}$ which would violate the masslessness of NG bosons and
no-interaction conditions like Eq.~(\ref{no-interact}). Instead, there
are higher-derivative counter\-terms such as the scale-invariant
four-point interaction \vspace{-2mm}  
\begin{equation}
F_\sigma^{-4}\bigl(\del\sigma\bigr)^4
\end{equation}
which is $O(q^2)$ in NG-boson momenta $q$ relative to leading order.
In the presence of explicit scale breaking, as in Eq.~(\ref{FN}),
$\sigma$ propagators  carry a small mass $m_\sigma = O\bigl(\sqrt{\epsilon}\bigr)$.
Then there can be a $d=4$ counterterm in $V(\sigma)$, but the
correction to $\lambda$ is clearly $O(\epsilon)$. Therefore Zumino's
condition (\ref{Bruno}) is \emph{stable}  under NG-mode RG flow.

So far, the discussion has been restricted to the NG-boson sector.
The result is an expansion in powers of \vspace{-3mm}  
\begin{equation}
\{q \mbox{ or } m_{\textsc{NG}}\}\bigl/\{4\pi F_\sigma \mbox{ or } 4\pi F_\pi\}
\end{equation}
with coefficients depending on logarithms $\ln(m_{\textsc{NG}}/\mu)$;
the renormalization scale $\mu$ provides the sole UV cutoff for
integrals. For dimensional regularization in $n$ complex dimensions,
include the $O(\epsilon)$ terms (otherwise all loop integrals vanish),
and in ${\cal L}_0$, replace
\begin{equation}
e^{2\sigma/F_\sigma}\ \longrightarrow\ e^{(n-2)\sigma/F_\sigma}\,.
\end{equation}

The inclusion of non-NG particles such as fermions with mass
$M \not= 0$ for $\epsilon \to 0$ presents difficulties already familiar
from baryonic chiral perturbation theory \cite{Gass88,Scherer}: for
fermion fields, the expansion is in $(i\slashed{\del} - M)$, not $i\del$, so
higher-derivative fermionic terms can be of leading order. Consequently,
extending the NG-mode renormalization procedure to massive fermions
is not obvious. Special techniques have been invented to deal with loops
containing at least one NG boson \cite{Jenkins91,Becher99,Fuchs03},
but little can be said about pure non-NG particle dynamics such as
effects due to closed fermion loops. Instead, it must be assumed that
all non-NG dynamics can be contained in the low-energy constants
of loop expansions involving NG bosons, where chiral and (in
our case) conformal symmetry provide some guidance.

We mention closed fermion loops because it might be thought that they
should be part of the renormalization procedure. Could they produce
counterterms which give NG bosons mass and violate Zumino's condition?
If so, non-NG dynamics would force the theory out of the NG mode.

Consider a toy model such as the $\sigma \sim 0$ expansion of
the scale-invariant  Lagrangian
\begin{equation}\label{Ltoy}
{\cal L}_{\text{toy}}  = \ltextfrac{1}{2}
\bigl(\del\sigma\bigr)^2 e^{2\sigma/F_\sigma}
+ \bar{\psi}\Bigl(\ltextfrac{i}{2}
\overset{\leftrightarrow}{\raisebox{0pt}[0.82\height]{$\slashed{\del}$}}
- Me^{\sigma/F_\sigma}\Bigr)\psi  \,.
\end{equation}
In the tree approximation, for which ${\cal L}_{\text{toy}}$ is designed,
one can read off relations such as the scalar analogue of the
Goldberger-Treiman  relation [Eq.~(\ref{sigmaPP}) and Fig.~\ref{fig:g_sigPP}].
If ${\cal L}_{\text{toy}}$ is supposed to produce a renormalizable
perturbation series in the Yukawa coupling  $-M/F_\sigma$, closed
fermion loops certainly do produce divergent self-energy,  triangle
and box diagrams.

The flaw in this picture is the assertion that, for momenta $\gtrsim M$,
non-NG particle dynamics can be represented by the perturbative series of a
local renormalizable theory for baryon and meson fields or their
TC counterparts. There is no hint of this from QCD or experiment.
Interactions between non-NG hadrons are strong and produce higher
resonances which could not all be represented by separate fields.

Instead, it must be recognized that there can be nonrenormalizable
higher-derivative fermionic terms in leading order, as in the modified
toy example
\begin{align}\label{Lmod}
{\cal L}_{\text{mod}} &= \ltextfrac{1}{2}\bigl(\del\sigma\bigr)^2 e^{2\sigma/F_\sigma}
+ c_1\del^\mu\del^\nu\bar{\Psi}\ltextfrac{i}{2}
\overset{\leftrightarrow}{\raisebox{0pt}[0.82\height]{$\slashed{\del}$}}
   \del_\mu\del_\nu\Psi   \nn \\
&\quad + c_2\bar{\Psi}\Bigl(\ltextfrac{i}{2}
\overset{\leftrightarrow}{\raisebox{0pt}[0.82\height]{$\slashed{\del}$}}
  - Me^{\sigma/F_\sigma}\Bigr)\Psi e^{4\sigma/F_\sigma}  \,,
\end{align}
where $c_1M^4 + c_2 = 1$ and we have chosen a new fermion variable
\begin{equation}
\Psi(x) = \exp\Bigl\{-2\sigma(x)\bigl/F_\sigma\Bigr\}\psi(x)
\end{equation}
which carries dimension $-\frac{1}{2}$. In the tree approximation, this
model also produces Eq.~(\ref{sigmaPP}), but the corresponding fermion
propagator has asymptotic behavior
\begin{equation}
S_F(p)
= i\frac{\slashed{p}(c_1 p^4 + c_2) + M}{p^2(c_1 p^4 + c_2)^2 -  M^2}
\sim \frac{i\slashed{p}}{c_1 p^6} \ , \quad p \to \infty
\end{equation}
which makes all closed fermion loops converge.

Of course, this procedure is arbitrary, but that is the point: nothing
can be said about dynamics in the non-NG sector. We must follow the
example of chiral perturbation theory, and start from the basic
hypothesis, well supported by experiment in the chiral case, that non-NG
particle dynamics does not force the theory out of the NG mode.

\subsection{Digression: Fubini's ``new approach''}
\label{Fubini}

Modern investigators of light Higgs bosons often cite Fubini's 1976
paper \cite{Fub76} as evidence that the NG scaling mode cannot be
realized in the limit of conformal symmetry. A cursory reading of
Ref.~\cite{Fub76} can easily produce this wrong conclusion,
especially if earlier work leading to Zumino's condition (\ref{Bruno})
(to which Fubini does not refer) is not known.

Fubini's approach was not just ``new'': it was radically different from
the standard theory of dilaton Lagrangians described above. Conformal
invariance is imposed on $\lambda\phi^4$ theory and, more generally,
on polynomial scalar-field Lagrangians in $D$ space-time dimensions
with no dependence on dimensionful constants. Scale breaking due to
renormalization is ignored. All fields are unconstrained: nonlinear
chiral or scale fields depending on $F_\pi$ or $F_\sigma$ are not
present. Then Fubini considered introducing a fundamental scale $a$ via
a state $|0\rangle_F$ which he called the ``vacuum'' but which looks
more like a coherent state; it corresponds to a classical field $B(x)$:
\begin{equation}
{}_F\langle 0|\phi(x)|0 \rangle_F = B(x) \,.
\label{class_B}\end{equation}
He observed (correctly) that $B(x)$ cannot be constant for $\lambda
\not= 0$, and so $|0\rangle_F$ does not preserve translation
invariance. Instead, it preserves a linear combination
\begin{equation}
R_\mu = \ltextfrac{1}{2}\bigl(aP_\mu + a^{-1}K_\mu\bigr)
\end{equation}
of the momentum components $P_\mu$ and special conformal
generators  $K_\mu$. To restore translation invariance, Fubini proposed 
 a ``statistical'' average over the continuum of degenerate ``vacua''%
\begin{equation}
|h \rangle_F = \exp\bigl(iP_\mu h^\mu\bigr) |0 \rangle_F \,,
\end{equation}
but the properties of the resulting theory and its true vacuum (if it has
one) are not known.

Fubini's conclusions do \emph{not} exclude the existence of dilaton
Lagrangians which preserve translation invariance, because his choice of
conformal models excludes the set of known dilaton Lagrangians,
all of which obey Zumino's condition (\ref{Bruno}). Fubini considered
Eq.~(\ref{trial-L}) but \emph{not} Eq.~(\ref{trialL}): he \emph{avoided} the
error of assuming Eq.~(\ref{param}) for $\lambda \not= 0$. His analysis
leaves $\phi$ unconstrained, contrary to Eq.~(\ref{phi_con}), and so yields an
$x$-dependent result (\ref{class_B}). In contrast, genuine dilaton
Lagrangians involve constrained scale fields (\ref{param}) with constant
vacuum expectation values
\begin{equation}
\langle\text{vac}|F_\sigma\exp\bigl(\sigma(x)\bigl/F_\sigma\bigr)|\text{vac}\rangle
 = F_\sigma \,.
\end{equation}

Fubini's interests were semiclassical, with apparently no intention
that his work be compared with the literature on nonlinear dilaton
Lagrangians of six years earlier
\cite{Nambu68,Isham70b,Ell70,Zum70,Ell71,Carr71}.
He was not known to be against the existence of the NG mode for global
scale transformations, nor was his work seen in that light when
it was published.

\subsection{Changing field variables}
\label{change-field}
Unlike nonlinear chiral Lagrangians, dilaton Lagran\-gians can be
linearized%
\footnote{This terminology is standard, but what is really meant is
   that the Lagrangian becomes a polynomial in the field variables.
   Similarly, read ``nonpolynomial'' for ``nonlinear''.}
by a change of variable consistent with the equivalence theorem if
renormalization is ignored and noninteger  dimensions are absent. On
dimensional grounds, the nonlinear Lagrangian necessarily depends on a
dimensionful quantity, the dilaton decay constant $F_\sigma$, but that
dependence tends to be hidden in the linear version. This may mask the
presence of an NG scaling mode; if so, it certainly obscures NG-mode
renormalization. Alternatively, in the absence of other fields such as
chiral bosons, it may indicate a theory actually in the WW scaling mode
with all $F_\sigma$ dependence transformed away.

The equivalence theorem%
\footnote{In statements of the theorem, a Lagrangian theory is defined by the
all-order loop expansion due to small-field fluctuations about a local
minimum of the potential. Modulo renormalization, Lagrangians related
by an invertible point transformation mapping one fluctuation region
to the other, as in Eq.~(\ref{invert}) below, are equivalent: their
$S$ matrices agree. The mapping $\sigma \leftrightarrow \phi$
of Eq.~(\ref{param}) is forbidden because the constraint
(\ref{phi_con}) disallows fluctuations $\phi \sim 0$.
\label{equivalence}}
was originally derived \cite{Chis61,Kame61,Coleman69} without
regard to renormalization, so it was explicitly valid only in the
tree approximation. Subsequently, a renormalized version of the
theorem was proven for renormalizable theories \cite{Lam73a,Bergere75},
but not generally for NG-mode renormalization of nonlinear chiral
models \cite{Lam73b,Arzt93}. We believe that an equivalence
theorem can be formulated and proven for nonlinear NG-boson
Lagrangians with derivative interactions in the limit of exact symmetry,
all renormalized in the NG mode as outlined in Sec.~\ref{Zumino}, but
an all-order analysis remains to be done.

As an example of the equivalence theorem in the tree approximation,
consider the toy Lagrangian (\ref{Ltoy}). The field $\sigma$ can be
expanded about a point $\sigma_0$ determined by the limit $\epsilon \to 0$
of a scale-violating perturbation $\sim \epsilon(\sigma -\sigma_0)^2$.
If we choose $\sigma_0 = 0$, the fermion $\psi$ has mass $M$ in lowest
order, so clearly, ${\cal  L}_{\text{toy}}$ is a dilaton Lagrangian: its amplitudes
exhibit the NG scaling mode in the limit $\epsilon \to 0$. Is it
equivalent to a polynomial Lagrangian? The answer is ``yes,'' but
\emph{only} if the new field variable is constrained, e.g.
\begin{equation}\label{Lex1}
\sigma \to \phi_c = F_\sigma\bigl(e^{\sigma/F_\sigma} - 1\bigr) \ ,
\quad \phi_c \geqslant - F_\sigma \,.
\end{equation}
This change of variables is permitted by the equivalence theorem
because the constraint on $\phi_c$ does not interfere with
fluctuations
$\phi_c \sim 0$ corresponding to $\sigma \sim 0$:
\begin{equation}
\phi_c = \sigma + \sum_{n>1} \frac{\sigma^n}{n!\, F_\sigma^{n-1}}
= \sigma + O(\sigma^2) \,, \hspace{2mm} |\phi_c| \ll F_\sigma \,.
\label{invert}\end{equation}
The result is a polynomial Lagrangian in the constrained field $\phi_c$
\begin{equation}
{\cal L}_{\text{toy}}'
 = \ltextfrac{1}{2}\del_\mu\phi_c\del^\mu\phi_c
+ \bar{\Psi}\bigl(i\slashed{\del} - M - MF^{-1}_\sigma\phi_c\bigr)\Psi
\end{equation}
giving the same tree-diagram $S$ matrix as ${\cal L}_{\text{toy}}$. As
noted for ${\cal L}_{\text{toy}}$ at the end of Sec.~\ref{Zumino},
${\cal L}_{\text{toy}}'$ is not a good basis for NG-mode renormalization.

When renormalizing in the NG mode, it is not \emph{a priori} obvious that
parametrizations of the chiral matrix field $U$ and the scalar field
(\ref{param}) in terms of unconstrained NG fields survive the process.
Furthermore, not all Lagrangians equivalent at tree level are equally
amenable, because the process can be upset by terms proportional
to the equations of motion. The most undesirable scenario is having to
subtract convergent as well as divergent loop diagrams by hand to
enforce the masslessness of NG bosons and the no-interaction
requirement (\ref{no-interact}) generalized to amplitudes with many
NG-boson legs:
\begin{equation}
\left.{\cal A}^{}_{\pi\ldots\pi\sigma\ldots\sigma}\right|_{\text{all\,$q=0$}}
= 0\ ,\quad m_\pi = 0 = m_\sigma\ ,\quad \epsilon \to 0\,.
\end{equation}
In each order of the loop expansion, that would require an infinite
set of counterterms, i.e.\ the renormalization procedure would be nonlocal.

Note that by itself,
\begin{equation}
{\cal L}_0 = \ltextfrac{1}{2}(\del\sigma)^2\exp(2\sigma/F_\sigma)
\label{Kreimer}
\end{equation}
is \emph{not} a dilaton Lagrangian. The theory appears to be interacting,
with a loop expansion which requires renormalization. However, when
renormalized by subtracting about any point in momentum space which is
not IR singular, seemingly complicated sets of diagrams at each loop
order sum to zero on shell \cite{Kreimer16}. Evidently, ${\cal L}_0$
for $\sigma \sim 0$ is equivalent to $\frac{1}{2}(\del\phi_c)^2$ for
$\phi_c \sim 0$, so tree-level amplitudes sum to zero on shell;
then cutting rules can be used to extend the result to loops. The conclusion
is that all dependence on $F_\sigma$ is absorbed by the change
of variable (\ref{Lex1}). This shows that merely writing a scalar field as
$f\exp(\sigma/f)$ is \emph{not enough} to ensure the existence of
dilatons: it must be shown that \emph{amplitudes} of the scale-invariant
theory depend on dimensionful constants.

\subsection{Scalons are not dilatons}
\label{GilWein}
In their influential work on scalons, Gildener and Weinberg
\cite{scalon} considered a scale-invariant limit for polynomial
Lagrangians, but unlike Fubini, they \emph{wanted} to produce
amplitudes with no dependence on a dimensionful constant. They
did this by retaining translation invariance and assuming the tree
approximation for unshifted fields. All dependence on dimensionful
constants would be generated by an \emph{explicit} breaking of scale
invariance due to renormalization corrections depending on a scale $\mu$.

Scalon theories are constructed as follows. First, a polynomial Lagrangian
${\cal L}_{\text{gauge}}$ is constructed for a scale-invariant gauge theory
involving one \cite{Cole73} or more \cite{scalon,Meiss07,Chang07,Foot07,Gold08}
scalars. In the tree approximation, all of these scalars are massless, but none
of them can be a dilaton because, by construction, amplitudes do
not depend on dimensionful constants. So scale invariance is realized in
the WW mode, which (as for ${\cal L}_{\text{free}}$ above) is entirely consistent
with the presence of flat directions. Then one-loop quantum corrections
${\cal V}_{\text{CW}}$ \cite{Cole73} are calculated and used to perturb
${\cal L}_{\text{gauge}}$:
\begin{equation}
{\cal L}_{\text{one-loop}}
= {\cal L}_{\text{gauge}} - {\cal V}_{\text{CW}}
= {\cal L}_{\text{K.E.}} - {\cal V}_{\text{eff}}  \,.
\end{equation}
The explicit breaking of scale invariance by logarithmic factors
$\ln(\phi^2/\mu^2)$ in ${\cal V}_{\text{CW}}$ gives rise to two
scale-violating effects, viz.\ a compact set of chiral- (not scale-)
degenerate minima of ${\cal V}_{\text{eff}}$, and masses for one or more
scalons. Despite the third paragraph of Ref.~\cite{scalon}, none of these
scalons can be a pseudodilaton because, in the scale invariant limit
${\cal V}_{\text{CW}} \to 0$, amplitudes have no scales and hence
\emph{there are no dilatons.} Scalon theories deserve to be studied in
their own right, but must not be confused with dilaton theories.

This may be the origin of a pervasive belief that the NG mode for scaling
is possible only in the presence of explicit scale violation
\cite{Csa15}, as in oft-repeated references to ``spontaneous breaking of
\emph{approximate} scale invariance.'' This sounds odd because it is not
correct: only in the limit of \emph{exact} scale invariance can the
distinction between the NG and WW scaling modes be made. The
most obvious cause of this is the misunderstanding of Fubini's work
\cite{Fub76} discussed in Sec.~\ref{Fubini}. In walking TC or
scalon theory, which is generally not dependent on Ref.~\cite{Fub76}, it
may stem either from the third paragraph of Ref.~\cite{scalon} or simply
from an implicit assumption that ``conformality'' is always in the WW mode.

A key element of this belief is that the way to elevate any theory to
dilaton status is to write $f\exp(\sigma/f)$ for a scalar field close to
a fixed point and avoid discussing what this means for the fixed point
itself. In the scale-invariant limit, there are four main
possibilities:
\begin{enumerate}[label=(\arabic*)]
\item The WW mode is produced because $f \to 0$. That is the origin
of the ``fine-tuning'' problem of scalon theories
\cite{Gold08,Vecc10,Bell13,Bell14,Cor13}, where $f^2$ is proportional
to the magnitude of explicit scale breaking. Approximate scale invariance
requires $f/v \ll 1$ contrary to $f \sim v$ experimentally. More generally,
the expansion
\begin{equation}
\hspace{0.9cm}
f\exp(\sigma/f) = f + \sigma + \sigma^2/(2f) + \sigma^3/(6f^2) + \ldots
\end{equation}
fails: it would produce singularities $\sim f^{-p}$ in effective
Lagrangian vertices.
\item A phase transition causes the scale-violating expansion to fail. 
In walking TC, the walking coupling $\alpha$ is separated from a
WW-mode fixed point $\alphaWW$ by a chiral phase transition
\cite{Appel_LSD10} at the sill of the conformal window.  Nevertheless, the
small value of the Higgs mass is claimed to be a first-order consequence
of the expansion in $\alpha$ about $\alphaWW$. That creates severe
conceptual difficulties \cite{Golt16,Golt16a} for ``dilatonic''
walking TC theories (Sec.\ \ref{EFT}). \label{chiralphase}
\item The constant $f$ can be transformed away via the equivalence theorem,
allowing the fixed point to be in the WW mode. That may circumvent the
fine-tuning or phase-transition problems, but then there would be no
soft-dilaton theorems: any effective Lagrangian could be rendered
independent of $f$, as in the example (\ref{Kreimer}) above.
\item At the fixed point, $f$ is the decay constant $F_\sigma\not= 0$ given
by Eq.~(\ref{Fsigma}), so soft-dilaton theorems (Appendix \ref{A.3}) exist
and amplitudes do not scale at the fixed point (Sec.~\ref{NGsolutions}).
Then the fixed point is in the NG mode, which excludes walking TC and scalons.
\end{enumerate}

Theoretical ambiguity about whether the fixed point is in the NG or WW mode is
popular but untenable: a choice must be made. Physically, the NG mode is
far closer to reality and hence a far better candidate for theories of
approximate scale invariance: the particle spectrum in the scale-invariant
limit (Appendix \ref{app:ward}) resembles that of the real world. Compare
that with the WW mode, where there are no thresholds except for branch cuts
and poles at zero momentum, and particles may not even exist \cite{Geo07}.

\section{Comments on phenomenology}
\label{comments}
Since our Higgs-boson  theory differs fundamentally from all others (they are
not expansions about a scale-invariant theory with a scale-dependent
vacuum), its phenomenology cannot be inferred from a subclass of
existing theories: a new analysis is necessary. We begin with
remarks about the width of pseudodilatons, the relative magnitudes of
pNG boson decay constants in QCD and crawling  TC, and the electroweak
$S$ parameter \cite{STU90,STU}.

QCD and crawling TC borrow an idea from broken scale invariance
for strong interactions that a chiral condensate can also act as a scale
condensate \cite{Ell70,Cre70}, implying a relation for the $\sigma
\to \pi\pi$ coupling
\begin{equation}
f_\sigma g_{\sigma\pi\pi} \simeq -m^2_\sigma
\label{sigpipi}
\end{equation}
which remains valid in chiral-scale perturbation theory \cite{CT1,CT2,CT3}.
Equation (\ref{sigpipi}) implies a width of a few hundred MeV for $\sigma$,%
  \footnote{The dilaton-Higgs of crawling TC is relatively narrow
  because (unlike the case of QCD), the pions are eaten, and there
  is no phase space for $\sigma$ to decay strongly into
  other particles.  This is consistent with the current \cite{PDG}
  upper bound $\Gamma_h/m_h \lesssim 10^{-4}$.}
which is consistent with data for the $f_0(500)$ resonance, the obvious
candidate for the QCD pseudodilaton.%
\footnote{This provides a clear counterexample to the claim
\cite{Bell13,Bell14,Cor13} that no light dilaton is expected in QCD.}
Here $f_\sigma$ and $f_\pi$ are observed to have similar orders of
magnitude within a factor of $\sim 2$. Given that both arise from having
$\langle\bar{q}q\rangle_{\text{vac}} \not= 0$ in the scale-invariant
limit, this was to be expected. Note that we could not use a symmetry
argument to fix the ratio $f_\sigma/f_\pi$, because the Coleman-Mandula
theorem \cite{Cole67} does not permit internal chiral and space-time
scale symmetry to be unified.

  Since this works for QCD, there is good reason to let
  $\langle\bar{\psi}\psi\rangle_\text{vac}$ be a condensate
  for both chiral and scale transformations in crawling TC, with
  similar orders of magnitude for the electroweak scale $v = F_\pi$
  and the TC dilaton decay constant $F_\sigma$. This avoids the
  fine-tuning problem of scalon theories noted above, where
  the strength of explicit scale breaking $f$ must be artificially adjusted
  to match the scale $v$ of the chiral condensate
  \cite{Gold08,Vecc10,Bell13,Bell14}.

  It is often suggested that TC theories have trouble generating a small 
  enough value of the $S$ parameter (defined such that $S=0$ in the SM) 
  that is compatible with the experimental number $S = 0.05 \pm 0.10$
  \cite{PDG}. Quoted values of $S$ typically include the estimates
  $S\approx 0.32$ obtained originally by Peskin and Takeuchi \cite{STU}
  and $S=0.42(2)$ in recent two-flavor lattice calculations \cite{LSD1,LSD2}.
  But the prescription \cite{STU} used to obtain these estimates involves
  subtracting the contribution of a heavy SM Higgs boson, and must be
  amended \cite{Foad12} if the TC spectrum contains a light scalar. In
  Ref.~\cite{Pich13}, TC scenarios which include a generic light scalar
  resonance were confronted with electroweak precision data. Figure 6 of
  Ref.~\cite{Pich13}, which plots the deviation $\kappa_W$ from the SM ($f/v$
  or $F_{\sigma}/v$ in our notation) against the technirho mass $M_V$,
  shows that the experimental constraints on $S$ require
  $v\simeq F_{\sigma}$ and $M_V \simeq 1$ TeV. Both requirements are
  naturally satisfied in crawling TC.

\section{Electroweak EFT}
\label{EFT}

By analogy with QCD, where at energies below the confinement scale one can
use EFT methods to describe pion dynamics, an EFT for dynamical electroweak
symmetry is the most efficient way to describe physics at energies ranging
from a few GeV to several hundred GeV. In this range, all SM interactions are
relatively weak. Perturbation theory is possible not only in the electroweak
couplings $g_w$ and $g_w'$ but also in the gluon coupling constant $g_s$
because of asymptotic freedom for QCD. The upper limit of several hundred
GeV is chosen so that interactions presumed to be strong at the TeV scale
\begin{equation}\label{Lambda_v}
\Lambda_v \sim 4\pi v = 4\pi F_\pi
\end{equation}
become sufficiently weak in the SM sector to justify a perturbative EFT
approach.  At energies $\sim \Lambda_v$, hadronic bound states from the TC
interactions are expected to populate the spectrum and be responsible for
the Higgs sector seen at lower energies. The EFT is constructed by requiring
$SU(3)_c\times SU(2)_L\times U(1)_Y$ gauge invariance and including the
currently observed particle content, with the Higgs identified as a
pseudodilaton instead of a weak doublet.  The resulting theory is an effective
chiral Lagrangian (augmented with gauge bosons and fermions), which
for crawling TC is extended \cite{Ell70} to include the NG mode of scale
invariance.

Electroweak EFT was originally developed \cite{Appel80,Long80} with a heavy
Higgs boson in mind. Although no longer valid, some basic features of that work
survived subsequent developments \cite{Feru92,Bagg94,Koul93,Burg99,Wang06}
and remain in low-energy EFTs for light Higgs bosons
\cite{Grin07,Contino10,Buch12,Alon12,Buch13}. In all of these theories, the
effective Lagrangian has a chiral component for the would-be NG bosons which
give (conveniently in Landau gauge) mass to the weak $W^\pm$ and $Z^0$ bosons.
The standard procedure is to choose a nonlinear chiral Lagrangian
\cite{Wein68,Coleman69,Callan69,Weinberg79} based on (say) a unitary matrix
field $U$ \cite{Gass85,Georgi_book}; linear models are inconvenient because they depend
on extraneous non-NG fields such as the sigma field of the linear sigma model.
The advantage of the effective Lagrangian formalism is that, with symmetries
implemented at an operator level, radiative corrections are easily computed,
and contact can be made with the SM Lagrangian in order to spot potential
deviations in the phenomenology.

As noted in Sec.~\ref{change-field}, the extension to dilatons is
\emph{necessarily} nonlinear: the spin-0 field which transforms with
scale dimension 1 enters linearly but produces the NG scaling mode only if
it is suitably \emph{constrained} and hence a nonlinear function of
unconstrained fields. In analogy with Eq.~(\ref{Lex1}), we use a special
notation $\chi_c$ to distinguish our $\chi$ field from the WW-mode
fields implicitly used in walking TC or scalon theories. The key feature of
our  theory is that $\chi_c$ is constrained in the \emph{exact} limit of scale
invariance as well as when there is explicit scale symmetry breaking.

By definition, the fields $U$ and $\chi_c$ transform linearly under the
electroweak gauge group and scale transformations. It is convenient to choose
constraints which are manifestly symmetry preserving
\begin{equation}
U = SU(2)\mbox{ matrix} \quad\text{and}\quad \chi_c > 0
\label{constr}
\end{equation}
and for which there are standard parametrizations in terms of
unconstrained Goldstone fields $\varphi^a$ \cite{Gass85,Georgi_book}
and $\sigma$ \cite{Ell70,Zum70,Salam69}:
\begin{equation}
U = e^{i\varphi^a\tau^a/v}  \quad \mbox{and} \quad
\chi_c = F_\sigma e^{\sigma/F_\sigma} \,.
\label{params}
\end{equation}
Here $\tau^a$ are Pauli matrices.

The next step is to specify the theory responsible for crawling TC and how
its effects are to be incorporated into our EFT. As for all TC theories, we
assume it to be a gauge theory which exhibits asymptotic freedom in the UV
limit, i.e.\ well above $\Lambda_v$. Since the range of energies being
considered is well below the strongly interacting TeV scale, the result is
controlled by the IR limit of whatever TC theory is held responsible
for those effects. In that limit, the TC coupling $\alpha$ either runs to
a fixed point $\alphaIR$, as in the left diagram of Fig.~\ref{fig:beta}
(crawling TC), or it runs to $\infty$.

In crawling TC, the Higgs boson is light because it corresponds to a
small $O(\epsilon)$ term in the IR expansion of the continuous variable
$\epsilon = \alphaIR - \alpha > 0$ about the NG-mode fixed point
$\alphaIR$. This is a great advantage over walking TC, where the
small value of $\alpha -\alphaWW$ in the walking region is said to be
responsible for the small Higgs mass. That assumes that the walking
region of the solid curve in the right diagram of Fig.~\ref{fig:beta} can
be approximated by the dashed curve in that diagram near $\alphaWW$.
The problem is that the solid and dashed curves are separated by a
strong phase discontinuity \cite{Appel_LSD10} at the critical number
of flavors $N^*_f$ defining the sill of the conformal window; (see
footnote \ref{phase}, and item \ref{chiralphase} on page \pageref{chiralphase}).
Confinement, a light scalon and a large chiral condensate are presumed
to exist in the walking region for $N_f < N_f^*$, but suddenly disappear
for $N_f \geqslant N_f^*$, where amplitudes do not depend on dimensionful
constants and where many analyses even rely on two-loop perturbation
theory \cite{Caswell,BZ}. Why should the Higgs mass be continuous
at the phase discontinuity when everything else is not?

It has been suggested \cite{Golt16,Golt16a} that these contradictions
can be circumvented by apply\-ing Veneziano's version \cite{Gabriele}
of the large-$N_c$ limit ($N_f/N_c$ fixed) without crossing the sill.
But the logical difficulty remains that, no matter what limits are
taken, a region cannot be found where the theory is ``chirally
broken and confining'' and, at the same time, in the conformal WW mode.
Another problem for walking TC is that $N_f$ is large with $N_f^2-4$
physical light technipions, which is hard to reconcile with phenomenology.
All of these problems go away if the possibility of an NG-mode IRFP for
small $N_f$ is acknowledged.

We consider crawling TC for a QCD-like $SU(3)$ gauge theory
but with only $N_f =2$ flavors of massless Dirac techniquarks so that,
at low energies, all technipions are eaten giving SM gauge
bosons and fermions their masses.%
\footnote{A fully realistic version of our model would avoid stable,
fractionally charged technibaryons \cite{Chiv89} e.g.\ by including
a fourth generation of leptons to allow the techniquarks to carry
SM-like hypercharges.  We assume that any additional matter fields
are heavier than the electroweak scale and are therefore excluded as
dynamical degrees of freedom in the EFT.}
We stress that the form of the EFT to be derived below does not 
depend on $N_f$, as long as one is outside the conformal window. 
The choice $N_f=2$ simply avoids having to justify the absence of 
light physical technipions.

As noted in Sec.~\ref{WWorNG}, the possibility that IR fixed points
occur at small values of $N_f$ has been studied extensively \cite{Deur16},
but currently there is little direct evidence for or against their existence
(see Sec.~\ref{lattice}).  If present, they are almost certainly in the NG
scaling mode, as indicated in the left diagram of Fig.~\ref{fig:beta}.
That is because they lie outside the conformal window: dimensional
transmutation can occur, with the WW-mode scaling laws (\ref{CS WW}) replaced
by the soft-dilaton theorems (\ref{result}) and (\ref{result2}). 

We make the standard assumption that TC theory mimics massless QCD.
At the TeV scale and below, the technigluon coupling $\alpha$ is
strong, techniquarks and technigluons are confined and bound states
and resonances are expected to be produced. All technihadrons in the
non-NG sector are heavy, i.e.\ in the TeV range. Unlike QCD, the would-be
technipions are unphysical, but in crawling TC there is a pseudodilaton
(the Higgs particle), which plays a role similar to that of the QCD
resonance $f_0(500)$ in chiral-scale perturbation theory \cite{CT1,CT2}.
At energies well below $\Lambda_v$, one can build an EFT where the dynamical
degrees of freedom are the quarks, leptons and gauge fields of the SM and the
unconstrained Goldstone fields $\phi^a$ and $\sigma$.
Effects due to TC fields such as $\langle\bar{\psi}\psi
\rangle_\text{vac} \not=0$ are still present, but hidden inside the
low-energy coefficients of the EFT. The gauge potentials are
$G^A_\mu$, $W^a_\mu$ and $B_\mu$ with field-strength tensors
$G_{\mu\nu}^A$, $W_{\mu\nu}^a$ and $B_{\mu\nu}$ for $SU(3)_c$ gluons
and $SU(2)_L$ and $U(1)_Y$ electroweak bosons, respectively. The SM
fermions have the usual charge assignments under the SM gauge
group,
\begin{center}
\begin{tabular}{c|cccc}
\rule{0pt}{1em}%
& $SU(3)$           & $SU(3)_c$ & $SU(2)_L$ & $U(1)_Y$ \\[0.5mm]
\hline \rule{0pt}{4.2mm}$q_L$
& ${\bf 1}$         & ${\bf 3}$ & ${\bf 2}$ & $\hphantom{-}\frac{1}{6}$ \\[0.5mm]
$u_R$ & ${\bf 1}$ & ${\bf 3}$ & ${\bf 1}$ & $\hphantom{-}\frac{2}{3}$\\[0.5mm]
$d_R$ & ${\bf 1}$ & ${\bf 3}$ & ${\bf 1}$ & $-\frac{1}{3}$ \\[0.5mm]
$\ell_L$ & ${\bf 1}$       & ${\bf 1}$ & ${\bf 2}$ & $-\frac{1}{2}$ \\[0.5mm]
$e_R$  & ${\bf 1}$         & ${\bf 1}$ & ${\bf 1}$ & $-1$ \\
\end{tabular}
\end{center}
where generation indices $i=1,2,3$ on the matter fields are understood and
the $SU(2)_L$ doublets take the usual form
\begin{equation}\label{doublet}
q_L = \begin{pmatrix} u  \\ d \end{pmatrix}_L\qquad \mbox{and} \qquad
\ell_L = \begin{pmatrix} \nu_\ell \\ e \end{pmatrix}_L\,.
\end{equation}

In crawling TC, the SM Higgs doublet is replaced with a chiral-singlet
dilaton field $\sigma$ and a triplet of Goldstone fields $\phi^a$, so the
EFT combines the loop expansion of a renormalizable theory with that of an
effective Goldstone Lagrangian. This is in close analogy with, e.g., what
happens when pion dynamics is coupled to QED. It is understood that all mass
is to be produced by a Higgs-style mechanism, so the relevant renormalizable
Lagrangian is that for a massless version of the SM with terms depending on
massive constants like $v$ omitted. It is convenient to postpone including
the dilaton field $\sigma$; first we add to the massless SM Lagrangian the
lowest-order nonlinear chiral and Yukawa terms constructed from $U$ such
that SM gauge invariance is preserved. Under $SU(2)_L\times U(1)_Y$, $U$
must transform to a new matrix $\widetilde{U}$ which also satisfies the
constraint (\ref{constr}), i.e.\ it is unitary and obeys the condition
$\det\widetilde{U} = 1$:
\begin{equation}
U \to \widetilde{U} = V_L U V_Y \quad\mbox{with}\quad
\det V_L = 1 = \det V_Y\,.
\end{equation}
It follows that $V_Y$ is not proportional to $I$ and so $U$ does not have
a unique value of $Y$. Instead, $V_Y$ must belong to the $U(1)$ subgroup of
$SU(2)$ generated by $\tau_3$ [for consistency with the charge assignments
in Eq.~(\ref{doublet})]. That yields a familiar result 
\begin{align}
U \to \widetilde{U} = e^{i\bm{\omega\cdot\tau}} U e^{i\eta\tau_3}
\label{Long}\end{align}
originally obtained \cite{Long80} from the gauge property for the matrix
field for a heavy Higgs boson. Our presentation shows that there is no need
to introduce a Higgs field to determine the gauge property of $U$.

Then invariance under the SM gauge group gives the well-known
EFT Lagrangian for Higgsless dynamical electroweak symmetry in
leading order (LO) \cite{Appel80,Long80,Buch12}
\begin{widetext}
\begin{align}
{\cal L}_\text{no\,Higgs}
=\ &- \ltextfrac{1}{4} G_{\mu\nu}^A G^{A\mu\nu}
- \ltextfrac{1}{4}W^a_{\mu\nu}W^{a\mu\nu}
- \ltextfrac{1}{4}B_{\mu\nu}B^{\mu\nu}
+ \bar q_L i\slashed{D} q_L + \bar u_R i\slashed{D} u_R
+ \bar d_R i\slashed{D} d_R + \bar \ell_L i\slashed{D} \ell_L
+ \bar e_R i\slashed{D} e_R\,  \nn\\
&- v\Big\{\bar q_L \hat Y_u U {\cal U}_R + \bar q_L  \hat Y_d U {\cal D}_R
+ \bar \ell_L  \hat Y_e U {\cal E}_R + \text{H.c.} \Big\}
+\ltextfrac{1}{4}v^2\text{tr}( D_\mu U D^\mu U^\dagger )\,,
\label{LO Higgsless}
\end{align}
\end{widetext}
where the doublet notation
\begin{equation}
\bar{{\mathcal{U}}}_R = \begin{pmatrix}\bar{u}_R\ &\ 0\end{pmatrix}
\,,\
\bar{{\mathcal{D}}}_R = \begin{pmatrix}0\ &\ \bar{d}_R\end{pmatrix} \
\,,\
\bar{{\cal E}}_R = \begin{pmatrix}0\ &\ \bar{e}_R\end{pmatrix}
\end{equation}
for right-handed fermions matches the $2 \times 2$ matrix $U$, and
$\hat Y_{u,d,e}$ are $3\times 3$ Yukawa matrices in generation space. The
masses for the gauge bosons and fermions are contained in the last line
when $U = I$ (unitary gauge). In terms of the $U(1)_Y$ hypercharges
$Y_f$ tabulated above, the gauge-covariant derivatives of quark fields 
$q = u\mbox{ or }d$ are
\begin{align}\label{covD}
D_\mu q_L
&= (\partial_\mu + ig_s G_\mu + ig_w W_\mu + ig_w' Y_{q_L} B_\mu) q_L \,,
\nn \\
D_\mu q_R &= (\partial_\mu + ig_s G_\mu + ig_w' Y_{q_R} B_\mu) q_R \,,
\end{align}
with analogous expressions for leptons obtained by omitting the $SU(3)_c$
terms. The covariant derivative associated with the gauge property
(\ref{Long}) is \cite{Long80}
\begin{equation}
D_\mu U = \partial_\mu U + ig_w W_\mu U - \tfrac{i}{2}g_w' B_\mu U\tau^3\,.
\end{equation}

Equation (\ref{LO Higgsless}) can be made scale invariant by multiplying
each operator by an appropriate power of the dimension-1 field
$e^{\sigma/F_\sigma}$ and adding a dilaton kinetic term \cite{Ell70}
\begin{equation}
\ltextfrac{1}{2}
F^2_\sigma\bigl(\del e^{\sigma/F_\sigma}\bigr)^2
= \ltextfrac{1}{2}
e^{2\sigma/F_\sigma} \del_\mu\sigma\del^\mu\sigma \, .
\label{ke}
\end{equation}
More generally,  approximate scale invariance implies that a chiral
Lagrangian operator ${\cal Q}$ with dynamical dimension $d_{\cal Q}$
is replaced by
\begin{align}\label{mod}
&{\cal Q}_\sigma = {\cal Q} \times \Big\{c_{\cal Q} e^{(4-d_{\cal Q})\sigma/F_\sigma}
  + (1-c_{\cal Q}) e^{(4-d_{\cal Q}+\beta')\sigma/F_\sigma}\Big\} \nn \\
&\phantom{{\cal Q}_\sigma} 
  = c_{\cal Q}{\cal{Q}}_\text{inv}+(1-c_{\cal Q}){\cal{Q}}_{\beta'} \,.
\end{align}
Here ${\cal{Q}}_\text{inv}$ has dimension 4 (the scale-invariant
part), while ${\cal{Q}}_{\beta'}$ accounts for explicit scale symmetry breaking
by the trace anomaly near $\alphaIR$ and so has dimension $4 + \beta'$
(Appendix \ref{app:scaling}). The coefficient of ${\cal{Q}}_{\beta'}$ is fixed
by requiring that the original operator ${\cal Q}$ be recovered in the absence
of dilaton interactions. The dimensions $d_{\cal Q}$ take the naive values
implied by canonical dimensions, i.e.\ $1$ and $\frac{3}{2}$ for gauge and
fermion fields and $0$ for the unitary field $U$.%

The values of the low-energy constants $c_{\cal Q}$ depend on dynamics and
are not fixed by symmetry arguments alone. However, scale invariance imposes
constraints on them. For a Lagrangian of the form $\sum_j {\cal Q}_\sigma^j$,
the trace of the improved energy-momentum tensor is
\begin{align}
\theta^{\mu}_{\,\,\mu} \big|_\text{eff}
&= \sum_j \bigl(d_{{\cal Q}^j_\sigma} - 4\bigr)\Bigl\{ {\cal Q}_\sigma^j
- \bigl\langle {\cal Q}_\sigma^j\bigr\rangle_{\text{vac}}\Bigr\} \nn \\
&= \beta' \sum_j (1- {c_{{\cal{Q}}j}})\Big\{ {\cal{Q}}_{\beta'}^j
- \langle{\cal{Q}}_{\beta'}^j\rangle_\text{vac}\Big\} \,,
\end{align}
where only operators ${\cal{Q}}^j_{\beta'}$ with dynamical dimension
$\neq 4$ contribute. The requirement that this expression vanish in the
scale-invariant limit $\theta_\mu^\mu \to 0$ implies \cite{CT3}
\begin{align}
c_{{\cal Q}j} = 1 + O(\epsilon)\,,
\end{align}
where the correction $O(\epsilon)$ is due to the explicit breaking 
of scale invariance by the trace anomaly in the low-energy region
$\alpha \lesssim \alphaIR$.

In the limit $\epsilon \to 0$, a potential $\sim e^{4\sigma/F_\sigma}$
is forbidden by Zumino's consistency condition (\ref{Bruno}). If
its coefficient is $O(\epsilon)$,  $e^{4\sigma/F_\sigma}$ by itself is
still not acceptable because it has no minimum for finite variations of
the unconstrained field $\sigma$ (Sec.~\ref{Zumino}).
However, a dilaton potential $V$ of first order in $\epsilon$ is
possible because there can be a term of dimension $4 + \beta'$ as
well, 
\begin{equation}\label{VLOa}
V(\sigma) = c_{1V} \Bigl\{e^{4\sigma/F_\sigma} - 1\Bigr\}
+ c_{2V} \Bigl\{e^{(4+\beta')\sigma/F_\sigma} - 1\Bigr\} \,,
\end{equation}
\noindent \pagebreak
where both $c_{1V}$ and $c_{2V}$ are $O(\epsilon)$, with constant
terms subtracted  as in Zumino's example (\ref{FN}). The function
$V(\sigma)$ has a minimum for $c_{1V} < 0$ and $c_{2V} > 0$, which we
assume. The value $\langle\sigma\rangle_{\text{vac}}$ of $\sigma$ at the
minimum is a matter of convention because field translations
\begin{equation}
\sigma \to \sigma + \text{constant}
\end{equation}
merely affect which of the physically equivalent scale-invariant worlds is
chosen as $\epsilon \to 0$ (Appendix \ref{app:ward}). Our convention for
$\sigma$ is  $\langle\sigma\rangle_{\text{vac}} = 0$. Minimizing $V$ at
$\sigma = 0$, we have
\begin{equation}
4c_{1V} + (4+\beta')c_{2V} = 0
\end{equation}
so (say) $c_{1V}$ can be eliminated in favor of $c_{2V}$. In turn,
$c_{V2}$ can be eliminated in terms of the pseudodilaton mass
$M_\sigma$ by equating the second-order term of $V$ to
$\frac{1}{2}M_\sigma^2\sigma^2$. The result is an explicit formula
for the LO Higgs potential in crawling TC:
\begin{align}
&V(\sigma) = \frac{M_\sigma^2F_\sigma^2}{\beta'}
 \nonumber \\
& \times\left[-\frac{1}{4}e^{4\sigma/F_\sigma}
  + \frac{1}{4+\beta'}e^{(4+\beta')\sigma/F_\sigma}
  + \frac{\beta'}{4(4+\beta')}\right]  \,.
\label{VLO}
\end{align}
The form of this potential is fixed solely by the presence of an IR fixed
point and the fact that the explicit breaking of scale symmetry occurs
through the operator $\hat{G}^2$.

The full LO Lagrangian ${\cal L}_{\textsc{lo}}$ is obtained by collecting
from above all modifications of the Higgsless Lagrangian (\ref{LO Higgsless})
and discarding terms considered to be next-to-leading order (NLO). 
It is at this point that consistent rules for the expansion into LO,
NLO, next-to-NLO, \ldots must be
adopted. As for any EFT with an underlying strongly coupled dynamics, the
expansion is organized by the number of loops, with each order absorbing
the divergences of the previous one. This ensures that the EFT is renormalized
order by order. In the following we will rely on $v \sim F_\sigma$ (see the
end of Sec.\ \ref{comments}).

The pure-dilaton part of ${\cal L}_{\textsc{lo}}$ is easily found,
being so similar to a standard nonlinear chiral Lagrangian. We seek
a simultaneous expansion in momenta $p$ and masses of pNG bosons
(just $M_\sigma$ in our case):
\begin{equation}
p \sim M_\sigma \ll 4\pi F_\sigma \sim \Lambda_v \,.
\end{equation}
The scale-invariant kinetic term (\ref{ke})  is already $O(p^2)$, so
extra $O(p^2\epsilon)$ terms generated by Eq.~(\ref{mod}) for ${\cal Q} \to
\frac{1}{2}(\del\sigma)^2$ are NLO. The $O(p^2)$ term (\ref{ke}) is of
the same order as the $O(M^2_\sigma)$ dilaton potential $V$, so the LO
contribution is
\begin{equation}
{\cal L}_{\sigma,\,\textsc{lo}}
= \ltextfrac{1}{2}
   e^{2\sigma/F_\sigma} \del_\mu\sigma\del^\mu\sigma
 - V(\sigma)  \,.
\label{Lsigma}
\end{equation}
This Lagrangian is suitable for the tree approximation: the $p, M_\sigma$
dependence of each propagator $i/(p^2 - M_\sigma^2)$ is compensated by the
$O(p^2)$ or $O(M_\sigma^2)$ behavior of the next vertex.

The remaining terms in ${\cal L}_{\textsc{lo}}$ are obtained by making the
Higgsless Lagrangian (\ref{LO Higgsless}) scale invariant. The result
\begin{widetext}
\begin{align}\label{LOchiD}
{\cal L}_{\textsc{lo}} 
&= - \ltextfrac{1}{4} G_{\mu\nu}^A G^{A\mu\nu}
- \ltextfrac{1}{4}W^a_{\mu\nu}W^{a\mu\nu} - \ltextfrac{1}{4}B_{\mu\nu}B^{\mu\nu}
 + \bar q_L i\slashed{D} q_L + \bar u_R i\slashed{D} u_R
  + \bar d_R i\slashed{D} d_R + \bar \ell_L i\slashed{D} \ell_L
  + \bar e_R i\slashed{D} e_R   \nn\\
 &\ + \ltextfrac{1}{2}e^{2\sigma/F_\sigma} \del_\mu\sigma\del^\mu\sigma
       - V(\sigma) +\ltextfrac{1}{4}v^2
       \text{tr}( D_\mu U D^\mu U^\dagger)e^{2\sigma/F_\sigma}
       - v\Big\{ \bar q_L \hat Y_u U {\mathcal{U}}_R
  + \bar q_L  \hat Y_d U {\mathcal{D}}_R
  + \bar \ell_L \hat Y_e U {\cal E}_R +
      \text{H.c.}\Big\}e^{\sigma/F_\sigma}
\end{align}
\end{widetext}
describes the low-energy behavior of strongly interacting TC plus
weak interactions of the dilaton with the SM gauge bosons and fermions.
As we saw for the dilaton kinetic energy, not all terms from Eq.~(\ref{mod})
are needed, but the reasons for this are less obvious and require a
discussion.

Equation (\ref{LOchiD}) contains covariant derivatives $D_\mu$ given
by Eq.~(\ref{covD}), so gauge invariance requires products like
$g_w W_\mu$ in Eq.~(\ref{covD}) to be counted like $\del_\mu$, i.e.\ as
$O(p)$. In the original version of this rule \cite{Gass84, Gass85},
the gauge field was taken to be $O(p)$. That choice works if the field
is external, but it is not suitable when gauge propagators appear; then
it is necessary to require \cite{Urech94} 
\begin{equation} \mbox{gauge field } = O(1)\ , \ \mbox{ charge } = O(p)
\label{urech}
\end{equation}
so that gauge-boson kinetic energies are $O(p^2)$, corresponding
to $O(p^{-2})$ behavior for the gauge propagators. In bosonic diagrams,
this $O(p^{-2})$ behavior is compensated by the next vertex being
$O(p^2)$ because of the $O(p)$ rule (\ref{urech}) for gauge
coupling constants.%
\footnote{In Sec.~VI and Appendices A and B of Ref.~\cite{CT2}, the old $O(p)$
rule for an external photon field $A_\mu$ was used. To adapt that discussion
to fit Eq.~(\ref{urech}), simply regard $eA_\mu$ as the photon source, where
$-e$ is the electron's charge. Then everything works, including chiral
power counting for UV-convergent one-loop amplitudes such as $K_S \to
\gamma\gamma$ \cite{d'A_Espriu86,Goity87}.}
\pagebreak
There are \mbox{separate} rules for fermions because propagators are $O(p^{-1})$
\cite{Knecht99}:
\begin{equation}
\mbox{fermion field } = O(p^{1/2})\,, \mbox{Yukawa coupling }
= O(p) .
\label{knecht}
\end{equation}

The rules (\ref{urech}) and (\ref{knecht}) should be understood purely as
tagging devices to ensure correct power counting for $\ell$-loop chiral
amplitudes. Numerical estimates for the gauge couplings $g_w,g_w',g_s$
in various energy regimes should not be inferred beyond the requirement
that perturbation theory remains applicable.

The rules for the tree approximation and beyond can be efficiently described
in terms of chiral dimensions \cite{Buchalla:2013eza,Buchalla:2016sop}.
Fields or coupling constants counted as $O(p^{[\ldots]})$ above are said to
have chiral dimension $[\ldots]$:
\begin{alignat}{2}\label{chidim}
[G_\mu, W_\mu, B_\mu, \sigma, \phi^a] &= 0\,,  & \quad
 [\psi]& = \tfrac{1}{2}\,, \nn \\
[g_s, g_w, g_w', \hat{Y}_{u,d,e}, \del_\mu] &= 1\,, & \quad
[M_{\sigma}^2]& = 2\,.
\end{alignat}
The construction of the LO Lagrangian is summarized by the rule that
it be homogeneous in chiral dimension with $[{\cal L}_{\textsc{lo}}] = 2$.

The utility of Eq.~(\ref{chidim}) becomes evident beyond LO,
where rules for the low-energy expansion of EFT loop diagrams are
needed. Despite the complications presented by the structure of loop
diagrams, the rule for the N$^\ell$LO Lagrangian at $\ell$-loop order
implied by low-energy power counting is \mbox{simple}:  
construct all operators with chiral dimension $2\ell + 2$, i.e.
\begin{equation}\label{chidim2}
{\cal L}_{\textsc{eft}} = \sum_{\ell \geqslant 0}
{\cal L}_{\textsc{n}^\ell\textsc{lo}}\ \mbox{ with }
\bigl[{\cal L}_{\textsc{n}^\ell\textsc{lo}}\bigr] = 2\ell + 2.
\end{equation}
At each order, the set of operators includes renormalization counterterms
needed to render all loop integrals UV convergent. Similarly, on-shell
N$^\ell$LO amplitudes are $O(p^{2\ell + 2})$ modulo logarithms, where
$p \sim 0$ refers to momenta, NG boson masses and the rule (\ref{chidim})
for coupling constants.

NLO amplitudes are interesting \cite{Buchalla:2017jlu,Alonso:2017tdy},
especially for electroweak processes like $h\to\gamma\gamma$. The QCD
analogues of these are two-photon reactions such as $f_0(500) \to
\gamma\gamma$ in chiral-scale perturbation theory
(the forerunner of crawling TC). However, as noted in Appendix A of
Ref.~\cite{CT2}, a general NLO analysis depends on the next order of Taylor
expansions of $\beta$ and $\gamma$ functions in $\alpha$ about $\alphaIR$,
which would take us far afield. Therefore, for the remainder of this
article, we restrict our attention to the LO approximation.

An immediate comparison of the Higgs sector of crawling TC with that of
the SM is obscured by the complicated $\sigma$ dependence of the formula
(\ref{LOchiD}) for ${\cal L}_{\textsc{lo}}$. However, in the tree approximation,
there is a field redefinition \cite{Buch13}
\begin{align}
h=\int_0^\sigma \! e^{\sigma'/F_\sigma} d\sigma'=F_{\sigma}(e^{\sigma/F_\sigma}-1)\,,
\quad h \geq -F_\sigma
\end{align}
which simplifies the structure of Higgs vertices and importantly, satisfies
the requirements of the equivalence theorem%
\footref{equivalence};
indeed, $h$ is just the constrained field $\phi_c$ already discussed in
Eqs.~(\ref{Lex1}) and (\ref{invert}). The constraint on $h$ refers to the
fact that its scale transformations
\begin{equation}
h \to \rho^{-1}h + F_\sigma\bigl(\rho^{-1} - 1\bigr)\ ,\quad x \to \rho x
\end{equation}
are restricted to the region $-F_\sigma \leqslant h < \infty$. The change
of field variables $\sigma \to h$ is permitted by the theorem because
fluctuations $\sigma \sim 0$ about the minimum of $V$ are mapped to
$h \sim 0$.

With this redefinition, the EFT Lagrangian (\ref{LOchiD}) becomes
\begin{widetext}
\begin{align}\label{eff1}
{\cal L}_{\textsc{lo}} =\ &
- \ltextfrac{1}{4} G_{\mu\nu}^A G^{A\mu\nu}
- \ltextfrac{1}{4}W^a_{\mu\nu}W^{a\mu\nu}
- \ltextfrac{1}{4}B_{\mu\nu}B^{\mu\nu}
+\bar q_L i\slashed{D} q_L + \bar u_R i\slashed{D} u_R
+ \bar d_R i\slashed{D} d_R + \bar \ell_L i\slashed{D} \ell_L
+ \bar e_R i\slashed{D} e_R
\nn \\[1mm]
& +\ltextfrac{1}{2}(\partial h)^2 - V(h)
+\ltextfrac{1}{4}v^2\text{tr}( D_\mu U D^\mu U^\dagger )
\bigl(1 + h/F_\sigma\bigr)^2 \nn \\[1mm]
& -v\Big\{ \bar q_L \hat Y_u U \mathcal{U}_R + \bar q_L \hat Y_d U \mathcal{D}_R
+ \bar \ell_L \hat Y_e U \mathcal{E}_R + \text{H.c.}\Big\}
\bigl(1 + h/F_\sigma\bigr) \,.
\end{align}
\end{widetext}
Apart from the Higgs potential $V$, this LO result resembles the
SM, where the factors $(1 + h/F_\sigma)^2$ and $(1 + h/F_\sigma)$
in the last two terms become $(1 + h/v)^2$ and $(1 + h/v)$
respectively. Similar results are often quoted for ``dilatonic''
or scalon-type theories, despite the WW mode being chosen in the
limit of scale invariance. The difference
is that the experimental result $F_\sigma \sim v$ indicated by
measurements \cite{Aad16} of the decays $h \to \tau\tau,\ WW,\ ZZ$
and $b\bar{b}$ is expected in crawling TC but requires ``fine
tuning'' in scalon theories (Secs. \ref{GilWein} and \ref{comments}).

What clearly distinguishes our result from the SM and other theories
is the unique dependence of the Higgs potential $V$ on the nonperturbative
constant $\beta'$:
\begin{widetext}
\begin{align}\label{pot}
&V(h) = \frac{M_{\sigma}^2F_{\sigma}^2}{\beta'}
\left[-\frac{1}{4}
\left(1+\frac{h}{F_\sigma}\right)^4+\frac{1}{4+\beta'}
\left(1+\frac{h}{F_\sigma}\right)^{4+\beta'} \!+ \frac{\beta'}{4(4+\beta')}\right] \,.
\end{align}
In the expansion
\begin{equation}
V(h) = \sum _{n \geqslant 2} a^{}_n\bigg(\frac{h}{F_\sigma}\bigg)^n\,,
\end{equation}
the coefficients $a_n$ are given by
\begin{align}\label{coeff}
a_n =\ &\frac{M_\sigma^2 F_\sigma^2}{\beta'} \frac{1}{n!} \Bigg\{
\frac{\Gamma[4+\beta']}{\Gamma[5+\beta'-n]} - \frac{3!}{(4-n)!}\Bigg\}
\ , \mbox{ with}\quad \frac{1}{(4-n)!} \equiv \left\{\begin{array}{cl} 1\;,& n = 4\,, \\
0\;,& n > 4\,. \end{array}\right. 
\end{align}
\end{widetext}
We have retained the notation $M_{\sigma}$ for the LO mass of $h$ as a
reminder that our Higgs boson is a genuine pseudodilaton. Expanding the
gamma functions in Eq.~(\ref{coeff}) yields
\begin{align}
V(h) =\ M_\sigma^2&F_\sigma^2 \Bigg\{ \frac{1}{2}\bigg(\frac{h}{F_\sigma}\bigg)^2
   + \frac{5+\beta'}{3!}\bigg(\frac{h}{F_\sigma}\bigg)^3  \nn\\
 &+ \frac{11+\beta'(\beta'+6)}{4!}\bigg(\frac{h}{F_\sigma}\bigg)^4
   + O(h^5) \Bigg\}\,.
\label{VHiggs}
\end{align}
The corresponding SM formula is
\begin{equation}
V_{\textsc{SM}}(h)
= \ltextfrac{1}{2}m_h^2 h^2\Bigl\{1 + h\bigl/(2v)\Bigr\}^2\,.
\end{equation}
Unlike the SM, our effective theory is not renormalizable, so powers
$\sim h^5$ and higher are present in Eq.~(\ref{VHiggs}). Although a
determination of the Higgs quartic self-coupling appears to be out of the
LHC's reach and very challenging even for future colliders \cite{Fuks17},
measurements of Higgs double production during the high-luminosity phase
of the LHC should make it possible to place bounds on the cubic coupling
\cite{Dolan15,Cao15}. Currently, this is the most promising way to
determine $\beta'$.

Subject to the NG-mode requirements for $h$ noted above, it is evident
that our Lagrangian (\ref{eff1}) belongs to a class of Lagrangians
\cite{Feru92,Bagg94,Koul93,Burg99,Wang06,Grin07,Contino10,Alon12,Buch13}
proposed for dynamical electroweak theories with the Higgs field treated as
a generic scalar. General formulas are given in Sec.~2 of Ref.~\cite{Buch13}.

Given the explicit form of the potential, we can check the relation
(\ref{MsigPCDC}) between the dilaton mass and the technigluon condensate.
Let us compare the trace anomaly in the EFT
\begin{equation}
\theta_\mu^\mu\big|_\text{eff} = -\frac{ M_\sigma^2 F_\sigma^2 }{4+\beta'}
\left\{\biggl(1+\frac{h}{F_{\sigma}}\biggr)^{4+\beta'} -1\right\}
\label{EFTtrace}
\end{equation}
with that of the underlying theory
\begin{equation}
\theta_\mu^\mu = -\frac{\epsilon \beta'}{4\alphaIR}
\Big\{
 \hat G^2 - \langle\hat G^2 \rangle_\text{vac}
\Big\} + O(\epsilon^2)\,.
\end{equation}
As specified in Appendices \ref{technigluon} and \ref{app:scaling}, the
operator $\hat G^2$ scales homogeneously (no mixing with the identity operator
$I$) and with dynamical dimension $4 + \beta'$. That is also true for the
operator $(1 + h/F_\sigma)^{4 + \beta'}$ in Eq.~(\ref{EFTtrace}), so we can conclude
that $\langle\hat G^2 \rangle_{\text{vac}}$ corresponds to the remaining term
in Eq. (\ref{EFTtrace}):
\begin{equation}
\frac{\epsilon \beta'}{4\alphaIR}
\bigl\langle\hat G^2 \bigr\rangle_\text{vac}
= \frac{ M_\sigma^2 F_\sigma^2 }{4+\beta'}  + O(\epsilon^2)\,.
\end{equation}
This yields the LO formula
\begin{equation}
M_\sigma^2 = \frac{\epsilon\beta'(4+\beta')}{4\alphaIR F_\sigma^2}
\bigl\langle\hat G^2\bigr\rangle_\text{vac}
 + O(\epsilon^2)\,,
\label{Msig-EFT}
\end{equation}
in agreement with the result (\ref{MsigPCDC}) derived from the CS equations.
At NLO, the correction to Eq.~(\ref{Msig-EFT}) is of the form
$\sim M_\sigma^4\ln(M_\sigma/\Lambda_{\sigma,v})$; scale invariance forbids
quadratic dependence on the scales $\Lambda_{\sigma,v}$.

\section{Signals for crawling TC on the lattice}
\label{lattice}

\begin{figure}[h]
\centering\includegraphics[scale=0.6]{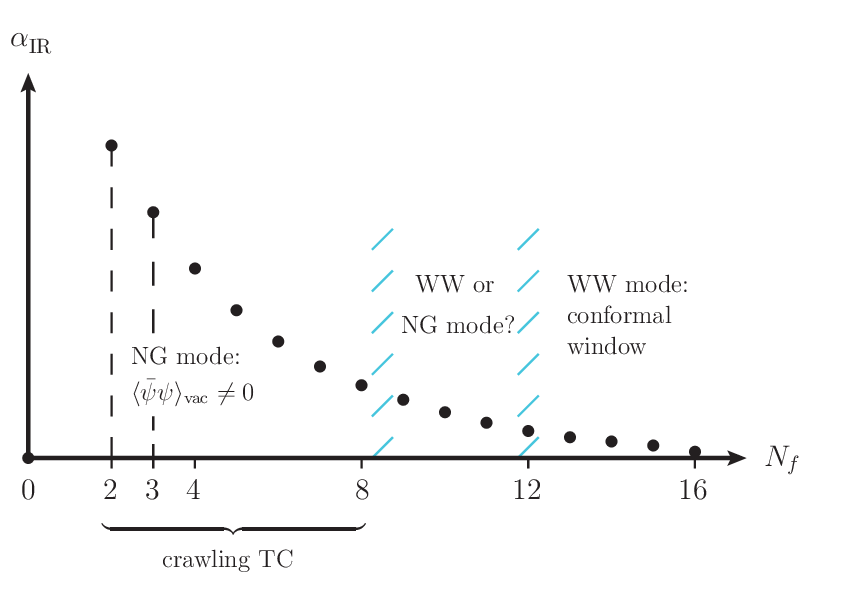}
\caption{Search for IR fixed points in $SU(3)$ gauge theories with $N_f$ Dirac
fermions in the triplet representation.  The diagram assumes that IR fixed
points of some kind exist from $N_f = 2$ to $N_f = 16$, with
chiral-scale perturbation theory $\chi$PT$_\sigma$ at $N_f=3$.  
Crawling TC with $N_f^2 - 4$ physical technipions (none for $N_f = 2$)
is possible anywhere outside the conformal window.
For simplicity of presentation, we choose
to have $\alphaIR$ (or $\alphaWW$ for $N_f \gtrsim 12$) be a decreasing
function of $N_f$ for a given renormalization scheme.}
\label{fig:IRFP_search}
\end{figure}

Throughout this paper we have been careful to distinguish IR fixed points
according to whether scale invariance is realized in the WW or NG mode.  This
distinction is especially important for the search for IR fixed points on the
lattice, since these investigations typically rely on criteria or techniques
that apply to the WW mode only.  In particular, the existence of NG-mode IR
fixed points cannot be inferred from spectral studies based on hyperscaling
relations \cite{Del10}, because these scaling laws are forbidden by the
soft-dilaton results of Eqs.~(\ref{result}) and (\ref{result2}): conformal
symmetry is \emph{hidden}.  Another key point [noted below Eq.~(\ref{IRlogs}) in
Sec.~\ref{NGsolutions}] is that NG-mode IR fixed points are theoretically
possible for any value of $N_f$ \emph{outside} the conformal window.  When
conformal symmetry is hidden, there is no need to tune $N_f$ to lie on the edge
of the conformal window (which is standard practice in ``dilatonic'' walking gauge
theories \cite{Appel10,Yam11,Appel13,Yam14,Golt16}).  This suggests that the
set of IR fixed points may be larger than previously envisioned; a sketch for
$SU(3)$ gauge theories is shown in Fig.~\ref{fig:IRFP_search}.

The most direct ways to look for candidate theories of crawling TC on
the lattice are as follows:
\begin{enumerate}[label=(\arabic*)]
\item  Search for the ``freezing'' of $\alpha$ \cite{Deur16} for $N_f$ values
outside the conformal window, where dimensional transmutation and chiral
condensation can occur. Note that the UV expansions of $\alpha$ typically
used to measure nonperturbative corrections to asymptotic freedom for QCD
amplitudes are not applicable. The energy scale must be lowered beyond the
region where UV expansions break down and into the far IR region
$\alpha \lesssim \alphaIR$.
\item  Test whether amplitudes exhibit the singular behavior displayed in
Eq.~(\ref{sing}).
\item  Confirm the presence of a pseudodilaton with a small value of
$M_\sigma^2$, especially if Eq.~(\ref{MsigPCDC}) or Eq.~(\ref{Msig-EFT})
can be tested.
\end{enumerate}
Lattice calculations for $N_f=8$ triplet fermions 
\cite{Aoki14,Aoki16,Appel16,Appel19}
and $N_f=2$ sextet fermions \cite{Fod14} suggest that, in the IR region,
$\alpha$ varies slowly and chiral condensation produces technipions (63 for
$N_f=8$ triplets and 3 for $N_f=2$ sextets) plus a light scalar boson. These
examples have been taken as support for walking gauge theories, but
they could actually point to crawling scenarios. At present, definite 
conclusions cannot be drawn because current lattice investigations can
be matched to chiral constraints only for large fermion masses
\cite{LSD2,Gasb17,Gasb18,Appel19a}. Consequently these investigations  
may be too far from the chiral limit for soft dilaton theorems to apply.

If a candidate pseudodilaton $\sigma$ is seen on the lattice for a small
number $N_f$ of fermion triplets, it may be possible to determine
$F^2_\sigma$ from the $\sigma$-pole residue of a component of
$\langle\theta_{\alpha\beta}(x)\theta_{\mu\nu}(0)\rangle_{\text{vac}}$.
(We say ``may,'' because explicit breaking of space-time symmetries 
by the lattice regulator makes $\theta_{\alpha\beta}$ hard to study on the
lattice \cite{Suz16}.) Since lattice simulations are performed with
massive techniquarks $m_\psi \not= 0$, what is actually measured is
the $m_\psi \not= 0$ version $\tilde{F}_\sigma$ of the decay constant 
[see Eq.~(\ref{Msigfull})], from which $F_\sigma$ is found by extrapolating 
in $m_\psi$ to $m_\psi = 0$.

Then the anomalous dimension $\gamma_m$ of $\bar{\psi}\psi$ at $\alphaIR$
can be deduced from the LO soft-$\sigma$ theorem
\begin{equation}
3-\gamma_m
 = F_\sigma\frac{\bigl\langle\sigma\bigl|\bar{\psi}\psi\bigr|\text{vac}\bigr\rangle}%
 {\bigl\langle\text{vac}\bigl|\bar{\psi}\psi\bigr|\text{vac}\bigr\rangle}
 + O(m_\psi)  \,.
 \end{equation}
 This involves the $\sigma$-pole residues of
 $\langle\bar{\psi}\psi(x)\bar{\psi}\psi(0)\rangle_{\text{vac}}$ and
 $\langle\theta_{\alpha\beta}(x)\bar{\psi}\psi(0)\rangle_\text{vac}$, where
 the latter is required to check magnitudes and fix the sign of $3-\gamma_m$.

 An analysis \cite{Golt16} of $m_\psi$ dependence in effective Lagrangians
 has led to proposals \cite{Appel17,Appel17a} that ``dilaton-based'' potentials
 for walking TC and conformally deformed theories be tested on the lattice.
 Can a similar approach be applied to our Higgs potential (\ref{VLOa})
 and hence determine $\beta'$?  

 The first step of Refs.~\cite{Golt16,Appel17,Appel17a} is to account for
 $m_\psi \not= 0$ by including in the EFT a chiral mass operator 
 which  shifts the VEV of the dimension-$1$ scalar field $\chi$ from 
 $\langle\chi \rangle_{m_\psi = 0}$ to $\langle\chi\rangle_{m_\psi \not= 0}$. Constraints
 on decay constants and spin-$0$ masses are found by minimizing the
 $m_\psi \not= 0$ potential and evaluating its curvature at the minimum.
 The $m_\psi$ dependence of the results appears to be entirely determined
 by dependence on the shift
 $\langle\chi\rangle_{m_\psi = 0} \to \langle\chi\rangle_{m_\psi \not= 0}$.

 Applying the same procedure to Eq.~(\ref{VLO}), one finds that physical
 results do not depend on the value chosen for $\langle\chi\rangle > 0$, or
 equivalently, for $\langle\sigma\rangle$. Invariance under $\sigma \to \sigma$ +
 constant is expected for a true dilaton because the equivalence theorem  
 allows us to choose any real value of $\langle\sigma\rangle$: indeed,
 $\langle\sigma\rangle$ does not have to vanish even for $m_\psi \to
 0$. The equations are simplest if we use this freedom to retain the
 choice $\langle\sigma\rangle = 0$ as $m_\psi$ is turned on. Even with that
 choice, it is necessary to distinguish observable masses and decay constants
 such as $\tilde{M}_\sigma$ and $\tilde{F}_\sigma$ for values of $m_\psi \not= 0$
 from those for $m_\psi = 0$ ($M_\sigma$ and $F_\sigma$). Similarly, we replace the
 $m_\psi = 0$ coefficients $c_{1V}$ and $c_{2V}$ of Eq.~(\ref{VLOa}) by their
 counterparts $\tilde{c}_{1V}$ and $\tilde{c}_{2V}$ for $m_\psi \not= 0$.
 Since $c_{1V}$ and $c_{2V}$ are counted as LO, we must also count $O(m_\psi)$
 corrections to them as LO, as in chiral-scale perturbation theory
 \cite{CT1,CT2,CT3}:
 \begin{align}\label{tilde}    
 \tilde{c}_{nV} = c_{nV} + m_\psi d_{nV}
 + O(m_\psi^2,m_\psi\epsilon,\epsilon^2)\ , \quad n= 1,2.
 \end{align}
 Here $d_{nV}$ do not depend on the scale-breaking parameters $m_\psi$ or
 $\epsilon$. In terms of $\sigma$, the effective LO mass operator for $N_f$
 degenerate flavors is \cite{CT1,CT2,CT3}
 \begin{align}
 {\cal L}_\text{mass} =\ &\ltextfrac{1}{2}
  m_\psi B_\pi F_\pi^2
 \text{Tr}(U + U^\dagger)e^{(3-\gamma_m)\sigma/\tilde{F}_\sigma} \nn \\
 &+ O(m_\psi^2,m_\psi\epsilon) \,,
 \end{align}
 where to first order in $m_\psi$, we can use the $m_\psi = 0$ values of the decay
 constants $F_\pi$ and the condensate constant $B_\pi$ appearing in
 $\langle\bar\psi^j\psi^i\rangle_\text{vac} = - F_\pi^2
 B_\pi\delta^{ij}$.

 The $m_\psi$-dependent potential to be  minimized is 
 \begin{align}
 &\tilde{V}(\sigma) =
 \tilde{c}_{1V}e^{4\sigma/\tilde{F}_\sigma}
 + \tilde{c}_{2V}e^{(4+\beta')\sigma/\tilde{F}_\sigma} \nn\\
 &\ - N_f m_\psi B_\pi F_\pi^2 e^{(3-\gamma_m)\sigma/\tilde{F}_\sigma}
 + O(m^2_\psi, m_\psi\epsilon, \epsilon^2) \,.
 \end{align}
 Keeping just LO terms $O(\epsilon)$ and $O(m_\psi)$, the result of
 minimizing $\tilde{V}$ at $\sigma = 0$ is
 \begin{equation}\label{first}
 0\ \underset{\textsc{lo}}{=}\ 4\tilde{c}_{1V} + (4+\beta')\tilde{c}_{2V}
 - N_f m_\psi B_\pi F_\pi^2 (3-\gamma_m) \,.
 \end{equation}
 We now set the $O(\sigma^2)$ term equal to $\frac{1}{2}\tilde{M}_\sigma^2\sigma^2$
 and find
 \begin{align}\label{second}                
 \tilde{M}_\sigma^2\tilde{F}_\sigma^2 \underset{\textsc{lo}}{=}
 16\tilde{c}_{1V} + (4+\beta')^2\tilde{c}_{2V}
 - N_f m_\psi B_\pi F^2_\pi(3 - \gamma_m)^2  
 \end{align}
 which, from Eqs.~(\ref{tilde}) and (\ref{first}), corresponds to $m_\psi$
 dependence
 \begin{align}
 &\tilde{M}_\sigma^2\tilde{F}_\sigma^2\ \underset{\textsc{lo}}{=}\
 M_\sigma^2 F_\sigma^2 \nn \\
 &\hspace{0.5cm}+ m_\psi\left[d_{2V}\beta'(4+\beta') + N_f B_\pi
 F_\pi^2(3-\gamma_m)(1+\gamma_m)\right]  \,.
 \end{align}
 Equations (\ref{first}) and (\ref{second}) can be solved for $\tilde{c}_{1V}$
 and $\tilde{c}_{2V}$:
 \begin{align}
 &4\beta'\tilde{c}_{1V}  \underset{\textsc{lo}}{=}
 - \tilde{M}_\sigma^2\tilde{F}_\sigma^2
 + N_f m_\psi B_\pi F^2_\pi(3 - \gamma_m)(\beta' + \gamma_m + 1) \,,
 \nn \\
 &\beta'(4+\beta')\tilde{c}_{2V}  \underset{\textsc{lo}}{=}
 \tilde{M}_\sigma^2\tilde{F}_\sigma^2
 - N_f m_\psi B_\pi F^2_\pi(3 - \gamma_m)(\gamma_m + 1) \,.
 \label{coeffs}
 \end{align}
 Unlike Refs.~\cite{Golt16,Appel17,Appel17a}, we do not find any additional
 constraints at this stage. Without input from higher-order terms
 in $\tilde{V}$, we have no independent information about
 $\tilde{c}_{1V}$ or $\tilde{c}_{2V}$.

 A determination of $\beta'$ from $\tilde{V}$ is difficult because it
 involves calculating a three-point function on the lattice and then
 going on shell to measure the cubic Higgs coupling. Given Eq.
 (\ref{coeffs}), we find 
\begin{widetext}
 \begin{align}
 g_{\sigma\sigma\sigma} & \underset{\textsc{lo}}{=}\
 (1/3!)\bigl\langle -(\del\sigma)^2\sigma/\tilde{F}_\sigma +
 O(\sigma^3) \mbox{ terms in } \tilde{V}\bigr\rangle_{\text{on shell}}
 \nn \\[2mm]        
 & \underset{\textsc{lo}}{=}\
 (5 + \beta')\tilde{M}_\sigma^2/\tilde{F}_\sigma
 - N_f m_\psi B_\pi F_\pi^2
 (3-\gamma_m)(\gamma_m+1)(\gamma_m+\beta'+1)/\tilde{F}_\sigma^3  \,.
 \end{align}
\end{widetext}
In the limit $m_\psi \to 0$, this agrees (as it should) with the $O(h^3)$
coupling in Eq.~(\ref{VHiggs}). 

Otherwise, if $\alphaIR$ can be isolated, it may be easier to obtain $\beta'$
directly from the running of $\alpha$ near $\alphaIR$. If an independent value of
the technigluon condensate becomes available (Appendix
\ref{technigluon}), the $m_\psi \not= 0$ version (\ref{Msigfull}) of
the dilaton mass formula may be tested.  

\section{Final remarks: \protect\\ 
Suppression of FCNC\lowercase{s}}

Many papers have been written about the idea that the Higgs
boson is some sort of ``dilaton,'' but unlike crawling TC, there is a  
lack of commitment to the NG-mode requirement that the limit of exact 
scale invariance produce scale-dependent amplitudes  (Appendix
\ref{app:ward}).   As explained in Sec.~\ref{GilWein}, schemes like
walking TC and deformed conformal potentials are not dilaton theories:
they follow the example of scalon theory \cite{scalon} by assuming
manifest scale invariance (i.e.\ no scaling-NG mechanism) in the
scale-invariant limit. Only in crawling TC, where there is a genuine
dilaton with a \emph{nonzero} decay constant in the scale-invariant
limit, can it be argued that approximate scale invariance protects the
small mass of the Higgs boson.

To satisfy the NG-mode requirement, crawling TC assumes the existence of
a nonperturbative IR fixed point $\alphaIR$ at which conformal invariance
is exact, and a condensate $\langle\bar\psi\psi\rangle_\text{vac}$ for
both electroweak and scale transformations that is nonvanishing in the
conformal limit $\alpha \to \alphaIR$ (left diagram of Fig.~\ref{fig:beta}).
Both of the decay constants $v$ and $F_\sigma$ arise from this condensate
in the limit of scale invariance, so their ratio $v/F_\sigma$ is allowed to
be of order unity without the fine-tuning problem of scalon-type theories.

Hidden scale invariance corresponds to new solutions for
the CS equations near $\alphaIR$, with scaling laws for Green's
functions replaced by the soft-dilaton theorems (\ref{result}) and
(\ref{result2}). Since the scaling-law criteria used to find IR fixed
points inside the conformal window are not valid for NG-mode
fixed points, they may appear at small $N_f$ values 
(Fig.~\ref{fig:IRFP_search}).

The distinctive feature of our theory is the dependence of the
Higgs potential of Eqs.~(\ref{pot})--(\ref{VHiggs}) on $\beta'$, the
slope of $\beta(\alpha)$ at $\alphaIR$ (left diagram of 
Fig.~\ref{fig:beta}). We look forward to determinations of $\beta'$ via
experiment (Sec.~\ref{EFT}) or the lattice (Sec.~\ref{lattice}).

Finally, we note that standard explanations of the mass hierarchy
of quarks and leptons and the suppression of FCNCs can be naturally
adapted to fit crawling TC.

According to the theory of extended technicolor (ETC) \cite{ETC1,ETC2,
Lane02,Hill03}, there is a unification scale $\Lambda_V \gg \Lambda_v$ at
which the SM and TC gauge groups combine to form an ETC gauge group with
an intermediate boson $X$ of mass
\begin{equation}
M_X \sim \Lambda_V/(4\pi) \gg W,Z \mbox{ masses}.
\end{equation}
The ETC coupling $g_X$ of SM fermions $\psi_\textsc{sm} = q, \ell$
to TC fermions%
\footnote{Previously denoted as $\psi$, as in Fig.~\ref{fig:beta}
and Eq.~(\ref{TCtrace}). ETC fermions $\psi_\textsc{etc}$ do not play a
major role in this analysis.}
$\psi_\textsc{tc}$ and directly or indirectly to other SM fermions induces
FCNCs via effective four-fermion interactions such as
\begin{align}
{\cal L}_{q_i \leftrightarrow q_j}
&= c_{ij}\bigl(g_X\bigr/M_X\bigr)^2{\bar{q}{}_i}_L \gamma^\mu{\psi_\textsc{tc}}_R
\bar{\psi}_\textsc{tc}{}^{}_L \gamma_\mu {q{}_j}_R  + \mbox{H.c.},
\label{S=1} \\
{\cal L}_{|\Delta S| = 2}
&= c_{\Delta S = 2}\bigl(g_X\bigr/M_X\bigr)^2\bar{d}^{}_L \gamma^\mu s_R
\bar{s}^{}_L \gamma_\mu d^{}_R  + \mbox{H.c.},
\label{S=2}
\end{align}
where $c_{ij}$ and $c_{\Delta S = 2}$ are $O(1)$ numerical coefficients.
The observed bound on $K^0 \leftrightarrow \bar{K}^0$ requires $M_X/g_X
\gtrsim 10^3$ TeV in Eq.~(\ref{S=2}), but then, rough estimates of the
contributions of ${\cal L}_{q_i \leftrightarrow q_j}$ and its leptonic
analogue ${\cal L}_{\ell_i \leftrightarrow \ell_j}$ to the SM-fermion mass
matrix tend to be orders of magnitude too small to fit the observed
quark-lepton spectrum. This assumes vacuum insertion for
the $\psi_\textsc{tc}$-dependent part of Eq.~(\ref{S=1}) renormalized at
the ETC scale,
\begin{equation}
{\psi_\textsc{tc}}_R \bar{\psi}_\textsc{tc}{}^{}_L
\to \bigl\langle\text{vac}\bigl|\bar{\psi}_\textsc{tc}{}^{}_L
{\psi_\textsc{tc}}_R\bigr|\text{vac}\bigr\rangle_\textsc{etc} \,.
\end{equation}
The conclusion then follows from the relation
\begin{align}\label{enhance}
&\bigl\langle\text{vac}\bigl|\bar{\psi}_\textsc{tc}{}^{}_L
{\psi_\textsc{tc}}_R\bigr|\text{vac}\bigr\rangle_\textsc{etc} \nn \\
&\quad =  \bigl\langle\text{vac}\bigl|\bar{\psi}_\textsc{tc}{}^{}_L
{\psi_\textsc{tc}}_R\bigr|\text{vac}\bigr\rangle_\textsc{tc}
\exp\int_{\Lambda_v}^{\Lambda_V}
\!\frac{d\mu}{\mu}\,\gamma_m\bigl(\alpha(\mu)\bigr)
\end{align}
between ETC- and TC-scale amplitudes implied by the CS equation
(\ref{CS}) for ${\cal O} = \bar{\psi}_\textsc{tc}{}^{}_L {\psi_\textsc{tc}}_R$,
and from the observation that
\begin{equation}
\bigl\langle\text{vac}\bigl|\bar{\psi}_\textsc{tc}{}^{}_L
{\psi_\textsc{tc}}_R\bigr|\text{vac}\bigr\rangle_\textsc{tc}
= O(\Lambda_v^3)
\end{equation}
is very small compared with the ETC amplitude. If asymptotic
freedom in TC sets in above $\Lambda_v$ as rapidly as it does in QCD,
the exponential factor in (\ref{enhance}) is at most logarithmic in
$\Lambda_V\bigl/\Lambda_v$ and thus much too small to fit the
SM-fermion spectrum.

Walking TC dispenses with part of the QCD/TC analogy by assuming that
the TC $\beta$ function is close to zero over the large range of energies
between the TC and ETC scales \cite{Hold81,Hold85,Yamawaki,Aki86,Appel86,
Appel87a,Appel87b}. That corresponds to the walking region of the right
diagram of Fig.~\ref{fig:beta}. Since $\beta \approx 0$ implies
$\gamma_m(\alpha) \approx$ constant, $\gamma_m(\alpha)$ can be approximated
by a constant value $\gamma_m^*$ in the integral. The result is
power enhancement
\begin{equation}
\exp\int_{\Lambda_v}^{\Lambda_V}
\!\frac{d\mu}{\mu}\,\gamma_m\bigl(\alpha(\mu)\bigr)
\approx \bigl(\Lambda_V\bigl/
\Lambda_v\bigr)^{\mbox{\footnotesize $\gamma_m^*$}}
\end{equation}
in Eq.~(\ref{enhance}). A minimal enhancement $\gtrsim 10^2$ is
obtained for $\gamma_m^* \approx 1$. That gives an order-of-magnitude fit
to the SM-fermion spectrum (apart from the top quark and neutrinos, which
require special treatment).

In crawling TC, power enhancement can occur in the crawling region
(left diagram of Fig.~\ref{fig:beta}) if $\beta'$ for  
TC ($N_f \not= 3$) is much smaller than $\beta'$ for QCD where%
\footnote{Precocious asymptotic freedom for QCD is observed in the
$N_f = 3$ region, i.e.\ for momenta $Q \gg m_s$ up to the charm threshold.
It has nothing to do with the candidate $N_f = 2$ theory for crawling TC;
in particular, $s$ does not decouple in the chiral $SU(2)_L \times SU(2)_R$
limit taken at fixed $m_s$. On the lattice, the cases $N_f = 2+1$ and
$N_f = 2$ are clearly distinct.}
$N_f = 3$ after decoupling $t,b,c$. That would allow TC resonances to
appear up to an energy $M_\text{max}$ much larger than $\Lambda_v$
and explain the delay in the onset of asymptotic freedom in TC compared
with QCD. The IR end of the integral in Eq.~(\ref{enhance}) is sensitive to
the proximity of $\alpha(m_h)$ to the fixed point $\alphaIR$, so $m_h$ is
the relevant lower limit. Given that $\alpha$ varies little between
$\alphaIR$ and $\alpha(M_\text{max})$, we have $\gamma_m(\alpha) \simeq
\gamma_m(\alphaIR)$, so the exponential factor in Eq.~(\ref{enhance}) becomes
\begin{equation}
\exp\int_{m_h}^{M_\text{max}}
\!\frac{d\mu}{\mu}\,\gamma_m\bigl(\alpha(\mu)\bigr)
\approx \bigl(M_\text{max}
\bigl/m_h)^{\mbox{\footnotesize $\gamma_m(\alphaIR)$}}  \,.
\end{equation}
Like $\gamma_m^*$,  $\gamma_m(\alphaIR)$ is a nonperturbative number. If we
take (say) $\gamma_m(\alphaIR) = 1$, an enhancement of $10^2$ corresponds to
$M_\text{max} = 12.5$ TeV.

In conclusion, crawling TC is a consistent theory which avoids the
conceptual difficulties of walking TC while sharing its benefits.
Crawling TC (left diagram of Fig.~\ref{fig:beta}) does
not suffer from walking TC's phase discontinuity (between the solid and
dashed lines of the right diagram), and allows the number of physical
technipions to be minimized by taking $N_f$ small.

\label{conclusions}

\begin{acknowledgments}
We thank Tom Appelquist, Georg Bergner, Simone Biondini, Luigi Del Debbio,
C\'esar G\'omez, Kieran Holland, Andrei Kataev, David Mesterh\'azy,
Germano Nardini, Elisabetta Pallante, Mannque Rho, Francesco Sannino,
David Schaich, Koichi Yamawaki, and Roman Zwicky for useful correspondence
and discussions. Eliezer Rabinovici and Paolo Di Vecchia are thanked for
comments on Fubini's work \cite{Fub76} which helped us to reformulate 
Sec.~\ref{notscalons}. The work of L.C.T.\ was supported by the Swiss National
Science Foundation. The work of O.C.\ is supported in part by the
Bundesministerium for Bildung und Forschung (BMBF FSP-105), and
by the Deutsche Forschungsgemeinschaft (DFG FOR 1873).
\end{acknowledgments}

\appendix

\section{Decoupling: The difference between current and constituent masses}
\label{decouple}
While writing this paper and its predecessors \cite{CT1, CT2, CT3},
we were puzzled by a reluctance in the literature to consider the NG
mode for scale invariance, especially at an IR fixed point of a gauge
theory. Possible reasons for this are suggested in the text of this
paper. However, our attention has just been drawn to a reason we did
not take seriously: a belief that condensates decouple in the IR limit
because they give their constituent fields ``mass'', effectively making
them heavy relative to IR scales. Fermion condensates are usually
mentioned in this regard \cite{Appel97}, presumably because a gluonic
analogue of the Lagrangian mass term $-m\bar{\psi}\psi$ cannot be
constructed without introducing extra field variables (perturbative
Higgs mechanism).

The gap in this argument is that it does not distinguish between
a light fermion's small current mass $m$ and its large constituent mass
$\Sigma(0)$. Here $\Sigma(q^2)$ is the fermion's self-energy dynamically
generated by the nonperturbative mechanism responsible for fermionic
condensation $\langle\bar{\psi}\psi\rangle_\text{vac} \not= 0$ and (in
QCD) the $SU(3)_{L+R}$-invariant part of the masses of non-NG hadrons.
While $m$ violates chiral symmetry explicitly, $\Sigma$ does not: it
remains nonzero in the limit $m \to 0$ of chiral $SU(N_f)_L \times
SU(N_f)_R$ symmetry. In general, $m$ and $\Sigma$ are matrices in
flavor and spinor space. The current mass appears in the covariant
operator $iD\hspace{-2.2mm}\slash\,-m$ in the Schwinger-Dyson equation
\begin{equation}\label{SDE}
\text{T}^*\bigl\langle\bigl(i{D\hspace{-2.2mm}\slash}_x - m\bigr)
\psi(x)\bar{\psi}(y)\bigr\rangle_\text{vac} = i\delta^4(x-y) \,,
\end{equation}
where the time-ordering operation T$^*$ preserves chiral $SU(N_f)_L
\times SU(N_f)_R$ and gauge covariance, and T$^*\{\del_\mu\ldots\} \equiv
\del_\mu\text{T}^*\{\ldots\}$ by definition. The self energy $\Sigma$
appears in the solution for the dressed propagator
\begin{align}\label{dressed}
\int\!\!d^4x &\, e^{iq.x}\,
\text{T}^*\langle\psi(x)\bar{\psi}(0)\rangle_\text{vac}
 = \frac{i}{A(q^2)q\hspace{-1.7mm}\slash - m - \Sigma(q^2)} \,.
\end{align}
Equations ({\ref{SDE}) and (\ref{dressed}) imply the standard
self-consistent conditions for $\frac{1}{2}(1\pm\gamma_5)\Sigma$ shown
in Fig.~\ref{fig:self}.

\begin{figure}[b]
\centering\includegraphics[scale=0.69]{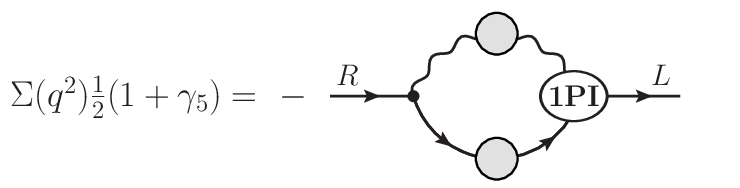}
\caption{Equation (\ref{SDE}) written as a relation between
one-particle-irreducible (1PI) two-point and three-point functions
in momentum space. The fermion and gauge-boson propagators
within the loop are fully dressed.}
\label{fig:self}
\end{figure}

The argument above contains an implicit assumption that $\Sigma$ sets
the scale for decoupling, i.e.\ that for momenta $q \ll \Sigma$, the
fermions are very ``heavy'' relative to $q$ and so decouple. We show
below that QCD is not tenable if that assertion is believed.

Fortunately for QCD, a natural extension of the perturbative
Appelquist-Carazzone theorem \cite{AC75} to include dimensionally
transmuted scales such as $\Sigma(0)$ indicates that it is the
\emph{current} mass which matters. In chiral perturbation theory,
both $m$ and $q$ tend to zero such that $m/q^2$ is finite, so
decoupling does not occur. It is also reassuring that the existence
of gluonic condensates does not cause gluons to decouple. More
generally, dimensionally transmuted scales $\cal M \not=$ 0 are
allowed in the extreme IR limit of QCD and (by analogy) TC. These
observations hold irrespective of whether an IR fixed point exists
or not, so if it exists, nothing prevents it from realizing scale
invariance in the NG mode.

Consider QCD at low energies, with the $t$, $b$ and $c$ quarks
decoupled. In the limit $m_{u,d,s} \to 0$ of chiral $SU(3)_L
\times SU(3)_R$ symmetry, low-energy theorems can be derived for
amplitudes involving the eight chiral NG bosons $\pi, K, \eta$ and
local operators such as chiral currents and $\bar{q}_L q_R$.
For amplitudes with non-NG states excluded, soft-meson theorems
are derivable when all external momenta $q$ tend to zero. The key
point is that this soft-meson limit and the IR limit of the
running of $\alpha_s$ for $N_f = 3$ are \emph{indistinguishable}:
the results of applying chiral perturbation theory and the RG
\emph{must match}. An assumption that $u,d,s$ decouple in the
IR limit would imply decoupling of $\pi$, $K$ and $\eta$ and so
contradict the well-known soft-meson theorems for these NG
bosons required by chiral perturbation theory. That would be a
disaster for QCD.

Instead, what happens is that $\pi$, $K$ and $\eta$ exist in the
IR limit of QCD. There are two main possibilities:
\begin{enumerate}[label=(\alph*)]
\item There is no IR fixed point; rather, $\alpha_s$ runs to $\infty$
with scale invariance explicitly broken. Standard chiral $SU(3)_L
\times SU(3)_R$ perturbation theory is applicable.
\item There is an IR fixed point $\alpha_s{}^{}_\textsc{ir}$ which
is necessarily in the scaling NG mode because of the quark condensate
responsible for chiral NG bosons. There are nine NG bosons, a
dilaton $\sigma$ as well as $\pi$, $K$ and $\eta$, and chiral-scale
perturbation theory \cite{CT1,CT2,CT3} is applicable.
\end{enumerate}

It is important to check that these conclusions are consistent with
the Appelquist-Carazzone theorem. It states that, to all orders in a
perturbative gauge theory, a field with a large Lagrangian mass $M$
decouples in the limit $M \to \infty$ taken at fixed renormalization
scale $\mu$, with finite changes in other renormalization constants
such as the gauge coupling $\alpha$. Since dimensionally transmuted
constants $\cal M$ such as those associated with fermion condensation
are nonperturbative, as is evident from the discussion of Eq.~(\ref{M_inv}),
the proof of Appelquist and Carazzone effectively assumes that all
$\cal M$ vanish. So at first sight, the theorem is irrelevant, and
nothing needs to be checked.

However, it is reasonable to suppose that the Appelquist-Carazzone
theorem could be extended to include $\cal M$ constants with the
conditions of the theorem otherwise unchanged. First, it is necessary
to identify RG invariants $\cal M$ associated with the field with a
large current mass $M$; either they tend to $\infty$ with $M$, such
as particle thresholds for heavy-quark production in QCD, or they
vanish because they involve Green's functions like
$\langle\bar{b}b\rangle_\text{vac}$ which depend on that field. Then,
for RG invariants ${\cal M }_\text{res}$ that remain in the residual
theory after decoupling, there would be a finite change due e.g.\
to the finite renormalization of $\alpha$ and $\beta$ in
Eq.~(\ref{M_inv}). This is just a consistency argument, not a
derivation, but we suggest that a reasonable derivation may be
possible using Landau's diagrammatic analysis of the nonperturbative
behavior of vertex functions \cite{Landau59}. Although this extension
of the theorem appears not to have been stated, much less derived,
it is implicit in nonperturbative applications such as the decoupling
of $t$, $b$ and $c$ in the presence of light-quark condensates such
as $\langle\bar{u}_L u_R\rangle_\text{vac} \not= 0$.

Following Appelquist and Carazzone \cite{AC75}, let us scale the
theorem's conditions such that the limit $M \to \infty$ for finite
momenta $q$ is replaced by the IR limit $q \to 0$ at fixed $M$, with
$\mu$ fixed in both cases (mass-independent renormalization). In
perturbation theory, that works if all masses $\not= M$ are scaled
with $q$, so that, in the IR-limit version of the theorem, they tend
to zero with $q$.

The problem is that the same argument applied to the $\cal M$-dependent
extension of the theorem would require $\cal M_\text{res}$ to scale with
$q$, and \emph{not} with the heavy current mass $M$. In the finite-$q$
version of the theorem, all $\cal M$ remain finite or vanish as
$M \to \infty$,
\begin{equation}
{\cal M}_\text{res}\bigl/M \to 0\,.
\end{equation}
so decoupling in the IR limit $q \sim 0$ with $M$ fixed can be concluded
only if it is \emph{assumed} that all ${\cal M}_\text{res}$ vanish.

Clearly, decoupling is a consequence of a \emph{current} mass $M$
being \emph{large} relative to dimensionally transmuted scales
${\cal M}_\text{res}$ in the residual theory. Light-quark condensates
and their TC analogues do not decouple in the chiral IR limit.

\section{Non-Lagrangian methods for chiral and conformal NG bosons}
\label{chiral}
The NG mode for a symmetry is usually explained in terms
of symmetric Lagrangians with potential functions that have
flat directions. This obscures the more general understanding
developed in the 1960s \cite{Nambu60,AdlerDashen} that currents $J_{\mu 5}$
and their divergences are all that is needed. As preparation for
the scaling application in Sec.~\ref{NGsolutions},
we present a brief review of the analysis for chiral
$SU(N_f)_L \times SU(N_f)_R$ symmetry, and extend it to
scale and conformal invariance, where the currents are
nonlocal.

The aim is to derive theorems for NG mesons carrying soft
momenta $q \to 0$ as the symmetry limit $\del^\mu J_{\mu 5} \to 0$
for current divergences is taken. There are two ways of proceeding
(Appendices \ref{A.1} and \ref{A.2} below); the choice depends on the order in
which these limits are taken. This matters because there can be
factors involving the pseudo-NG mass $m$ for which the limits do not
commute, e.g.
\begin{equation}
\lim_{q \to 0} \lim_{m \to 0}\frac{m^2}{m^2 - q^2}  = 0 \quad,\quad
\lim_{m \to 0} \lim_{q \to 0}\frac{m^2}{m^2 - q^2}  = 1
\end{equation}
but as long as this lack of uniformity is respected, the answer ends
up being the same. Scale and conformal invariance are
considered in Appendix \ref{A.3}.

\subsection{Chiral-symmetric theory}
\label{A.1}
Consider a chiral $SU(N_f)_L\times SU(N_f)_R$-symmetric TC theory with
conserved axial-vector currents $J_{\mu 5}^a=\bar \psi\gamma_\mu\gamma_5T^a\psi$
and axial charges
\begin{align}
Q_5^a = \int\!\! d^{3} x\, J_{05}^a(x)\,, \quad \mbox{Tr }T^a = 0\,,
\quad a = 1,\ldots, N_f^2 -1\,.
\end{align}
For any operator (or operator product) ${\cal O}$ which is not chiral invariant,
there is another operator
\begin{equation}
\delta^a_5{\cal O} = i\bigl[Q_5^a,{\cal O}\bigr] \not= 0 \,.
\label{chiral_comm}
\end{equation}
A nonzero VEV of $\delta^a_5{\cal O}$ can occur only if
$|\text{vac}\rangle$ is not chiral invariant. Then the amplitude
\begin{equation}
\bigl\langle\delta^a_5{\cal O}\bigr\rangle_\text{vac}
\not= 0
\end{equation}
is called a chiral condensate. In the standard case, ${\cal O}$ is the pseudoscalar
operator $\bar{\psi}\gamma_5 T^b \psi$:
\begin{align}
\bigl\langle\delta_5^a\bigl\{\bar{\psi}\gamma_5 T^b \psi\bigr\}\bigr\rangle_\text{vac}
&= - \bigl\langle\bar{\psi}\bigl(T^a T^b + T^b T^a\bigr)\psi\bigr\rangle_\text{vac} \nn \\
&\neq 0 \,.
\end{align}
Let $\cal O$ be a local spin-0 operator ${\cal O}(x)$. Then
the Ward identity for the time-ordered amplitude
\begin{equation}
{\cal A}_{\mu 5}^a(q) = \int\!\! d^{4} x\,e^{i q\cdot x}\, \text{T}
\bigl\langle J_{\mu 5}^a(x){\cal O}(0)\bigr\rangle_\text{vac}
\end{equation}
is given by
\begin{equation}
q^\mu {\cal A}_{\mu 5}^a(q) = i \int\!\! d^{4} x\, e^{i q\cdot x} \delta(x_0)
\bigl\langle\bigl[J_{05}^a(x),{\cal O}(0)\bigr]\bigr\rangle_\text{vac}\,.
\label{Ward}
\end{equation}
At zero momentum $q \to 0$, Eq.~(\ref{Ward}) reduces to
\begin{align}
\lim_{q\to 0} q^\mu {\cal A}_{\mu 5}^a(q) =  i \bigl\langle
\bigl[Q_5^a,{\cal O}(0)\bigr]\bigr\rangle_\text{vac}
= \bigl\langle\delta_5^a {\cal O}(0)\bigr\rangle_\text{vac} \not= 0  \,,
\label{Goldstone}
\end{align}
which is possible only if ${\cal A}^a_{\mu 5}(q)$ has an $O(1/q)$ singularity.
This implies Goldstone's theorem: such a singularity can arise only if
there are $N_f^2-1$ massless technipions $\pi^a$ coupled to
$J_{\mu 5}^a$,
\begin{align}
{\cal A}_{\mu 5}^a(q) =
&- \frac{q_\mu}{q^2} F_\pi \bigl\langle \pi^a(q=0)|{\cal O}(0)|\text{vac}\bigr\rangle
\nn \\
&+ \mbox{ terms finite at } q=0\,,
\label{pole1}
\end{align}
where the decay constant $F_\pi$ is defined by
\begin{equation}
\bigl\langle\text{vac}\bigl|J_{\mu 5}^a(0)\bigr| \pi^b(q)\bigr\rangle
= i \delta^{ab}F_\pi q_\mu \,.
\label{Fpi}
\end{equation}
Equation (\ref{Goldstone}) fixes the residue of the $q^2 = 0$
pole in Eq.~(\ref{pole1}).
The result is a standard soft-$\pi$ theorem 
\begin{equation}
F_\pi \bigl\langle \pi^a(q=0)\bigl|{\cal O}(0)\bigr|\text{vac}\bigr\rangle
= -\bigl\langle\delta_5^a {\cal O}(0)\bigr\rangle_\text{vac}
\,.
\label{soft}
\end{equation}

\subsection{Chiral currents partially conserved}
\label{A.2}
The alternative 1960s procedure is to give the currents small divergences
\begin{equation}
\del^\mu J^a_{\mu 5} = D^a_5 = 2i\sum_\psi m_\psi \bar{\psi}\gamma_5 T^a\psi
\label{div}
\end{equation}
by letting each techniquark have a small renormalized mass $m_\psi$ and
then take the symmetry limit $m_\psi \to 0$. This is the forerunner of our
approach in Sec.~\ref{NGsolutions}, where the scale-breaking divergence
$\theta^\mu_\mu$ tends to zero as $\alpha$ approaches the fixed point
$\alphaIR$.

For massive TC fermions, the Ward identity (\ref{Ward}) is replaced by
\begin{align}
q^\mu {\cal A}_{\mu 5}^a(q) =\ &i \int\!\! d^{4} x\, e^{i q\cdot x} \delta(x_0)
\bigl\langle\bigl[J_{05}^a(x),{\cal O}(0)\bigr]
 \bigr\rangle_\text{vac}    \nonumber \\
&+\ i\int\!\! d^{4} x\,e^{i q\cdot x}\, \text{T}
\bigl\langle D^a_5(x)
{\cal O}(0)\bigr\rangle_\text{vac}  \,.
\label{Ward2}
\end{align}
The traditional derivation of the soft-$\pi$ theorem then runs
as follows~\cite{AdlerDashen}.
The NG  bosons acquire mass, so ${\cal A}_{\mu 5}^a(q)$ cannot have a
$1/q$ singularity, and the $q \to 0$ limit of Eq.~(\ref{Ward2}) is
\begin{equation}
0 = \bigl\langle\delta_5^a {\cal O}(0)\bigr\rangle_\text{vac}
+ i \int\!\! d^{4} x\, \text{T} \bigl\langle D^a_5(x)
{\cal O}(0)\bigr\rangle_\text{vac}  \,,
\end{equation}
where the commutators (\ref{chiral_comm}) are now understood to be taken at
equal times. The second term is a zero-momentum insertion of the current
divergence $D^a_5$. It can be nonzero in the limit $D^a_5 \to 0$
only if there exists a single-particle intermediate state which becomes
massless as
\begin{equation}
i m_\psi\bigl/\bigl(q^2 - M_\pi^2\bigr)\bigr|_{q = 0}
= -i m_\psi\bigl/M_\pi^2  \to \mbox{ finite.}
\end{equation}
When the residue of this $\pi$ pole is evaluated via
$\langle\text{vac}|D^a_5(0)|\pi^b\rangle = M_\pi^2F_\pi\delta^{ab}$,
the soft-$\pi$ result (\ref{soft}) is recovered. Note that pole
dominance is \emph{not} assumed: in the symmetry limit, branch cuts
are less singular than poles.

\subsection{Soft-dilaton theorems for scale and conformal invariance}
\label{A.3}
Goldstone's theorem, that the number of NG bosons equals the number of
independent group generators which transform the vacuum, is generally
valid only  for local currents. A separate analysis is necessary for
nonlocal operators such as the dilatation and conformal currents
\begin{equation}
{\cal D}_\nu = x^\mu\theta_{\mu\nu}(x) \,, \
{\cal K}_{\mu\nu} = \bigl(2x_\mu x_\lambda -
x^2g_{\mu\lambda}\bigr)\theta^\lambda_\nu (x)
\end{equation}
which correspond to generators
\begin{equation}
D(t) = \int\!\!d^3x\,{\cal D}_0(t,\bm{x}) \,, \
K_\mu(t) = \int\!\!d^3x\, {\cal K}_{\mu 0}(t,\bm{x}) \,.
\end{equation}
Given these definitions, the partial conservation equations
\begin{equation}
\del^\nu{\cal D}_\nu = \theta^\lambda_\lambda \quad\mbox{and}\quad
\del^\nu{\cal K}_{\mu\nu} = 2x_\mu \theta^\lambda_\lambda
\label{partial}
\end{equation}
show that scale invariance $\theta^\lambda_\lambda \to 0$ ensures
conformal invariance.

The result that only one NG boson is needed---the dilaton of scale
invariance---was not obvious at first \cite{Salam69, Isham70a}, but it
was quickly realized \cite{Isham70b} that conformal-invariant
Lagrangians can be constructed by having the derivatives
$\del_\mu\sigma$ of the dilaton field $\sigma(x)$ act as Goldstone
fields for the four conformal generators $K_\mu$. Also,
Eq.~(\ref{partial}) shows that a pseudo-NG boson for either scale or
conformal invariance must have spin 0. The absence of extra NG bosons
has been attributed \cite{Volkov73, Ivanov75, Low02} to the failure
of $K_\mu$ to commute with the translation generators $P_\mu$ in the
limit of conformal invariance,
\begin{equation}
\bigl[K_\mu, P_\nu\bigr] = - 2i\bigl(g_{\mu\nu}D + M_{\mu\nu}\bigr)
\not= 0 \ ,  \quad \theta^\lambda_\lambda \to 0
\label{[K,P]}
\end{equation}
where $M_{\mu\nu}$ generate Lorentz transformations.

The literature on the NG mode for scale and conformal invariance is
dominated by Lagrangian models of scale invariance. Unlike the chiral case
\cite{Coleman69, Callan69, Dashen+W69}, the model independence of their
predictions for multiple soft-dilaton emission has yet to be proven
explicitly, and they have not been used at all to obtain conformal
theorems. Instead, soft-dilaton results for special conformal transformations
\cite{RJC71, PdiV16, PdiV17} were obtained by the non-Lagrangian method,
which we consider now. It is model independent and resembles the chiral
version discussed in Appendices \ref{A.1} and \ref{A.2}, but there are some interesting
differences which are best seen for the symmetric case
$\theta^\lambda_\lambda = 0$.

We begin with the analogue of Appendix \ref{A.1} for scale symmetry
(${\cal D}_\nu$ conserved), excluding for a moment the special case of
a single spin-0 operator ${\cal O}$. Instead of ${\cal O}(0)$, let us
consider a T-ordered product of ${\cal O}_1(0)$  (not necessarily
spin-0) and Fourier transforms
\begin{equation}
\widetilde{\cal O}_n(p_n) = \int\!d^4x_n\, e^{ip_n\cdot x_n}
{\cal O}_n(x_n) \ , \quad n > 1
\end{equation}
of other local operators ${\cal O}_n(x_n)$, and hence connected
momentum-space amplitudes $\langle\ldots\rangle_c$ with the
$\delta^4$ function for momentum conservation removed.

\begin{widetext}
The general amplitude involving the dilatation current can be written as
\begin{align}
{\cal B}_\nu = \int\!d^4x\,e^{iq\cdot x} \text{T}\langle\text{vac}|
{\cal D}_\nu(x){\cal O}_1(0)\mbox{$\prod\limits_{n > 1}$}
\widetilde{\cal O}_n(p_n)|\text{vac}\rangle_c 
 = -i\frac{\del\ }{\del q^\mu}\Gamma_{\mu\nu}(q;\{p\})  \,,
\label{Bnu}
\end{align}
where $\Gamma_{\mu\nu}$ is constructed entirely from local operators,
including the traceless tensor $\theta_{\mu\nu}$:
\begin{align}
\Gamma_{\mu\nu} =& \int\!d^4x\, e^{iq\cdot x}
\mbox{\large $\prod\limits_{m > 1}$}\biggl\{\int\!d^4x_m\,e^{ip_m\cdot x_m} \biggr\} \text{T}\bigl\langle\text{vac}\bigl|\theta_{\mu\nu}(x){\cal O}_1(0)
\mbox{$\prod\limits_{n > 1}$}{\cal O}_n(x_n)\bigr|\text{vac}\bigr\rangle_c \,.
\end{align}
Then the scaling Ward identity for ${\cal B}_\nu$ takes the form
\begin{align}
iq^\nu{\cal B}_\nu &= q^\nu\frac{\del\ }{\del  q^\mu}\Gamma_{\mu\nu}(q;\{p\}) \nn\\
&= -\int\!d^4x\, e^{iq\cdot x}\mbox{\large $\prod\limits_{m > 1}$}
\Bigl\{\int\!d^4x_m\,e^{ip_m\cdot x_m} \Bigr\}
\text{T}\Bigl\langle\text{vac}\Bigl|
\mbox{\large $\sum\limits_{\ell\geqslant 1}$}
\delta(x^0 - x^0_\ell)
\Bigl[ {\cal D}_0(x),{\cal O}_\ell(x_\ell)\Bigr]
\raisebox{0.4mm}{$\prod\limits_{\substack{n \not= \ell \\ n \geqslant 1}}$}
{\cal O}_n(x_n)\Bigr|\text{vac}\Bigr\rangle_{\!c\,}
\raisebox{-1.3mm}{$\biggr|$}_{x_1 = 0} .
\label{scaleWI}
\end{align}
\end{widetext}
For scale invariance in the NG mode, the vacuum state is not scale
invariant, so there will be operators $\{{\cal O}_n\}$
for which the right-hand side of Eq.~(\ref{scaleWI}) does not vanish in
the limit of zero momentum $q$:
\begin{align}
&\lim_{q \to 0} q^\nu\frac{\del\ }{\del q^\mu}\Gamma_{\mu\nu}(q;\{p\}) \nn \\
&= - \text{T}\bigl\langle\text{vac}\bigl|\bigl[D,{\cal O}_1(0)
\mbox{$\prod\limits_{n > 1}$}\widetilde{\cal O}_n(p_n)\bigr]\bigr|
\text{vac}\bigr\rangle_c = F(\{p\}) \not= 0 \,.
\label{Goldstone2}
\end{align}
Equation (\ref{Goldstone2}) can be satisfied \emph{only} if there is a
singularity $\sim q_\mu q_\nu/q^2$ in $\Gamma_{\mu\nu}$ as $q \to 0$:
\begin{equation}
\Gamma_{\mu\nu}(q;\{p\}) = \frac{q_\mu q_\nu}{3q^2} F(\{p\}) +
G_{\mu\nu}(\{p\}) + O(q) \,.
\label{pole}
\end{equation}
Since this result includes amplitudes $F(\{p\})$ where internal
momentum transfers are not light-like, the $q^{-2}$ pole must
be due to a massless spin-0 particle, the dilaton $\sigma$. The
residue of the pole can be determined from Eq.~(\ref{Fsigma}) for the
decay constant $F_\sigma$. Given that ${\cal O}_n$ has dynamical
dimension $d_n$,
\begin{widetext}
\begin{align}
i\bigl[D,{\cal O}_1(0)\bigr] = d_1 {\cal O}_1(0) \,, \quad
i\bigl[D,\widetilde{\cal O}_m(p_m)\bigr] = \bigl(d_m - 4  -
p_m\!\cdot\del/\del p_m\bigr)\widetilde{\cal O}_m(p_m) \,,
\ m > 1
\end{align}
the pole term in Eq.~(\ref{pole}) implies the result
\begin{align}
F_\sigma\text{T}\bigl\langle\sigma(q=0)\bigl|
{\cal O}_1(0)\mbox{$\prod\limits_{n > 1}$}\widetilde{\cal O}_n(p_n)
\bigr|\text{vac}\bigr\rangle_c
= \Bigl\{d_1 + \mbox{$\sum\limits_{m > 1}$}
\bigl(d_m - 4 - p_m\!\cdot\del/\del p_m\bigr)\Bigr\}\text{T}\bigl\langle\text{vac}\bigl|
{\cal O}_1(0)\mbox{$\prod\limits_{n > 1}$}\widetilde{\cal O}_n(p_n)
\bigr|\text{vac}\bigr\rangle_c 
\label{softsigma}
\end{align}
\end{widetext}
which is a standard soft-$\sigma$ theorem.%
\footnote{The earliest example appeared in Sec.~5 of \cite{Mack68}.
The model in Sec.~4 is an almost scale-invariant version of 
the linear sigma model \cite{GMLevy,MGM62} (clarified in the Appendix of
Ref.~\cite{Mack69}); it was not used to derive the soft-dilaton theorem.}

The case where T$\{{\cal O}_1\prod_{n > 1}
\widetilde{\cal O}_n\}$ is just a single spin-0 operator $\cal O$ is
special. Consider the \emph{unordered} amplitude
\begin{equation}
\Gamma^+_{\!{\cal O}\mu\nu}(q) = \int\!d^4x\, e^{iq\cdot x}
\bigl\langle\text{vac}\bigl|\theta_{\mu\nu}(x){\cal O}(0)
\bigr|\text{vac}\bigr\rangle  \,.
\label{unord1}
\end{equation}
It has the remarkable property that momentum conservation
$q^\mu\Gamma^+_{\!{\cal O}\mu\nu} = 0$ and scale invariance
$g^{\mu\nu}\Gamma^+_{\!{\cal O}\mu\nu} = 0$ determine its
nonperturbative dependence on $q$:
\begin{equation}
\Gamma^+_{\!{\cal O}\mu\nu}(q)
= 2\pi k\, q_\mu q_\nu \delta(q^2)\theta(q_0) \,,
\quad k = \text{const.}.
\label{unord2}
\end{equation}
Time ordering introduces a constant ambiguity%
\footnote{Notice that $c$ cannot be chosen such that $\Gamma_{\mu\nu}$ is
conserved \emph{and} has zero trace. That will not matter.\label{matter}}
$cg_{\mu\nu}$,
\begin{equation}
\Gamma^{}_{\!{\cal O}\mu\nu}(q)
= ik\, q_\mu q_\nu \bigl/(q^2 + i\epsilon) + c g_{\mu\nu} \,,
\end{equation}
but the dependence on $c$ drops out when ${\cal D}_\nu(x){\cal O}(0)$
is time ordered: \pagebreak
\begin{align}
i\frac{\del\ }{\del q_\mu}\Gamma^{}_{\!{\cal O}\mu\nu}(q)
&= \int\!d^4x\,e^{iq\cdot x} \text{T}\langle\text{vac}|
{\cal D}_\nu(x){\cal O}(0)|\text{vac}\rangle   \nn\\
&= 3ik\, q_\nu \bigl/(q^2 + i\epsilon) \,.
\end{align}
The Ward identity which follows
\begin{equation}
iq^\nu\frac{\del\ }{\del q_\mu}\Gamma^{}_{\!{\cal O}\mu\nu}(q)
= 3ik
= i\langle\text{vac}|[D,{\cal O}(0)]|\text{vac}\rangle
\label{Ward3}
\end{equation}
has a $q$-independent right-hand side, so there is no need
to expand about $q=0$. From%
\footnote{Spin-0 operators $\cal O$ are defined such that an extra
term on the right-hand side $\propto I$ does not appear.
See Appendix \ref{B.1}.}
\begin{equation}
i\bigl[D,{\cal O}(x)\bigr]
= \bigl(d_{\cal O} + x^\mu\del_\mu\bigr){\cal O}(x) \,,
\end{equation}
we see that Eq.~(\ref{Ward3}) fixes the constant $k$ in Eq.~(\ref{unord2}):
\begin{equation}
\Gamma^+_{\!{\cal O}\mu\nu}(q) = \ltextfrac{2}{3}\pi d_{\cal O}
\langle\text{vac}|{\cal O}(0)|\text{vac}\rangle  q_\mu q_\nu
\delta(q^2)\theta(q_0)  \,.
\label{unord3}
\end{equation}
When the completeness sum $I = \sum_n|n\rangle\langle n|$ is
inserted between $\theta_{\mu\nu}$ and $\cal O$ in Eq.~(\ref{unord1}),
\emph{only} single-dilaton states $|n\rangle = |\sigma\rangle$ can
reproduce this $q$ dependence. So \emph{without} approximating, we
can relate $\Gamma^+_{\!{\cal O}\mu\nu}$ to the dilaton decay constant
$F_\sigma$ of Eq.~(\ref{Fsigma}):
\begin{widetext}
\begin{align}
\Gamma^+_{\!{\cal O}\mu\nu}(q)
 &= \int\!\frac{d^3p}{2p_0(2\pi)^3} (2\pi)^4\delta^4(q-p) \langle\text{vac}|\theta_{\mu\nu}(0)|\sigma(p)\rangle
\langle\sigma(p)|{\cal O}(0)|\text{vac}\rangle
\nonumber \\
&= 2\pi\delta(q^2)\theta(q_0)
\langle\text{vac}|\theta_{\mu\nu}(0)|\sigma(q)\rangle
\langle\sigma(q)|{\cal O}(0)|\text{vac}\rangle
   = \ltextfrac{2}{3}\pi F_\sigma q_\mu q_\nu\delta(q^2)\theta(q_0)
\langle\sigma(q)|{\cal O}(0)|\text{vac}\rangle  \,.
\label{unord4}
\end{align}
\end{widetext}
Comparison of Eqs.~(\ref{unord3}) and (\ref{unord4}) yields the
soft-dilaton formula (\ref{soft_dil}).

If scale invariance is in the NG mode, so also is conformal invariance:
$K_\mu|\text{vac}\rangle \not= 0$ because of the identity
\begin{align}
\bigl\langle\bigl[K_\mu , \bigl[P_\nu ,
  {\cal O}_1\mbox{$\prod\limits_{n > 1}$}\widetilde{\cal O}_n\bigr]
  \bigr]\bigr\rangle_\text{vac} &= 2i g_{\mu\nu}\bigl\langle\bigl[D ,
{\cal O}_1\mbox{$\prod\limits_{n > 1}$}\widetilde{\cal O}_n
\bigr]\bigr\rangle_\text{vac} \nn \\ &\not= 0
\end{align}
implied by Eq.~(\ref{[K,P]}) and Poincar\'{e} invariance of the
vacuum. For the conformal current ${\cal  K}_{\mu\nu}$, the analogue
of Eq.~(\ref{Bnu}) is
\begin{align}
{\cal B}_{\mu\nu}
&= \int\!d^4x\,e^{iq\cdot x} \text{T}\langle\text{vac}|
{\cal K}_{\mu\nu}(x){\cal O}_1(0)\mbox{$\prod\limits_{n > 1}$}
\widetilde{\cal O}_n(p_n)|\text{vac}\rangle_c
\nonumber \\
&= \biggl\{g_{\mu\lambda}
\frac{\del^2\hspace{4mm} }{\del q^\alpha\del q_\alpha\!}
- 2\frac{\del^2\hspace{4mm} }{\del q^\mu\del q^\lambda\!}\biggr\}
\Gamma^\lambda_\nu (q;\{p\}) \,.
\label{Bmunu}
\end{align}
Then $q^\nu{\cal B}_{\mu\nu}$ gives a conformal Ward identity similar
to Eq.~(\ref{scaleWI})  but with ${\cal D}_0$ replaced by
${\cal K}_{\mu  0}$ in equal-time commutators. In the limit $q \to
0$, the result is
\begin{align}
\lim_{q \to 0} &q^\nu\biggl\{ g_{\mu\lambda}
\frac{\del^2\hspace{4mm} }{\del q^\alpha\del q_\alpha\!}
- 2\frac{\del^2\hspace{4mm} }{\del q^\mu\del q^\lambda\!}\biggr\}
\Gamma^\lambda_\nu (q;\{p\}) \nn\\
&= i\,\text{T}\bigl\langle\text{vac}\bigl|\bigl[K_\mu , {\cal O}_1(0)
\mbox{$\prod\limits_{n > 1}$}\widetilde{\cal O}_n(p_n)\bigr]\bigr|
\text{vac}\bigr\rangle_c \,,
\label{confWI}
\end{align}
which is nonzero only if there is a singular term $\sim q^\lambda
q_\nu q_\beta/q^2$ in $\Gamma^\lambda_\nu$:
\begin{equation}
q^\nu\biggl\{g_{\mu\lambda}
\frac{\del^2\hspace{4mm} }{\del q^\alpha\del q_\alpha\!}
- 2\frac{\del^2\hspace{4mm} }{\del q^\mu\del q^\lambda\!}\biggr\}
\frac{q^\lambda q_\nu q_\beta}{q^2} = - 6 g_{\mu\beta} \,.
\end{equation}
Therefore, Eqs.~(\ref{Goldstone2}) and (\ref{pole}) can be extended
to include the $O(q)$ pole term:%
\footnote{The connection between $O(q)$ terms and special conformal
transformations was noted in Ref.~\cite{Gross70} and used to derive
soft-dilaton theorems \cite{RJC71} long ago. The subject has been revived
very recently \cite{PdiV16, PdiV17}; note the important distinction
they made between NG-mode dilatons and ``gravitational dilatons''.}
\begin{align}
&\Gamma_{\mu\nu}(q;\{p\})_{\sigma\,\text{pole}}  \nn\\
&= -\frac{q_\mu q_\nu}{6q^2}\text{T}\bigl\langle\text{vac}
\bigl|\bigl[2D + iq^\beta K_\beta, {\cal O}_1(0)
\mbox{$\prod\limits_{n > 1}$}\widetilde{\cal O}_n(p_n)\bigr]\bigr|
\text{vac}\bigr\rangle_c  \nn\\
&\ \quad+ O(q^2) \,.
\label{pole2}
\end{align}
Although $K_\beta$ appears as a projection $q^\beta K_\beta$ in a
light-like direction ($q^2 = 0,\ q_0 > 0$), all space-like
directions $q'-q$ and hence individual $K_\beta$ components can
be obtained by comparing $\sigma$ states with small on-shell
momenta $q$ and $q'$. The general soft-$\sigma$ result for
special conformal transformations is therefore
\begin{align}
F_\sigma\bigl(&\langle\sigma(q)| - \langle\sigma(q')|\bigr)
\text{T}\bigl\{{\cal O}_1(0)\mbox{$\prod\limits_{n > 1}$}
\widetilde{\cal O}_n(p_n)\bigr\}|\text{vac}\rangle_c
\nonumber \\
&= \ltextfrac{1}{2}
  (q' - q)^\beta \text{T}\bigl\langle
\text{vac}\bigl|\big[K_\beta, {\cal O}_1(0)
\mbox{$\prod\limits_{n > 1}$}\widetilde{\cal O}_n(p_n)\bigr]\bigr|
\text{vac}\bigr\rangle_c \nn \\
&\quad + O\bigl((q\text{ or }q')^2\bigr) \,.
\label{softsigma2}
\end{align}

As is well known \cite{Carr71,Mack69,Gatto72}, the $K_\mu$ commutators
are best classified via the little group at $x=0$. Each time $K_\mu$ commutes
with a dimension-$d$ operator, it reduces the dimension to $d-1$. So there
are towers of local operators (mostly derivatives of other operators) above
familiar spin-$J$ operators of minimal dimension such as chiral currents
and $\theta_{\mu\nu}$ whose $K_\mu$ commutators necessarily vanish at $x=0$.
For all operators $O_n$ of this type, we have
\begin{align}\label{commutators}
\bigl[D,O_n(0)\bigr] &= -id_n O_n(0) \ , \nn \\
\bigl[M_{\mu\nu},O_n(0)\bigr] &= -{\Sigma_n}_{\mu\nu}O_n(0) \ , \nn\\
\bigl[K_\mu,O_n(0)\bigr] &= 0
\end{align}
where $M_{\mu\nu}$ generate the Lorentz group and ${\Sigma_n}_{\mu\nu}$ are the
corresponding spin matrices for $O_n$. Translating with $\exp(iP^\mu x_\mu)$
yields the standard formula
\begin{align}
&i\bigl[K_\mu,O_n(x)\bigr] \nn \\
&\ = \bigl\{2x_\mu(d_n + x^\rho\del_\rho)
- x^2\del_\mu - 2ix^\rho{\Sigma_n}_{\mu\rho}\bigr\}O_n(x) \,.
\end{align}
So for operators ${\cal O}_n \to O_n$, Eq.~(\ref{softsigma2}) implies
\begin{widetext}
\begin{align}
F_\sigma\bigl(&\langle\sigma(q)| - \langle\sigma(q')|\bigr)
\text{T}\bigl\{O_1(0)\mbox{$\prod\limits_{n > 1}$}
\widetilde{O}_n(p_n)\bigr\}|\text{vac}\rangle_c \nn \\
=\ &(q' - q)^\mu \sum_{m>1}
\biggl\{\Bigl(4 - d_m + p_m\cdot\smallfrac{\del\ \ }{\del p_m}\Bigr)
\smallfrac{\del\ \ }{\del p^\mu_m}
- \smallfrac{1}{2} {p_m}_\mu\smallfrac{\del^2\ }{\del p^2_m}
+ i{\Sigma_m}_{\mu\rho}\smallfrac{\del\ \ }{\del{p_m}_\rho}\biggr\}
\text{T}\bigl\langle\text{vac}
\bigl|O_1(0)\mbox{$\prod\limits_{n > 1}$}\widetilde{O}_n(p_n)\bigr|
\text{vac}\bigr\rangle_c \nn \\
&+ O\bigl((q\text{ or }q')^2\bigr) \,.
\label{softsigma3}
\end{align}
\end{widetext}
The soft-dilaton theorems (\ref{softsigma}) and (\ref{softsigma3})
will be needed in Appendix \ref{app:ward}.

\section{Nonperturbative definition of gluon and technigluon
condensates}
\label{technigluon}
Definitions of the gluon condensate $\langle G^2\rangle_{\text{vac}}$
and its TC analogue $\langle\hat{G}^2\rangle_{\text{vac}}$ are
problematic because
\begin{enumerate}[label=(\arabic*)]
\item  they are perturbatively divergent nonperturbative quantities, and
\item  they involve operators like $\hat{G}^2$ which are hard to separate
  from the identity operator $I$ under renormalization or within
  operator product expansions.
\end{enumerate}
The issue arises in the discussion following Eq.~(\ref{indep}), where
the spectator operator ${\cal O} = (\alpha/\pi)\hat{G}^2$ is used to
obtain soft-dilaton results such as Eq.~(\ref{result}) and hence
Eq.~(\ref{MsigPCDC}) for the technigluon condensate. To avoid ambiguity,
we require that $\cal O$ be multiplicatively renormalizable%
\footnote{\label{qcd} In QCD with $m_q \not= 0$, mixing with
$m_q\bar{q}q$ must also be considered \cite{Grin89,Tarr82}.}
and (in the sense of that discussion) $\alpha$ independent.

In this appendix, we explain the need for these requirements,
noting that, while they serve our purposes and are consistent
with various proposals to define $\langle{\cal O}\rangle_\text{vac}$,
the results still lack sufficient precision for unambiguous calculations,
e.g.\ on the lattice. The extent to which a definitive nonperturbative
definition is possible is then considered. The analysis refers to the
physical region, which is $0 < \alpha < \alphaIR$ if there is an IR
fixed point $\alphaIR$, and $0 < \alpha < \infty$ if not ($\alphaIR
\to \infty$).

When the gauge coupling $\alpha$ is finite, functional integrals for
Green's functions are dominated by nonperturbative gauge \cite{Sav77}
and fermion fields. Evidently these fields are hard to characterize
analytically, i.e.\ beyond numerical lattice methods. Instead,
attention is focused on a few physical operators $\cal O$ that
form nonperturbative condensates $\langle{\cal O}\rangle_\text{vac}$
which can be given theoretical and phenomenological meaning. The
problem is to define $\cal O$ without introducing ambiguities
proportional to the identity operator $I$. Only then does the
condensate acquire a physical meaning.

Even before QCD was invented, it was known how to do this for divergences
$D^a_5$ of partially conserved currents \cite{GMOR}: they belong to an
irreducible representation of an equal-time non-Abelian chiral group
which distinguishes them from the chiral-invariant operator $I$. For
example, if $D^a_5$ and $I$ appear in an operator product expansion, their
contributions to its VEV can be distinguished, provided that other operators
with poorly defined condensates are known to have less singular coefficient
functions.

In gauge theories, these chiral condensates are formed when
$\cal O$ is a fermion bilinear: $\bar{q}_i(1 \pm \gamma_5){q}_j$ for
QCD and $\bar{\psi}_i(1 \pm \gamma_5)\psi_j$ for TC. As long as
the renormalization procedure respects chiral $SU(N_f)_L \times SU(N_f)_R$
symmetry, these operators do not have counterterms proportional to $I$
and so belong to an irreducible chiral representation. The corresponding
condensates are necessarily nonperturbative, so they produce power
corrections to short-distance expansions of operator products.

However, any discussion of power corrections at short distances in QCD
would be incomplete without including terms induced by the gluon
condensate \cite{SVZ79, SVZ_JETPlett78} formed when $\cal O$ is
the operator $(\alpha_s/\pi)G^2$. Then group theory cannot be used to
distinguish $\cal O$ from $I$, so the definition of 
$\langle G^2\rangle_\text{vac}$ remains ambiguous due to counterterms 
proportional to $I$. The same problem arises for the technigluon 
condensate $\langle \hat{G}^2\rangle_\text{vac}$.

A practical approach still in use \cite{Lee10,Bali14,Lee15} is to impose a regulator
such as the lattice and then identify and subtract perturbative
contributions to $\langle\hat{G}^2\rangle_\text{vac}$ up to some high
but finite order. The series is not expected to converge because of
renormalons, but perturbative coefficients can be checked numerically to
see if their behavior is consistent with Borel summability. The theoretical
argument for this is that, if all orders of perturbation theory can in
principle be summed by a well-defined technique and the result is then
subtracted, the remainder will be the nonperturbative amplitude being
sought \cite{Shif98}. However, even if Borel summability can be proven
to all orders, Borel's method is not unique: nonperturbative dynamics
may choose another well-defined procedure. Generally, there is \emph{no}
guarantee that nonperturbative amplitudes can be deduced from purely
perturbative considerations \cite{Beneke94,Beneke99}, so it is not
surprising that these issues remain a source of unease \cite{Appel13}.

Another idea is to multiply $\hat{G}^2$ by $\beta(\alpha)/(4\alpha)$ and
use RG invariance. Since purely nonperturbative constants can have no
Taylor series about $\alpha \sim 0$ (as noted for the constant $\cal M$
in Eq.~(\ref{M_inv})), perturbative terms cannot be invariant. But that
is not sufficient, because nothing has been done to distinguish the
desired operator $\cal O$ from $I$. For example, if $\cal O$ and $I$
appear in the expansion of a product of physical currents,
$\langle{\cal O} \rangle_\text{vac}$ can mix with the VEV of
the $I$ term under RG transformations.

Our proposal is to consider the scaling properties of \emph{operators},
not just amplitudes.  We consider mainly the TC case where chiral
symmetry of the Lagrangian stops $\hat{G}^2$ from mixing with techniquark
bilinears.

Let $\mu$ set the scale for an arbitrary renormalization prescription
$R$ for composite operators $\cal O$, including the trace operator
\begin{equation}\label{T}
T =\bigl(\beta(\alpha)/4\alpha\bigr)\hat{G}^2 \,.
\end{equation}
We need to distinguish the
results of variations $\mu \del/\del\mu$ at fixed coupling $\alpha$
and the total variation
\begin{equation}
\mu\frac{d\ }{d\mu} \equiv \mu\frac{\del\ }{\del\mu}
+ \beta(\alpha)\frac{\del\ }{\del\mu} \,.
\end{equation}

Because subtractions necessarily include perturbative terms, we
do not expect the result $T_R$ to be exactly RG invariant --- there
must be mixing with $I$,
\begin{equation}
\mu\frac{d\ }{d\mu} T_R = F_R(\mu, \alpha_\mu)I \,,
\end{equation}
where $F_R$ is an ordinary function. However, the $I$-dependent term
can be absorbed into the definition of the trace operator
\begin{equation}
T_{R'} = T_R - \int_c^\mu\!\frac{d\mu'}{\mu'}\, F_R(\mu', \alpha_{\mu'})I
\end{equation}
where $c$ is a constant independent of $\mu$ and $\alpha$. Note
that the integral over $\mu'$ takes account of the $\mu'$ dependence of
$\alpha_{\mu'}$.  Evidently the resulting operator is RG invariant:%
\begin{equation}
\mu\frac{d\ }{d\mu} T_{R'} = 0 \,.
\label{R'}
\end{equation}

We are not done, because the solution of Eq.~(\ref{R'}) is not unique.
Any ordinary function $f({\cal M})$ whose dependence on $\mu$ and
$\alpha$ is carried solely by the RG invariant mass $\cal M$ of
Eq.~(\ref{M_inv}) is itself RG invariant. Therefore all operators%
\begin{equation}
T_{R''} = T_{R'} + f({\cal M})I
\label{R''}
\end{equation}
are RG invariant. To preserve engineering dimensions, $f({\cal M})$
can be chosen to be ${\cal M}^4$ times a constant independent
of $\mu$ and $\alpha$. This ambiguity does not affect the multiplicative
renormalizability of operators obtained by multiplying by $\alpha$-dependent
factors.

Nevertheless, it is necessary to eliminate the ambiguity (\ref{R''}) if
soft-dilaton results such as Eq.~(\ref{result}) are to be derivable. That
happens when we apply the $\alpha$-independence criterion to
\begin{equation}\label{O''}
{\cal O}_{R''} = \biggl(\smallfrac{\alpha}{\pi}\hat{G}^2
\biggr)_{R''}
= \smallfrac{4\alpha^2}{\pi\beta}T_{R''}
\end{equation}
for use as a spectator operator in the $\theta^\mu_\mu$ insertion rule
(\ref{RHS}), because
\begin{equation}
\beta(\alpha)\smallfrac{\del\ }{\del\alpha}{\cal M}
= - \mu\smallfrac{\del\ }{\del\mu}{\cal M}
= -{\cal M}
\end{equation}
implies
\begin{equation}
\smallfrac{\del\ }{\del\alpha}
\biggl(
\smallfrac{4\alpha^2}{\pi\beta}{\cal M}^4\biggr)
= - \smallfrac{4\alpha^4}{\pi\beta^2}{\cal M}^4
\smallfrac{d\ }{d\alpha}\biggl(\smallfrac{\beta}{\alpha^2}
- \smallfrac{4}{\alpha}\biggr)
\not= 0 \,.
\end{equation}
Then $\cal O$ has no ambiguity $\propto I$ --- in principle.

In practice, we would like to be able to test our soft-dilaton
results by comparison with experimental or lattice data. For that, our
minimal requirements on $\hat{G}^2$ are necessary but not sufficient. It
is not even clear how to implement them for prescriptions currently on
offer for $\langle\hat{G}^2\rangle_\text{vac}$.

Phenomenology is based on the original QCD prescription
\cite{SVZ_JETPlett78,SVZ79, SVZb79}, where $m_q$-independent power
corrections in the small-$x$ expansion of two electromagnetic currents
\begin{equation}\label{currents}
J_\mu(x)J_\nu(0) \sim {\cal C}_{\mu\nu}{}^{}_I(x) I
+ {\cal C}_{\mu\nu}{}^{}_{G^2}(x)\smallfrac{\alpha_s}{\pi}G^2
+ \ldots
\end{equation}
are by definition contained entirely within the gluonic coefficient
function ${\cal C}_{\mu\nu}{}^{}_{G^2}(x)$. Dispersive sum rules
for the operator product (\ref{currents}) are reasonably consistent with
each other for $\langle(\alpha_s/\pi)G^2\rangle_\text{vac} \approx
0.012\ \text{GeV}^4$, but there is no reason to suppose that a similar
definition of the gluon condensate for a different operator product
would be equivalent.

The idea of this definition is to suppose that the $I$ term in a
short-distance expansion is purely perturbative. That is a difficult
concept: even if ${\cal C}_{\mu\nu}{}^{}_I(x,\alpha_s,\mu)$ is truncated
to a polynomial in $\alpha_s$, the running of $\alpha_s$ depends on
dimensionally transmuted masses $\cal M$ which can produce power
corrections.%
\footnote{Sometimes the presence of nonperturbative \rule{0mm}{3.5mm}power corrections
is attributed to a ``breakdown'' of the Wilson expansion. It is true that a
fully rigorous proof \cite{Zimm70} has so far been possible only within
perturbation theory, but Wilson and Zimmermann \cite{KGW_Zimm} gave
\mbox{convincing} arguments for operator product expansions to be valid in any
nonperturbative theory consistent with axiomatic field theory.
See also Ref.~\cite{Novikov85}.}
And it is unclear how this proposal can be related to purely theoretical
definitions, where perturbative truncation is also a problem. Examples
are directly estimating the one-point function
$\langle\hat{G}^2\rangle_\text{vac}$ on the lattice (noted above), or
adding a heavy fermion $\Psi$ and taking its mass $M$ to $\infty$:
\begin{equation}
\lim_{M \to \infty}\langle\text{vac}|M\bar{\Psi}\Psi|\text{vac}\rangle^{}_M
\underset{\text{def}}{=} \frac{\beta(\alpha)}{12\pi\alpha\beta_1}
\langle\text{vac}|\hat{G}^2|\text{vac}\rangle  \,.
\end{equation}

A way around this impasse may be to use experimental data to extend
$\alpha$ beyond the UV region where asymptotic freedom is applicable.
If $\alpha$ can be measured at finite values where the small-$\alpha$
expansion is no longer valid, perturbative truncation would not be
needed. This would have to be done within a renormalization scheme
suitable for matching to lattice calculations at these intermediate
non-UV energies. Then, if the thermodynamic limit can be demonstrated
for the Euclidean partition function on the lattice,
\begin{equation}
Z \sim \exp\bigl\{-V_4\Gamma(\alpha,\mu)\bigr\} \ ,
\quad \text{Euclidean volume}\ V_4 \to \infty\,,
\end{equation}
a practical nonperturbative definition of the Euclidean condensate
would be
\begin{equation}\label{Eucl}
\langle\text{vac}|\hat{G}^2|\text{vac}\rangle^{}_\text{Eucl}
\underset{\text{def}}{=} -4\alpha\smallfrac{\del\Gamma}{\del\alpha} \,.
\end{equation}
Eq.~(\ref{Eucl}) is equivalent to a condition \cite{Del13,Pro13}
arising from the Feynman-Hellmann theorem.

In crawling TC, where there is a fixed point $\alphaIR$, the technigluon
condensate at $\alphaIR$ appears in results such as Eqs.~\ref{result4})
and (\ref{MsigPCDC}). It is obtained as a limit $\alpha
\rightharpoondown \alphaIR$ of the amplitude
\begin{equation}
\bigl\langle\text{vac}\bigl|\bigl(\hat{G}^2\bigr)_{R''}
\bigr|\text{vac}\bigr\rangle
= \{\beta(\alpha)/4\alpha\}^{-1}
\langle\text{vac}|T_{R''}|\text{vac}\rangle \ ,
\label{cond1}
\end{equation}
where $0 < \alpha < \alphaIR$. The operator $(\hat{G}^2)_{R''}$ is
local, so the amplitude (\ref{cond1}) is a \emph{one}-point function
where intermediate states such as $|\sigma\rangle$ cannot occur. It
follows that $\langle(\hat{G}^2)_{R''}\rangle_{{\text{vac}}}$ is
continuous in the scaling limit:
\begin{align}
\bigl\langle\text{vac}\bigl|\bigr(\hat{G}^2\bigr)_{R''}
\bigr|\text{vac}\bigr\rangle^{}_{\text{at}\;\alphaIR}
= \lim_{\theta^\mu_\mu \to 0} \bigl\langle\text{vac}\bigl|
\bigl(\hat{G}^2\bigr)_{R''}\bigr|\text{vac}
\bigr\rangle^{}_{0 < \alpha < \alphaIR} \,.
\label{cond2}
\end{align}

\subsection{Relation to commutators with the dilatation generator
\texorpdfstring{$\bm{D}$}{D}}
\label{B.1}

A conventional soft-dilaton theorem such as Eq.~(\ref{soft_dil})
is valid for operators $\cal O$ which scale \emph{homogeneously} with
operator dimension $d_{\cal O}$, i.e.
\begin{equation}\label{homog}
i\bigl[D,{\cal O}(x)\bigr]
= \bigl(d_{\cal O} + x^\mu\del_\mu\bigr){\cal O}(x)
\end{equation}
with other operators absent. As seen above, if $\cal O$ has spin $0$,
mixing with the identity operator $I$ is hard to control, producing
ambiguities such as Eq.~(\ref{R''}). Since Eq.~(\ref{homog}) is not
invariant under the shift
${\cal O} \to \check{\cal O} = {\cal O} + cI$ ($c$ = const.),
\begin{equation}
i\bigl[D,\check{\cal O}(x)\bigr]
= \bigl(d_{\cal O} + x^\mu\del_\mu\bigr)\check{\cal O}(x)
- c\,d_{\cal O}I \,,
\end{equation}
it could act as an alternative to $\alpha$ independence as a criterion
for resolving the ambiguity in Eq.~(\ref{O''}). These commutators (equal-time
for $\alpha < \alphaIR$ where $D$ is not conserved) are determined by
short-distance expansions of $\theta_{\mu\nu}(x){\cal O}(0)$. As noted in
Sec.~\ref{NGsolutions} [footnote \ref{Drell-Yan} and Eq.~(\ref{dim})],
short-distance behavior for $0 < \alpha < \alphaIR$ is determined by
the fixed point $\alpha = 0$ (asymptotic freedom), whereas, when
$\alpha$ is first fixed at $\alphaIR$, it cannot run: short-distance
behavior is then controlled by the nonperturbative world at $\alphaIR$
(Appendix \ref{app:ward}). Therefore the cases $\alpha = \alphaIR$ and
$\alpha < \alphaIR$ must be considered separately.

At $\alphaIR$, the condition (\ref{homog}) would resolve the
ambiguity in spin-$0$ operators $\cal O$, e.g.\ the operator $\hat{G}^2$
in the mass formula (\ref{MsigPCDC}) with $d_{\cal O}$ identified as
$4 + \beta'$ [Eqs.~(\ref{gamma}) and (\ref{4+beta'})]. However, we have
been unable to relate this to the $\alpha$-independence criterion for
the operator (\ref{O''}) which defines $\hat{G}^2$ in the mass formula.
In Eq.~(\ref{RHS}), we considered replacing $\theta^\mu_\mu$ by
$\del^\mu{\cal D}_\mu$ in order to obtain a scaling Ward identity, but
could not circumvent the facts that Eq.~(\ref{RHS}) is valid only for
$\alpha < \alphaIR$ and the limits $x \to 0$ and $\alpha \rightharpoondown
\alphaIR$ do not commute. In Appendices \ref{chiral} and \ref{app:ward},
we assume that spin-$0$ operators can be classified according to the
condition (\ref{homog}).

In all cases $\alpha \leqslant \alphaIR$, the relevant operator product
expansion for spin-$0$ operators $\cal O$ takes the form
\begin{widetext}
\begin{align}
\theta_{\mu\nu}(x){\cal O}(0) \sim
\ {\cal C}_{\mu\nu}{}^{}_I(x)I
 + {\cal C}_{\mu\nu}{}^{}_{\cal O}(x)
  \bigl\{d^{}_{\cal O}{\cal O}(0) + k_{\cal O}I\bigr\}
  + {\cal C}_{\mu\nu\alpha}{}^{}_{\del{\cal O}}(x)\del^\alpha{\cal O}(0)
  + \{\text{less singular, or other operators}\},
\label{expand}
\end{align}
where the constant $k_{\cal O}$ has mass dimension 4, and where
the leading singularities
\begin{align}
{\cal C}_{\mu\nu}{}^{}_{\cal O}(x)
 &= \frac{1}{12\pi^2}\del_\mu\del_\nu\frac{1}{x^2} \,,
\label{CO} \\
{\cal C}_{\mu\nu\alpha}{}^{}_{\del{\cal O}}(x)  &= -\frac{1}{4\pi^2}
 \biggl\{g_{\mu\alpha}\del_\nu + g_{\nu\alpha}\del_\mu
 - \ltextfrac{1}{3}g_{\mu\nu}\del_\alpha
- \ltextfrac{2}{3}\del_\mu\del_\nu\del^{-2}\del_\alpha\biggr\}\frac{1}{x^2} \,,
\label{CdelO}
\end{align}
\end{widetext}
have $x$ dependence determined by conservation and tracelessness in the
indices $\mu\nu$; [compare Eq.~(\ref{unord2})]. Here the $i\epsilon$
prescription $x^2 \to x^2 - i\epsilon x_0$ for unordered products should
be understood, so we have $\del^2(x^2)^{-1} = 0$ with no $\delta^4(x)$
term\footref{matter} and can define $\del^{-2}\del_\alpha(x^2)^{-1}$ to be
$\frac{1}{2}x_\alpha(x^2)^{-1}$. The coefficient function
${\cal C}_{\mu\nu\alpha}{}^{}_{\del{\cal O}}$ is normalized to produce
the correct commutators of ${\cal O}(0)$ with the Poincar\'{e} generators
$P_\mu$ and $M_{\mu\nu}$. It also corresponds to the term
$x\cdot\del\,{\cal O}(x)$ in Eq.~(\ref{homog}).

At $\alphaIR$ where $\theta^\mu_\mu = 0$, ${\cal C}_{\mu\nu}{}^{}_I$
is both traceless and conserved and hence proportional to
$\del_\mu\del_\nu\bigl(1\bigl/x^2\bigr)$. Therefore, given the presence
of the term $k_{\cal O}{\cal C}_{\mu\nu}{}^{}_{\cal O}I$ in Eq.~(\ref{expand}),
we can set ${\cal C}_{\mu\nu}{}^{}_I = 0$. The commutator condition
(\ref{homog}) is reproduced if we set
\begin{equation}
k_{\cal O} = 0\,.
\end{equation}
On the lattice, it would be hard to insulate a test of this condition from
$\alpha \not= \alphaIR$ effects.

For $\alpha < \alphaIR$, the problem is that asymptotic freedom and the
trace anomaly require the coefficient function
\begin{equation}
{\cal C}_{\mu\nu}{}^{}_I(x) = \ltextfrac{1}{3}\bigl(g_{\mu\nu}\del^2 -
  \del_\mu\del_\nu\bigr){\cal G}(x^2)
\end{equation}
to be far more singular than the $O(x^{-4})$ coefficient function
${\cal C}_{\mu\nu}{}_{\cal O}(x)$.

For example, let $\cal O$ be the RG-invariant trace operator $T$
discussed in Eqs.~(\ref{T}) and (\ref{R''}). Then asymptotic freedom
requires the trace amplitude
\begin{equation}
{\cal F}(x^2) = \bigl\langle\text{vac}\bigl|\theta^\lambda_\lambda(x)
T_\text{inv}(0)\bigr|\text{vac}\bigr\rangle = \del^2{\cal G}(x^2)
\end{equation}
to have the following short-distance behavior,
\begin{align}
{\cal F}(x^2) &\sim \ltextfrac{1}{16}\beta^2_1 \bigl(\ln\mu^2
x^2\bigr)^{-2\beta_1}\bigl\langle\text{vac}\bigl| \hat{G}^2(x)
\hat{G}^2(0)\bigr|\text{vac}\bigr\rangle^{}_{\alpha = 0} \nn \\
&\sim K \bigl(\ln\mu^2 x^2\bigr)^{-2\beta_1}\!\bigr/(x^2)^4 \,,
\end{align}
where $K \not= 0$ is a constant and $\beta_1 > 0$ is the one-loop
$\beta$-function coefficient (\ref{one-loop}). That corresponds to
\begin{equation}
{\cal G}(x^2)
\sim \ltextfrac{1}{12}K\bigl(\ln\mu^2 x^2\bigr)^{-2\beta_1}\!\bigr/(x^2)^3
\end{equation}
and hence
\begin{equation}
{\cal C}_{\mu\nu}{}^{}_I(x) = O\bigl(x^{-8}\ln^{-2\beta'}(x^2)\bigr) \,,
\end{equation}
which is $O\bigl(x^{-4}\ln^{-2\beta'}(x^2)\bigr)$ compared with Eq.~(\ref{CO})
for ${\cal C}_{\mu\nu}{}^{}_{{\cal O}=T}(x)$ as $x \sim 0$.

Since $\cal G$ is RG invariant, it can be written as a function of $x^2$ and
a dimensionally transmuted scale $\cal M$:
\begin{equation}
{\cal G} = (x^2)^{-3}f\bigr(x^2{\cal M}^2\bigr)
\end{equation}
It is therefore likely that an $x \sim 0$ expansion of $\cal G$ contains a
nonleading term $\propto {\cal M}^4\bigl/x^2$ whose contribution to
${\cal C}_{\mu\nu}{}^{}_I$ in Eq.~(\ref{expand}) cannot be distinguished from
$\bigl\{k_{\cal O}{\cal C}_{\mu\nu}{}^{}_{\cal O}\bigr\}_{{\cal O} \to  T}I$.      
Then the latter term should be absorbed into ${\cal C}_{\mu\nu}{}^{}_I$  
by setting $k_T = 0$. We conclude that a study of $x \sim 0$ behavior
for $\alpha < \alphaIR$ does not produce a criterion to resolve the
ambiguity (\ref{R''}). All we can say is that asymptotic freedom requires 
$d_{{\cal O}=T} = 4$, as noted in Eq.~(\ref{can_dim}), and hence
\begin{align}
\bigl[D(x_0 + \epsilon),T(x)\bigr] \sim \bigl(4 + x\cdot\del\bigr)T(x)
+ O\bigl(\epsilon^{-4}\ln^{-2\beta_1}(\epsilon)\bigr)I
\end{align}
in the equal-time limit $\epsilon \to 0$.

\section{NG-mode conformal-invariant world at 
\texorpdfstring{$\bm{\alphaIR}$}{alphaIR}}
  \label{app:ward}

  Unlike a scale-invariant theory in the WW mode, the world at $\alphaIR$ is
  somewhat similar to the physical world (TC or QCD) for $0 < \alpha <
  \alphaIR$. It has a particle spectrum with
  \begin{enumerate}[label=(\arabic*)]
  \item  non-NG masses close to their physical values, because in the
  physical world, scale invariance is approximate at low energies, and
  \item  an NG sector which is massless because at $\alphaIR$, scale
  and chiral invariance are \emph{exact}.
  \end{enumerate}
  This situation is allowed because amplitudes at $\alphaIR$ can have
  a complicated dependence on scales set by the dilaton decay constant
  ($F_\sigma$ or $f_\sigma \not= 0$) and other dimensionally transmuted
  masses $\cal M$ and condensates. As we emphasize in various sections
  of this paper, the effects of $\cal M$ dependence must be carefully
  distinguished from those due to an explicit breaking of scale invariance
  in the Lagrangian, e.g.\ by fermion mass parameters or
  Coleman-Weinberg potentials.

  In case this picture seems counterintuitive, recall the fact that
  the NG scaling mode for the ground state at $\alphaIR$ requires it to
  have a noncompact scaling degeneracy in addition to compact chiral
  degeneracies. Under a finite scaling transformation, the ground
  state of a scale-invariant world $\cal W$ is transformed to the
  ground state $|\text{vac}\rangle'$ of another scale-invariant 
  world $\cal W'$:
  \begin{equation}
  |\text{vac}\rangle \to |\text{vac}\rangle' =
  e^{iD\rho}|\text{vac}\rangle \ , \quad  x \to x' = e^{-\rho} x\,.
  \label{vac'}
  \end{equation}
  The same is true for all members of a complete set of states
  $\{|n\rangle\}$ for $\cal W$,
  \begin{equation}
  \{|n\rangle\} \to \{|n\rangle'\} \quad,\quad
  e^{iD\rho}|n\rangle = |n\rangle'\,,
  \end{equation}
  where $\{|n\rangle'\}$ span the state space of $\cal W'$. The well-known
  identity \cite{Mack69,Ell70}
  \begin{equation}
  e^{iD\rho} P^2 e^{-iD\rho} = e^{2\rho} P^2
  \end{equation}
  implies that if $|n\rangle$ has mass $\cal M$, the mass of
  $|n\rangle'$ is
  \begin{equation}
  {\cal M}' = e^\rho{\cal M} \,.
  \label{mass-scale}
  \end{equation}
  Clearly this applies generally: all dimensionally transmuted masses
  $\cal M$ associated with a given world $\cal W$ are scaled up or down to
  ${\cal M}'$ in the transformed world $\cal W'$.

  Identities such as Eq.~(\ref{mass-scale}) are often quoted as a
  reason for supposing that a scale- or conformal-invariant world
  must be entirely unphysical. How can there be a particle spectrum
  if for every massive particle, there is a continuum of particles
  \cite{Wess60} with the same quantum numbers except for their mass,
  which ranges from infinitesimal values to infinity? Must we conclude
  that we have an unparticle theory \cite{Geo07} with power-law branch
  cuts at zero-momentum thresholds, or that particles, if they exist,
  are necessarily massless, as in free-field theory?

  To answer these concerns, note that such arguments depend on an
  implicit assumption that the ground state is either unique or that,
  while it may exhibit a compact chiral degeneracy, it lacks the
  scaling degeneracy specified by Eq.~(\ref{vac'}). Scaling degeneracy
  changes the picture completely, because different members of each
  particle continuum belong to different worlds.

  Consider observers $O$ and $O'$ in their respective universes,
  $\cal W$ and $\cal W'$. Assume that these observers choose (say) natural units
  when making measurements. Since these units involve reference to a
  dimensionally transmuted mass, a scale transformation necessarily
  scales the units used in $\cal W$  to those in $\cal W'$, e.g.
  \begin{equation}
  \text{GeV} \to  \text{GeV} ' \,.
  \label{unit-scale}
  \end{equation}
  Since no observer is able to compare measurements in different
  worlds, all experimental data in one world would be exactly the same
  as in another world.  Therefore these worlds must be \emph{physically
  equivalent}, as is the case for any other symmetry with a vacuum
  degeneracy. Each observer could rely on the \emph{same} scale-invariant
  version of the PDG tables, i.e.\ with particle masses at $\alphaIR$
  differing slightly from those of our world $0 < \alpha < \alphaIR$,
  as described above.

  Crawling TC picks out one of these scale-degenerate vacua via tiny
  values of the parameter $\epsilon = \alphaIR - \alpha$ which cause
  scale invariance to be broken explicitly by the nonvanishing trace
  anomaly on $0 < \alpha < \alphaIR$.

  Clearly, conformal group theory fails for $\cal M$-dependent amplitudes
  at $\alphaIR$. However, it has always been accepted that
  symmetries of the Hamiltonian, whether hidden or approximate, become
  exact for coefficient functions in short-distance expansions.
  These rules were originally proposed \cite{KGW69} for approximate scale
  and chiral $SU(N_f)_L \times SU(N_f)_R$ symmetry (hidden in the chiral
  case), and later applied to hidden scale and conformal invariance
  \cite{RJC72}. Subsequently, asymptotic chiral invariance was derived
  from soft-pion identities \cite{Bernard75}. Here we extend this method
  to derive asymptotic scale and conformal invariance from the soft-dilaton
  theorems (\ref{softsigma}) and (\ref{softsigma3}).

  We begin with the coordinate-space version of the scaling identity
  (\ref{softsigma}) for a general operator product:
  \begin{align}
  F_\sigma\text{T}&\bigl\langle\sigma(q=0)\bigl|\mbox{$\prod\limits_m$}\,
  {\cal O}_m(x_m)\bigr|\text{vac}\bigr\rangle_c  \nn\\
  &= \mbox{\large $\sum\limits_\ell$}
  \bigl(d_\ell + x_\ell\ccdot\del_\ell\bigr)
  \text{T}\bigl\langle\text{vac}\bigl|\mbox{$\prod\limits_m$}\,
  {\cal O}_m(x_m)\bigr|\text{vac}\bigr\rangle_c \,.
  \label{softsigma4}
  \end{align}
In what follows, neighborhoods of coinciding points are excised to
avoid time-ordering ambiguities involving $\delta^4(x_m-x_{m'})$
and its derivatives, as in Ref.~\cite{KGW69}. For a subset $m \in S$ of the
operators ${\cal O}_m$ in Eq.~(\ref{softsigma4}), there is an
operator-product expansion 
\begin{equation}\label{OPE}
\mbox{\large $\prod\limits_{m \in S}$}{\cal O}_m(x_m) \sim
\mbox{\large $\sum\limits_n$}{\cal C}_n\bigl(\{x_{\ell\in S}\}\bigr){\cal O}_n(0)
\end{equation}
for the short-distance limit $x_{\ell \in S} \to 0$ with other coordinates
$x_{\ell \not\in S}$ held fixed at values $\not= 0$. When Eq.~(\ref{OPE}) is
inserted into each side of Eq.~(\ref{softsigma4}), the result is an 
equivalence between asymptotic expansions:
\begin{widetext}
\begin{align}\label{mess}
\mbox{\large $\sum\limits_n$}{\cal C}_n\bigl(\{x_{\ell\in S}\}\bigr)
F_\sigma\text{T}\bigl\langle&\sigma(q=0)\bigl|{\cal O}_n(0)
\mbox{$\prod\limits_{m \not\in S}$}
{\cal O}_m(x_m)\bigr|\text{vac}\bigr\rangle_c \nn \\
&\sim\ \mbox{\large $\sum\limits_n$}\mbox{\large $\sum\limits_{\ell \in S}$}
\bigl(d_\ell + x_\ell\ccdot\del_\ell\bigr)
{\cal C}_n\bigl(\{x_{\ell\in S}\}\bigr)
\text{T}\bigl\langle\text{vac}\bigl|{\cal O}_n(0)
\mbox{$\prod\limits_{m \not\in S}$}
{\cal O}_m(x_m)\bigr|\text{vac}\bigr\rangle_c
\nn \\
& \quad\ + \mbox{\large $\sum\limits_n$}{\cal C}_n\bigl(\{x_{\ell\in S}\}\bigr)
\mbox{\large $\sum\limits_{\ell \not\in S}$}
\bigl(d_\ell + x_\ell\ccdot\del_\ell\bigr)
\text{T}\bigl\langle\text{vac}\bigl|{\cal O}_n(0)
\mbox{$\prod\limits_{m \not\in S}$}
{\cal O}_m(x_m)\bigr|\text{vac}\bigr\rangle_c \,.
\end{align}
The soft-$\sigma$ amplitude on the left-hand side can be eliminated
via Eq.~(\ref{softsigma4})
\begin{align}
F_\sigma&\text{T}\bigl\langle\sigma(q=0)\bigl|{\cal O}_n(0)
\mbox{$\prod\limits_{m \not\in S}$}
{\cal O}_m(x_m)\bigr|\text{vac}\bigr\rangle_c
= \Bigl\{d_n + \mbox{$\sum\limits_{\ell \not\in S}$}
\bigl(d_\ell + x_\ell\ccdot\del_\ell\bigr)\Bigr\}
\text{T}\bigl\langle\text{vac}\bigl|{\cal O}_n(0)
\mbox{$\prod\limits_{m \not\in S}$}{\cal O}_m(x_m)\bigr|
\text{vac}\bigr\rangle_c \,,
\label{softsigma5}
\end{align}
with the result
\begin{align}
\mbox{\large $\sum\limits_n$}\Bigl\{- d_n + &\mbox{$\sum\limits_{\ell \in S}$}
\bigl(d_\ell + x_\ell\ccdot\del_\ell\bigr)\Bigr\}
{\cal C}_n\bigl(\{x_{\ell\in S}\}\bigr)
\text{T}\bigl\langle\text{vac}\bigl|{\cal O}_n(0)
\mbox{$\prod\limits_{m \not\in S}$}{\cal O}_m(x_m)\bigr|\text{vac}\bigr\rangle_c
\sim 0 \,.
\end{align}
\end{widetext}
Since $\mbox{$\prod_{m \not\in S}$}{\cal O}_m$ can be chosen at will,
this asymptotic ex\-pansion is valid only if all coefficients of
$\bigl\langle{\cal O}_n\mbox{$\prod_{m \not\in S}$}{\cal O}_m\bigr\rangle$
vanish:
\begin{equation}
\Bigl\{
- d_n + \mbox{$\sum\limits_{\ell \in S}$}
\bigl(d_\ell + x_\ell\ccdot\del_\ell\bigr)
\Bigr\}
{\cal C}_n\bigl(\{x_{\ell\in S}\}\bigr) = 0 \,.
\end{equation}
Therefore all coefficient functions ${\cal C}_n$ scale with dimension
$\mbox{$\sum_{\ell \in S}$}d_\ell - d_n$:
\begin{equation}\label{dimrule}
{\cal C}_n\bigl(\{\rho x_{\ell\in S}\}\bigr)
= \rho^{d_n - \mbox{\scriptsize $\sum_{\ell \in S}$}d_\ell}\,
{\cal C}_n\bigl(\{x_{\ell\in S}\}\bigr)\,.
\end{equation}
This is the same as Wilson's rule \cite{KGW69} for \emph{leading}
singularities in a theory of WW-mode scale invariance explicitly broken
by generalized mass terms such as \emph{current} fermion masses. The
difference is that our result is for \emph{all} coefficient functions
in a theory of exact scale invariance in the NG mode. All dependence on
dimensionally transmuted ``constituent'' masses arises from scale
condensates formed from vacuum amplitudes $\bigl\langle{\cal O}_n
\mbox{$\prod_{m \not\in S}$}{\cal O}_m\bigr\rangle_{\text{vac}}
\not= 0$.

The same procedure works for special conformal transformations. We give
details for operators $O_m$ of the type (\ref{commutators}) which commute
with $K_\mu$ at $x=0$:
\begin{align}\label{OPE2}
\mbox{\large $\prod\limits_{m \in S}$}O_m(x_m) \sim
\mbox{\large $\sum\limits_n$}&{\cal C}_n\bigl(\{x_{\ell\in S}\}\bigr)O_n(0) \nn \\
&+ \mbox{ operators} \not\in \{O_m\} \,.
\end{align}
Let the action of an infinitesimal Lorentz transformation on the
tensor or spinor indices of an operator $O_\ell$ inside a field product
be denoted as follows:
\begin{align}
{\Sigma_\ell}_{\mu\nu}
&\biggl\{\mbox{\large $\prod\limits_{m \in S}$}O_m(x_m)\biggr\} \nn \\
&\equiv
\biggl(\mbox{\large $\prod\limits_{m < \ell}$}O_m(x_m)\biggr)
{\Sigma_\ell}_{\mu\nu}O_\ell(x_\ell)
\biggl(\mbox{\large $\prod\limits_{n > \ell}$}O_n(x_n)\biggr) \,.
\end{align}
In coordinate space, the conformal soft-$\sigma$ theorem (\ref{softsigma3})
becomes
\begin{widetext}
\begin{align}
F_\sigma&\bigl(\langle\sigma(q)| - \langle\sigma(q')|\bigr)
\text{T}\Bigl\{\mbox{$\prod\limits_m$}\,O_m(x_m)\Bigr\}|
\text{vac}\rangle_c \nn \\
&= -\Ltextfrac{i}{2}(q' - q)^\mu
\sum_\ell\Bigl\{2{x_\ell}_\mu
\bigl(d_\ell + x_\ell\ccdot\del_\ell\bigr) - x_\ell^2{\del_\ell}_\mu
- 2ix_\ell^\rho{\Sigma_{\ell}}_{\mu\rho}\Bigr\}
\text{T}\bigl\langle\text{vac} \bigl|\mbox{$\prod\limits_m$}\,
O_m(x_m)\bigr|\text{vac}\bigr\rangle_c\ +\ O\bigl((q\text{ or }q')^2\bigr) \,.
\label{softsigma6}
\end{align}
When the expansion (\ref{OPE2}) is applied to both sides of
Eq.~(\ref{softsigma6}), the result for the coefficient function of
$O_n$ is
\begin{align}
&\sum_{\ell \in S}\Bigl\{2{x_\ell}_\mu
\bigl(d_\ell + x_\ell\ccdot\del_\ell\bigr) - x_\ell^2{\del_\ell}_\mu
- 2ix_\ell^\rho{\Sigma_{\ell}}_{\mu\rho}\Bigr\}
{\cal C}_n\bigl(\{x_{\ell\in S}\}\bigr) = 0 \,.
\end{align}
\end{widetext}

\noindent
Here we made use of the observation above Eq.~(\ref{softsigma2}) that
the four components of $q - q'$ can be chosen independently. Together
with Eq.~(\ref{dimrule}) and Poincar\'e symmetry, this shows that exact
conformal invariance can be applied to asymptotic coefficient
functions despite the $\cal M$ dependence of amplitudes outside the
short-distance region.

\section{Scaling dimension of the technigluon
 (field-strength operator)\texorpdfstring{$^{\bf 2}$}{sup2}}
\label{app:scaling}
This appendix concerns the origins and derivation of
formulas for the anomalous scaling function (\ref{gamma}) and
dimension $4 + \beta'$ of the trace anomaly. These formulas, derived
in Refs.~\cite{CT1,CT2}, were common private knowledge as far back as the
1970s. Only recently, we rediscovered the original version
\begin{align}
\bigl\{\mbox{scaling dimension}\bigr\}
 &= 4 + \bar{\beta}'\bigl(g_\infty\bigr)\,, \nn \\
\quad g_\infty &= g\, \mbox{ at fixed point}
\label{NKN}
\end{align}
below Eq.~(28) of Ref.~\cite{Niels77}, where $\bar{\beta}$ and $g$
are given by
\begin{equation}
\bar{\beta}(g) = \mu\frac{dg}{d\mu} \quad\mbox{and}\quad
\alpha = \frac{g^2}{4\pi}\,.
\end{equation}
For Eq.~(\ref{gamma}), see also Refs.~\cite{Spiri84} and
\cite{Grin89,Gubser08} (drawn to our attention by
R.~Zwicky and E.~Pallante, respectively).

First, let us review the derivation in Refs.~\cite{CT1,CT2} of Eq.~(\ref{gamma}),
simplified for the case of TC with massless fermions. Construct CS
equations for RG invariant amplitudes ${\cal A}$
\begin{equation}
\bigg\{ \mu\frac{\partial}{\partial \mu} + \beta(\alpha)
\frac{\partial}{\partial \alpha} \bigg\} {\cal A} = 0
\label{CS eqn}
\end{equation}
and apply the operator $\alpha\del/\del\alpha$:
\begin{equation}
\bigg\{ \mu\frac{\partial}{\partial \mu} + \beta(\alpha)
\frac{\partial}{\partial \alpha} + \beta'(\alpha)
- \frac{\beta(\alpha)}{\alpha} \bigg\}
\alpha\frac{\partial{\cal A}}{\partial\alpha} = 0\,.
\label{insert}
\end{equation}
Since  $\alpha\del{\cal A}\bigl/\del\alpha$ is proportional to the
amplitude ${\cal A}_{\hat{G}^2}$ where $\hat G^2$ is inserted at zero
momentum into $\cal A$, Eq.~(\ref{insert}) implies
\begin{equation}
\biggl\{\mu\frac{\partial}{\partial \mu}
+ \beta(\alpha) \frac{\partial}{\partial \alpha}
+ \gamma_{\hat{G}^2}(\alpha) \biggr\}{\cal A}_{\hat{G}^2} = 0 \,,
\end{equation}
where
\begin{equation}
\gamma_{\hat{G}^2}(\alpha) = \beta'(\alpha) -\frac{\beta(\alpha)}{\alpha}
\label{gamma again}
\end{equation}
is the anomalous scaling function (\ref{gamma}) in the form given
in Refs.~\cite{CT1,CT2} and confirmed in Eq.~\cite{Pallante17}.
[There is an incorrect factor of 2 in Eq.~(13) of \cite{Gubser08}.] This result
was given originally for massless QCD in the form \cite{Grin89}
\begin{equation}
\gamma_{G^2}(g) = g\frac{\del\ }{\del g}\biggl(\frac{\bar{\beta}(g)}{g}\biggr)
\end{equation}
which corresponds to Eq.~(\ref{gamma again}) because of the
fixed-$\mu$ identity
\begin{equation}
g\frac{\del\ }{\del g}\biggl(\frac{\bar{\beta}(g)}{g}\biggr)
= \alpha\frac{\del\ }{\del\alpha}\biggl(\frac{\beta(\alpha)}{\alpha}\biggr)\,.
\label{identity}
\end{equation}

The dynamical dimension of the technigluon operator in the IR limit
$\alpha \to \alphaIR$ is therefore
\begin{equation}\label{4+beta'}
d_{\hat G^2} = 4 + \gamma_{\hat G^2}(\alphaIR)
= 4 + \beta'(\alphaIR) \,.
\end{equation}
This corresponds to the rule $4 + \beta'$ found for QCD
\cite{CT1,CT2}.

The same rule holds for a UV fixed point $g_\infty$ or
$\alpha_\infty$, which was the context of the original
version (\ref{NKN}); the relation $\bar{\beta}'(g_\infty) =
\beta'(\alpha_\infty)$ is a consequence of Eq.~(\ref{identity}). In
the UV case, the plus sign in $d_{\hat G^2} = 4 + \beta'$ is crucial
\cite{Niels77} because $\beta'$ is negative: scaling corrections have
dimension $d_{\hat G^2} < 4$ and so do not upset the leading
short-distance behavior of operator product expansions.

For an IR fixed point, where $\beta' > 0$, such as in chiral-scale
perturbation theory or crawling TC, the expansion is at low energies,
so the dimension of scale-breaking terms can be either $d >4$
due to the trace anomaly or (if there are fermion mass terms)
$d_\text{mass} <4$. The main proviso is to ensure the condition
(mass)$^2 \geqslant 0$ for all particles appearing in the expansion.

\end{document}